%% file: main.tex
\documentclass[10pt,twocolumn,letterpaper]{article}

\usepackage[pagenumbers]{cvpr} % To force page numbers, e.g. for an arXiv version
\usepackage[accsupp]{axessibility}

\makeatletter
\@namedef{ver@everyshi.sty}{}
\makeatother 

\input{preamble}

\begin{document}

\title{Unraveling Normal Anatomy via Fluid-Driven Anomaly Randomization}

\author{Peirong Liu\textsuperscript{1} \quad Ana Lawry Aguila\textsuperscript{1} \quad Juan E. Iglesias\textsuperscript{1,2,3}
\vspace{0.3cm} \\ 
\textsuperscript{1}Harvard Medical School and Massachusetts General Hospital \quad
\textsuperscript{2}UCL \quad \textsuperscript{3}MIT
}

\twocolumn[{%
\renewcommand\twocolumn[1][]{#1}%
\maketitle
\begin{center}
    \input{sec/showcase}

\end{center}%
}]

%%%%%%%%%%%%%%%%%%%%%%%%%

\input{sec/abstract}
\input{sec/intro}

\input{sec/related_work}
\input{sec/method/fluid} 
\input{sec/method/trainer} 
\input{sec/exp/main}

\input{sec/con}

%\subsubsection*{Acknowledgments}

%%%%%%%%%%%%%%%%%%%%%%%%%
%\clearpage

{\small
\bibliographystyle{ieee_fullname}
\bibliography{reference}
}
%%%%%%%%%%%%%%%%%%%%%%%%%

\input{sec/appendix/main}

%%%%%%%%%%%%%%%%%%%%%%%%%

\end{document}

%% file: preamble.tex
\usepackage{booktabs}

\usepackage{graphicx}
\usepackage{amsmath}
\usepackage{amsthm}
\usepackage{amssymb} 
\usepackage{pifont}% http://ctan.org/pkg/pifont
\newcommand{\xmark}{\ding{55}}%

\usepackage{wrapfig}

\usepackage[dvipsnames]{xcolor}
\definecolor{cvprblue}{rgb}{0.21,0.49,0.74}
\usepackage[pagebackref,breaklinks,colorlinks,citecolor=cvprblue]{hyperref} 

\usepackage[capitalize]{cleveref}
\crefname{section}{Sec.}{Secs.}
\Crefname{section}{Section}{Sections}
\Crefname{table}{Table}{Tables}
\crefname{table}{Tab.}{Tabs.}

\def\confName{CVPR}
\def\confYear{2025}
\def\paperID{***} % *** Enter the CVPR Paper ID here

\usepackage{xcolor}
\usepackage{tikz}
\usepackage{changepage}
\usepackage{tikz-3dplot}
\usepackage{tikzscale}
\usetikzlibrary{plotmarks}
\usepackage{pgfplots}
\usetikzlibrary{fadings}
\usetikzlibrary{tikzmark,calc} 
\usetikzlibrary{shapes.arrows}
\usetikzlibrary{decorations.pathreplacing, decorations.markings}
\usepgfplotslibrary{colorbrewer}
\usepgfmodule{nonlineartransformations} 
\makeatletter
\def\latticetilt{%
\pgf@xa=\pgf@x%
\pgf@ya=\pgf@y%
\pgfmathsetmacro{\myx}{\pgf@xa+\pgfkeysvalueof{/tikz/lattice/amplitude}*sin((\pgf@ya/\pgfkeysvalueof{/tikz/lattice/spacing})*360/\pgfkeysvalueof{/tikz/lattice/superlattice period})}%
\pgf@x=\myx pt%
\pgfmathsetmacro{\myy}{\pgf@ya+\pgfkeysvalueof{/tikz/lattice/amplitude}*sin((\pgf@xa/\pgfkeysvalueof{/tikz/lattice/spacing})*360/\pgfkeysvalueof{/tikz/lattice/superlattice period})}%
\pgf@y=\myy pt}

\usepackage[capbesideposition=right]{floatrow}

\xdefinecolor{navyblue}{RGB}{19 41 75}%{0.14510 0.29020 0.64706} 
\xdefinecolor{carolinablue}{RGB}{123 175 212}
\xdefinecolor{cb}{RGB}{123 175 212}
\xdefinecolor{flatgrey}{RGB}{162 170 173}
\xdefinecolor{matcha}{RGB}{183 224 176}

\definecolor{mygreen}{HTML}{DCEDC8}
\definecolor{lightgrey}{HTML}{E1E1E1}
\definecolor{darkblue}{HTML}{00275D}
\definecolor{mint}{HTML}{99EDC3}

\xdefinecolor{myyellow}{HTML}{FFE082}
\xdefinecolor{myblue}{RGB}{181, 222, 255}
\xdefinecolor{mypurple}{RGB}{202, 184, 255}
\xdefinecolor{myorange}{RGB}{255, 171, 145}
\xdefinecolor{tomato}{RGB}{255, 99, 71}

\xdefinecolor{mgh}{RGB}{0 139 176}
\xdefinecolor{harvardcrimson}{RGB}{165 28 48}
\xdefinecolor{hc}{RGB}{165 28 48} 
\xdefinecolor{harvardred}{RGB}{237 27 52}
\xdefinecolor{hr}{RGB}{237 27 52}
\xdefinecolor{harvardsalmon}{RGB}{236 143 156}
\xdefinecolor{hs}{RGB}{165 28 48}
\xdefinecolor{harvardblack}{RGB}{30 30 30}
\xdefinecolor{hblack}{RGB}{30 30 30}
\xdefinecolor{harvardgrey}{RGB}{147 161 173}
\xdefinecolor{hgrey}{RGB}{147 161 173}
\xdefinecolor{harvardgreen}{RGB}{77 184 72}
\xdefinecolor{hgreen}{RGB}{77 184 72}
\xdefinecolor{harvardlime}{RGB}{203 219 42}
\xdefinecolor{hlime}{RGB}{203 219 42}
\xdefinecolor{harvardblue}{RGB}{78 136 199}
\xdefinecolor{hblue}{RGB}{78 136 199}
\xdefinecolor{harvardsky}{RGB}{149 181 223}
\xdefinecolor{hsky}{RGB}{149 181 223}
\xdefinecolor{harvardwarmyellow}{RGB}{252 179 21}
\xdefinecolor{hwy}{RGB}{252 179 21}
\xdefinecolor{harvardyellow}{RGB}{255 222 45}
\xdefinecolor{hy}{RGB}{255 222 45}
\xdefinecolor{harvardturq}{RGB}{0 170 173}
\xdefinecolor{ht}{RGB}{0 170 173}
\xdefinecolor{harvardaqua}{RGB}{80 196 221}
\xdefinecolor{ha}{RGB}{80 196 221}
\xdefinecolor{harvardpurple}{RGB}{148 110 183}
\xdefinecolor{hp}{RGB}{148 110 183}
\xdefinecolor{harvardlavender}{RGB}{187 137 202}
\xdefinecolor{hl}{RGB}{187 137 202}

\newcommand{\dashedLayer}[6]{
			\def\a{#1} % Used to distinguish input resolution for current layer.
			\def\b{0.02}
			\def\c{#2} % Width of the cube to distinguish number of input channels for current layer.
			\def\t{#3} % X offset for current layer.
			\def\d{#4} % Y offset for current layer.

			\draw[line width=0.15mm, dash pattern=on 2.25pt off 1.8pt](\c+\t,0,\d) -- (\c+\t,\a,\d) -- (\t,\a,\d);                                                      % back plane
			\draw[line width=0.15mm, dash pattern=on 2.25pt off 1.8pt](\t,0,\a+\d) -- (\c+\t,0,\a+\d) node[midway,below] {#6} -- (\c+\t,\a,\a+\d) -- (\t,\a,\a+\d) -- (\t,0,\a+\d); % front plane 
			\draw[line width=0.15mm, dash pattern=on 2.25pt off 1.8pt](\c+\t,0,\d) -- (\c+\t,0,\a+\d);
			\draw[line width=0.15mm, dash pattern=on 2.25pt off 1.8pt](\c+\t,\a,\d) -- (\c+\t,\a,\a+\d);
			\draw[line width=0.15mm, dash pattern=on 2.25pt off 1.8pt](\t,\a,\d) -- (\t,\a,\a+\d);
			
			\draw[line width=0.15mm] (\c+\t,0,\d) -- (\c+\t,\a,\d);
			\draw[line width=0.15mm] (\c+\t,0,\d) -- (\c+\t,0,\a+\d);
			\draw[line width=0.15mm] (\c+\t,\a,\d) -- (\c+\t,\a,\a+\d);
			\draw[line width=0.15mm] (\c+\t,0,\a+\d) -- (\c+\t,\a,\a+\d);
			
			\filldraw[#5] (\t+\b,\b,\a+\d) -- (\c+\t-\b,\b,\a+\d) -- (\c+\t-\b,\a-\b,\a+\d) -- (\t+\b,\a-\b,\a+\d) -- (\t+\b,\b,\a+\d); % front plane
			\filldraw[#5] (\t+\b,\a,\a-\b+\d) -- (\c+\t-\b,\a,\a-\b+\d) -- (\c+\t-\b,\a,\b+\d) -- (\t+\b,\a,\b+\d);
			\ifthenelse {\equal{#5} {}}
			{} % Do not draw colored slice if #4 is blank.
			{\filldraw[#5] (\c+\t,\b,\a-\b+\d) -- (\c+\t,\b,\b+\d) -- (\c+\t,\a-\b,\b+\d) -- (\c+\t,\a-\b,\a-\b+\d);} % Else, draw a colored slice.
		}

\newcommand{\networkLayer}[6]{
			\def\a{#1} % Used to distinguish input resolution for current layer.
			\def\b{0.02}
			\def\c{#2} % Width of the cube to distinguish number of input channels for current layer.
			\def\t{#3} % X offset for current layer.
			\def\d{#4} % Y offset for current layer.

			\draw[line width=0.15mm](\c+\t,0,\d) -- (\c+\t,\a,\d) -- (\t,\a,\d);                                                      % back plane
			\draw[line width=0.15mm](\t,0,\a+\d) -- (\c+\t,0,\a+\d) node[midway,below] {#6} -- (\c+\t,\a,\a+\d) -- (\t,\a,\a+\d) -- (\t,0,\a+\d); % front plane
			\draw[line width=0.15mm](\c+\t,0,\d) -- (\c+\t,0,\a+\d);
			\draw[line width=0.15mm](\c+\t,\a,\d) -- (\c+\t,\a,\a+\d);
			\draw[line width=0.15mm](\t,\a,\d) -- (\t,\a,\a+\d);

			\filldraw[#5] (\t+\b,\b,\a+\d) -- (\c+\t-\b,\b,\a+\d) -- (\c+\t-\b,\a-\b,\a+\d) -- (\t+\b,\a-\b,\a+\d) -- (\t+\b,\b,\a+\d); % front plane
			\filldraw[#5] (\t+\b,\a,\a-\b+\d) -- (\c+\t-\b,\a,\a-\b+\d) -- (\c+\t-\b,\a,\b+\d) -- (\t+\b,\a,\b+\d);
			\ifthenelse {\equal{#5} {}}
			{} % Do not draw colored slice if #4 is blank.
			{\filldraw[#5] (\c+\t,\b,\a-\b+\d) -- (\c+\t,\b,\b+\d) -- (\c+\t,\a-\b,\b+\d) -- (\c+\t,\a-\b,\a-\b+\d);} % Else, draw a colored slice.
		}
		
\usepackage{listings}[language=Python, caption=Python example]

\definecolor{codegreen}{rgb}{0,0.6,0}
\definecolor{codegray}{rgb}{0.5,0.5,0.5}
\definecolor{codepurple}{rgb}{0.58,0,0.82}
\definecolor{backcolour}{rgb}{0.95,0.95,0.92}

\lstdefinestyle{mystyle}{
    backgroundcolor=\color{white},   
    commentstyle=\color{codegreen},
    keywordstyle=\color{magenta},
    numberstyle=\tiny\color{codegray},
    stringstyle=\color{codepurple},
    basicstyle=\ttfamily\footnotesize,
    breakatwhitespace=false,         
    breaklines=true,                 
    captionpos=b,                    
    keepspaces=true,                 
    numbers=left,                    
    numbersep=5pt,                  
    showspaces=false,                
    showstringspaces=false,
    showtabs=false,                  
    tabsize=2
}

\usepackage[linesnumbered,ruled,vlined]{algorithm2e}

\SetCommentSty{mycommfont}
\SetKwInput{Input}{Input}                
\SetKwInput{Output}{Output}         
\SetKwInput{Dataset}{Dataset}     
\SetKwInput{Parameters}{Parameters}  
\SetKwInput{Settings}{Settings}      
\SetKwInput{Initialization}{Initialization}     

\SetAlCapNameFnt{\small}
\SetAlCapFnt{\small}
\SetAlgorithmName{Algorithm}{Algorithm}{List of Algorithms} 
\makeatletter
\newcommand{\algorithmfootnote}[2][\footnotesize]{%
  \let\old@algocf@finish\@algocf@finish% Store algorithm finish macro
  \def\@algocf@finish{\old@algocf@finish% Update finish macro to insert "footnote"
    \leavevmode\rlap{\begin{minipage}{\linewidth}
    #1#2
    \end{minipage}}%
  }%
}
\makeatother

\usepackage{floatrow}
\usepackage{caption} 
\usepackage{subcaption}
\usepackage{makecell}
\usepackage{booktabs}
\usepackage{multirow}
\newcolumntype{P}[1]{>{\centering\arraybackslash}p{#1}}

\usepackage{hyperref} 
  
\newtheoremstyle{bfnote}%
  {}{}
  {\itshape}{}
  {\bfseries}{.}
  { }{\thmname{#1}\thmnumber{ #2}\thmnote{ (#3)}}
\theoremstyle{bfnote}

\newtheorem*{notation*}{Notation}

\usepackage{url}

\definecolor{olive}{rgb}{0.42, 0.56, 0.14}

\pgfplotsset{compat=1.18}

%% file: sec/showcase.tex
%\vspace*{-0.7cm}
\centering 

\captionsetup{type=figure}

\resizebox{\linewidth}{!}{
	\begin{tikzpicture}
		\tikzstyle{myarrows}=[line width=0.8mm,draw=blue!50,-triangle 45,postaction={draw, line width=0.05mm, shorten >=0.02mm, -}]
		\tikzstyle{mylines}=[line width=0.8mm]
  
		% Main size box
		%\draw[thin, color = black] (20, -8) -- (-2, -8) -- (-2, 3.42) -- (20, 3.42) -- (20, -8); 

	\pgfmathsetmacro{\cubex}{0.5*3}
	\pgfmathsetmacro{\cubey}{0.5*3} 
	\pgfmathsetmacro{\cubez}{0.12}

	%%%%%%%%%%%%%%%%%%%%%%%%%%%%%%%%%%%%%%%%%%%%%%%%%

\draw[dashed, color = hwy, fill = hwy!5, line width=0.5mm] (-3.15, 4) -- (11, 4) -- (11, -4) -- (-3.15, -4) -- (-3.15, 4); 

        \draw[line width = 2.pt, color = hwy] (5.4, 2) -- (7.6, 2);
        \draw[line width = 2.pt, ->, color = hwy] (6.5, 2) -- (6.5, 0.5) -- (10, 0.5) -- (10, 1.2);

        \draw[line width = 2.pt, color = hwy] (5.4, -2) -- (7.6, -2);
        \draw[line width = 2.pt, ->, color = hwy] (6.5, -2) -- (6.5, -3.5) -- (10, -3.5) -- (10, -2.8);

\foreach \i/\j in {-0.25/3.75, -0.25/1.75, 2.55/2.75, 4.35/2.75, 6.15/2.75, 8.35/2.75, 10.75/2.75,       -0.25/-0.25, -0.25/-2.25, 2.55/-1.25, 4.35/-1.25, 6.15/-1.25, 8.35/-1.25, 10.75/-1.25}
{
\draw[black,fill=gray!30, line width = 0.02mm] (\i, \j, 0) -- ++(-\cubex,0,0) -- ++(0,-\cubey,0) -- ++(\cubex,0,0) -- cycle;
\draw[black,fill=gray!35, line width = 0.02mm] (\i, \j, 0) -- ++(0,0,-\cubez) -- ++(0,-\cubey,0) -- ++(0,0,\cubez) -- cycle;
\draw[black,fill=gray!35, line width = 0.02mm] (\i, \j, 0) -- ++(-\cubex,0,0) -- ++(0,0,-\cubez) -- ++(\cubex,0,0) -- cycle;
 }

\node at (-3.15+8.5, 4-0.6) {\Large\color{hwy}\textbf{Fluid-Driven Anomaly Randomization}};
\node[right] at (-3.15+3.5+0.8, -0.2-0.075) {\textbf{\textit{Initial condition:}}};
\node[right] at (-3.15+6.25+0.8, 0.-0.075) {\textit{(i)} a realistic pathology annotation map};
\node[right] at (-3.15+6.25+0.8, -0.5-0.075) {\textit{(ii)} a random map generated from Perlin noise};

    \node at (-1, 3) {\includegraphics[width=1.5cm]{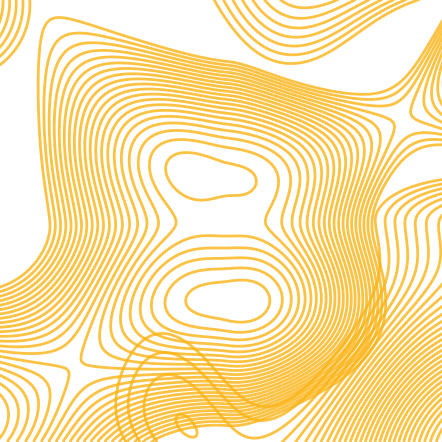}};
    \node at (-1, 3-2) {\includegraphics[width=1.5cm]{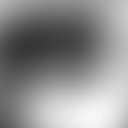}};

        \draw[line width = 2.pt, ->, color = hwy] (0.3, 2) -- (0.9, 2);
        \draw [decorate,decoration={brace,amplitude=5pt,raise=6ex},line width=2.pt,color = hwy] (-0.85, 3) -- (-0.85, 1);
        
    \node at (1.8, 2) {\includegraphics[width=1.5cm]{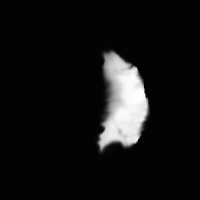}};
	\node at (3.6, 2) {\includegraphics[width=1.5cm]{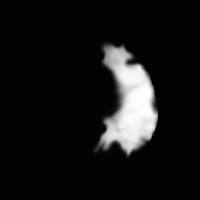}};
	\node at (5.4, 2) {\includegraphics[width=1.5cm]{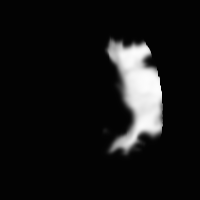}};
	\node at (7.6, 2) {\includegraphics[width=1.5cm]{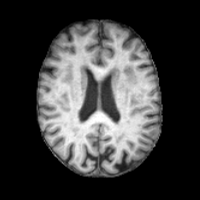}};
	\node at (10, 2) {\includegraphics[width=1.5cm]{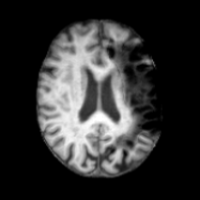}};  
	
	\node at (-2.75, 2)  {{\normalsize{\textit{(i)}}}};
	\node at (-2.25, 3)  {{\normalsize{{$\mathbf{V}$}}}};
	\node at (-2.25, 1)  {{\normalsize{$D$}}};
 
	\node at (1.8, 0.9)  {{\normalsize{$t = 0$}}};
	\node at (3.6, 0.9)  {{\normalsize{$t = 5$}}};
	\node at (5.4, 0.9)  {{\normalsize{$t = 10$}}};
	\node at (7.6, 0.9)  {{\normalsize{$I_0$}}};
	\node at (9.6, 0.9) {{\normalsize{$I$}}};

	\node at (-1, -1) {\includegraphics[width=1.5cm]{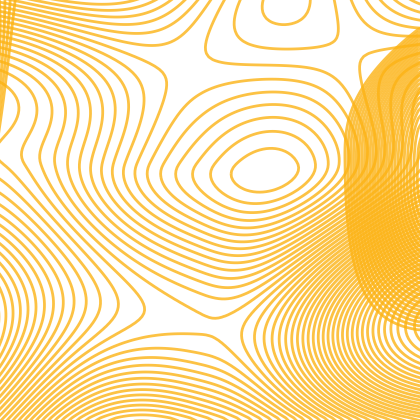}};
	\node at (-1, -3) {\includegraphics[width=1.5cm]{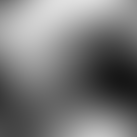}};

        \draw[line width = 2.pt, ->, color = hwy] (0.3, -2) -- (0.9, -2);
        \draw [decorate,decoration={brace,amplitude=5pt,raise=6ex},line width=2.pt,color = hwy] (-0.85, -1) -- (-0.85, -3);

	\node at (1.8, -2) {\includegraphics[width=1.5cm]{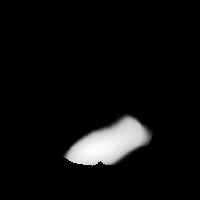}};
	\node at (3.6, -2) {\includegraphics[width=1.5cm]{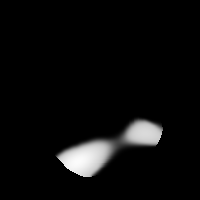}};
	\node at (5.4, -2) {\includegraphics[width=1.5cm]{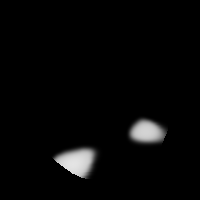}};
	\node at (7.6, -2) {\includegraphics[width=1.5cm]{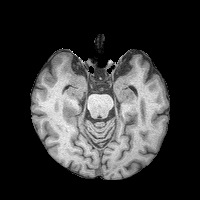}};
	\node at (10, -2) {\includegraphics[width=1.5cm]{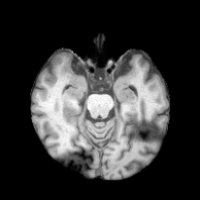}};  
	
	\node at (-2.75, -2)  {{\normalsize{\textit{(ii)}}}};
	\node at (-2.25, -1)  {{\normalsize{{$\mathbf{V}$}}}};
	\node at (-2.25, -3)  {{\normalsize{$D$}}};
 
	\node at (1.8, 0.9-4)  {{\normalsize{$t = 0$}}};
	\node at (3.6, 0.9-4)  {{\normalsize{$t = 5$}}};
	\node at (5.4, 0.9-4)  {{\normalsize{$t = 10$}}};
	\node at (7.6, 0.9-4)  {{\normalsize{$I_0$}}};
	\node at (9.6, 0.9-4) {{\normalsize{$I$}}};

	%%%%%%%%%%%%%%%%%%%%%%%%%%%%%%%%%%%%%%%%%%%%%%%%%
	%%%%%%%%%%%%%%   Diseased to Healthy   %%%%%%%%%%
	%%%%%%%%%%%%%%%%%%%%%%%%%%%%%%%%%%%%%%%%%%%%%%%%%

	%%%%%%%%%%%%%%%%%  Boxes  %%%%%%%%%%%%%%%%%%%%%%

        \pgfmathsetmacro{\dw}{6}
        
        \pgfmathsetmacro{\dx}{14.45}
        \pgfmathsetmacro{\dy}{-1.5}

\draw[dashed, color = hc, fill = hc!5, line width=0.5mm] (-3.15+\dx, 4) -- (-3.15+\dx+\dw, 4) -- (-3.15+\dx+\dw, -4) -- (-3.15+\dx, -4) -- (-3.15+\dx, 4);
	\node at (-2.5+\dx, 2+0.2)  {{\normalsize{T1w}}};
	\node at (-2.5+\dx, 2-0.2)  {{\normalsize{MRI}}};
	\node at (-2.5+\dx, 2+\dy+0.2-0.2)  {{\normalsize{T2w}}};
	\node at (-2.5+\dx, 2+\dy-0.2-0.2)  {{\normalsize{MRI}}};
	\node at (-2.5+\dx, 2+2*\dy+0.2-0.4)  {{\normalsize{FLAIR}}};
	\node at (-2.5+\dx, 2+2*\dy-0.2-0.4)  {{\normalsize{MRI}}};
	\node at (-2.5+\dx, 2+3*\dy-0.6)  {{\normalsize{CT}}};

        \node at (-3.15+\dx+0.5*\dw+0.3, 4-0.6) {\Large\color{hc}\textbf{Diseased} $\rightarrow$ \textbf{Healthy}};

        \node at (-3.15+\dx+0.5*\dw+0.4, 2) {\Large\color{hc}$\rightarrow$};
        \node at (-3.15+\dx+0.5*\dw+0.4, 2+\dy-0.2) {\Large\color{hc}$\rightarrow$};
        \node at (-3.15+\dx+0.5*\dw+0.4, 2+2*\dy-0.4) {\Large\color{hc}$\rightarrow$};
        \node at (-3.15+\dx+0.5*\dw+0.4, 2+3*\dy-0.6) {\Large\color{hc}$\rightarrow$};

        \pgfmathsetmacro{\dx}{14.5+6.25}
\draw[dashed, color = hblue, fill = hblue!5, line width=0.5mm] (-3.15+\dx, 4) -- (-3.15+\dx+\dw, 4) -- (-3.15+\dx+\dw, -4) -- (-3.15+\dx, -4) -- (-3.15+\dx, 4);

	\node at (-2.5+\dx, 2+0.2)  {{\normalsize{T1w}}};
	\node at (-2.5+\dx, 2-0.2)  {{\normalsize{MRI}}};
	\node at (-2.5+\dx, 2+\dy+0.2-0.2)  {{\normalsize{T2w}}};
	\node at (-2.5+\dx, 2+\dy-0.2-0.2)  {{\normalsize{MRI}}};
	\node at (-2.5+\dx, 2+2*\dy+0.2-0.4)  {{\normalsize{FLAIR}}};
	\node at (-2.5+\dx, 2+2*\dy-0.2-0.4)  {{\normalsize{MRI}}};
	\node at (-2.5+\dx, 2+3*\dy-0.6)  {{\normalsize{CT}}};

        \node at (-3.15+\dx+0.5*\dw+0.4, 4-0.6) {\Large\color{hblue}\textbf{Healthy} $\leftrightarrow$ \textbf{Healthy}};

        \node at (-3.15+\dx+0.5*\dw+0.4, 2) {\Large\color{hblue}$\leftrightarrow$};
        \node at (-3.15+\dx+0.5*\dw+0.4, 2+\dy-0.2) {\Large\color{hblue}$\leftrightarrow$};
        \node at (-3.15+\dx+0.5*\dw+0.4, 2+2*\dy-0.4) {\Large\color{hblue}$\leftrightarrow$};
        \node at (-3.15+\dx+0.5*\dw+0.4, 2+3*\dy-0.6) {\Large\color{hblue}$\leftrightarrow$};

	%%%%%%%%%%%%%%%%%%%%%%%%%%%%%%%%%%%%%%%%%%%%%%%%%

        \pgfmathsetmacro{\dx}{10.45}
        \pgfmathsetmacro{\dy}{-1.7}

\foreach \i/\j in {3.65+\dx/2.75, 4.75+\dx+1.65/2.75, 3.65+\dx/2.75+\dy, 4.75+\dx+1.65/2.75+\dy, 3.65+\dx/2.75+2*\dy, 4.75+\dx+1.65/2.75+2*\dy, 3.65+\dx/2.75+3*\dy, 4.75+\dx+1.65/2.75+3*\dy}
{ 
\draw[black,fill=gray!30, line width = 0.02mm] (\i, \j, 0) -- ++(-\cubex,0,0) -- ++(0,-\cubey,0) -- ++(\cubex,0,0) -- cycle;
\draw[black,fill=gray!35, line width = 0.02mm] (\i, \j, 0) -- ++(0,0,-\cubez) -- ++(0,-\cubey,0) -- ++(0,0,\cubez) -- cycle;
\draw[black,fill=gray!35, line width = 0.02mm] (\i, \j, 0) -- ++(-\cubex,0,0) -- ++(0,0,-\cubez) -- ++(\cubex,0,0) -- cycle;

        \pgfmathsetmacro{\x}{10.+3.35} 
        \pgfmathsetmacro{\dx}{2.75} 
	\node at (\x, 2) {\includegraphics[width=1.5cm]{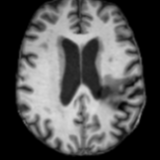}};  
	\node at (\x+\dx, 2) {\includegraphics[width=1.5cm]{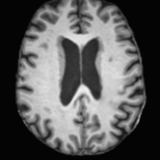}};  

	\node at (\x, 2+\dy) {\includegraphics[width=1.5cm]{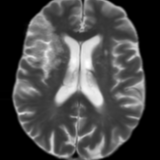}};  
	\node at (\x+\dx, 2+\dy) {\includegraphics[width=1.5cm]{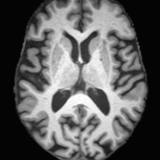}};  

	\node at (\x, 2+2*\dy) {\includegraphics[width=1.5cm]{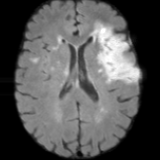}};  
	\node at (\x+\dx, 2+2*\dy) {\includegraphics[width=1.5cm]{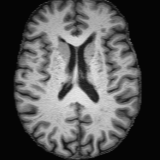}};  

	\node at (\x, 2+3*\dy) {\includegraphics[width=1.5cm]{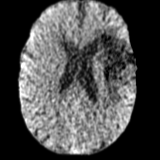}};  
	\node at (\x+\dx, 2+3*\dy) {\includegraphics[width=1.5cm]{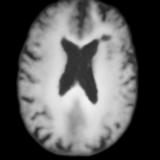}};

 }

	%%%%%%%%%%%%%%%%%%%%%%%%%%%%%%%%%%%%%%%%%%%%%%%%%
	%%%%%%%%%%%%%%   Healthy to Healthy   %%%%%%%%%%
	%%%%%%%%%%%%%%%%%%%%%%%%%%%%%%%%%%%%%%%%%%%%%%%%%

        \pgfmathsetmacro{\dx}{16.75} 
        
\foreach \i/\j in {3.65+\dx/2.75, 4.75+\dx+1.65/2.75, 3.65+\dx/2.75+\dy, 4.75+\dx+1.65/2.75+\dy, 3.65+\dx/2.75+2*\dy, 4.75+\dx+1.65/2.75+2*\dy, 3.65+\dx/2.75+3*\dy, 4.75+\dx+1.65/2.75+3*\dy}
{ 
\draw[black,fill=gray!30, line width = 0.02mm] (\i, \j, 0) -- ++(-\cubex,0,0) -- ++(0,-\cubey,0) -- ++(\cubex,0,0) -- cycle;
\draw[black,fill=gray!35, line width = 0.02mm] (\i, \j, 0) -- ++(0,0,-\cubez) -- ++(0,-\cubey,0) -- ++(0,0,\cubez) -- cycle;
\draw[black,fill=gray!35, line width = 0.02mm] (\i, \j, 0) -- ++(-\cubex,0,0) -- ++(0,0,-\cubez) -- ++(\cubex,0,0) -- cycle;
 }

        \pgfmathsetmacro{\x}{14.+5.65} 
        \pgfmathsetmacro{\dx}{2.75} 
	\node at (\x, 2) {\includegraphics[width=1.5cm]{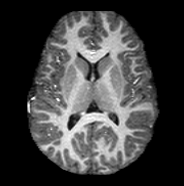}};  
	\node at (\x+\dx, 2) {\includegraphics[width=1.5cm]{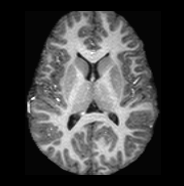}};  

	\node at (\x, 2+\dy) {\includegraphics[width=1.5cm]{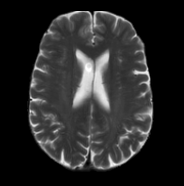}};  
	\node at (\x+\dx, 2+\dy) {\includegraphics[width=1.5cm]{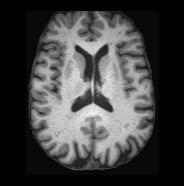}};  

	\node at (\x, 2+2*\dy) {\includegraphics[width=1.5cm]{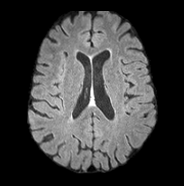}};  
	\node at (\x+\dx, 2+2*\dy) {\includegraphics[width=1.5cm]{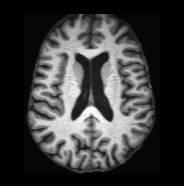}};  

	\node at (\x, 2+3*\dy) {\includegraphics[width=1.5cm]{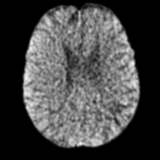}};  
	\node at (\x+\dx, 2+3*\dy) {\includegraphics[width=1.5cm]{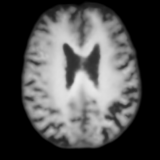}};  
        
	%%%%%%%%%%%%%%%%%%%%%%%%%%%%%%%%%%%%%%%%%%%%%%%%%

	\end{tikzpicture}
	}
	\vspace{-0.6cm}
\caption{Powered by the proposed fluid-driven anomaly randomization, \texttt{UNA} can handle a range of pathological patterns without requiring paired pathology annotations for training. \textit{(i)} By bridging the gap between healthy and diseased anatomy, \texttt{UNA} enables the use of general analysis models for images containing pathology; \textit{(ii)} By reconstructing anatomy in a modality-agnostic manner, \texttt{UNA} facilitates analysis with standard tools designed for high-resolution, healthy T1w MRI.} 
	%\vspace*{0.1cm}  
	 \label{showcase}

%% file: sec/abstract.tex
%\vspace*{-0.45cm}
\begin{abstract}
\vspace*{-0.35cm}
Data-driven machine learning has made significant strides in medical image analysis. However, most existing methods are tailored to specific modalities and assume a particular resolution (often isotropic). This limits their generalizability in clinical settings, where variations in scan appearance arise from differences in sequence parameters, resolution, and orientation. Furthermore, most general-purpose models are designed for healthy subjects and suffer from performance degradation when pathology is present.
We introduce \texttt{UNA} (Unraveling Normal Anatomy), the first modality-agnostic learning approach for normal brain anatomy reconstruction that can handle both healthy scans and cases with pathology.
We propose a fluid-driven anomaly randomization method that generates an unlimited number of realistic pathology profiles on-the-fly. \texttt{UNA} is trained on a combination of synthetic and real data, and can be applied directly to real images with potential pathology without the need for fine-tuning. We demonstrate \texttt{UNA}'s effectiveness in reconstructing healthy brain anatomy and showcase its direct application to anomaly detection, using both simulated and real images from 3D healthy and stroke datasets, including CT and MRI scans. By bridging the gap between healthy and diseased images, \texttt{UNA} enables the use of general-purpose models on diseased images, opening up new opportunities for large-scale analysis of uncurated clinical images in the presence of pathology. 
Code is available at \href{https://github.com/peirong26/UNA}{https://github.com/peirong26/UNA}.
%\footnote{Code is in \href{https://anonymous.4open.science/r/UNA/README.md}{this anonymous repository} and will be publicly available.}
\end{abstract}

%I would emphasise either here or in the abstract the benefits of healthy image reconstruction: for disease --> healthy, often image processing pipelines wont work on images with pathology/healthy reconstructions allow for anomaly detection, and healthy --> healthy allows for generation of missing modalities (great as theres often limited/missing data in medical datasets)

%% file: sec/intro.tex
\vspace{-0.2cm}
\section{Introduction}
\label{sec: intro}  
\vspace{-0.1cm}

%Magnetic resonance imaging (MRI) enables in vivo noninvasive imaging of the human brain with exquisite and tunable soft-tissue contrast~\cite{BrantZawadzki1992MPRA}. 

%I would emphasise either here or in the abstract the benefits of healthy image reconstruction: for disease --> healthy, often image processing pipelines wont work on images with pathology/healthy reconstructions allow for anomaly detection, and healthy --> healthy allows for generation of missing modalities (great as theres often limited/missing data in medical datasets)

Recent machine learning based methods have significantly advanced the speed and accuracy of brain image analysis tasks, such as image segmentation~\cite{Ronneberger2015UNetCN, Kamnitsas2016EfficientM3,Milletar2016VNetFC,Ding_2021_ICCV}, registration~\cite{Balakrishnan2018VoxelMorphAL,Yang2017QuicksilverFP,de2019deep}, and super-resolution~\cite{Tian2020ImprovingIV,Tanno2020UncertaintyMI}. Human brain imaging \emph{in vivo} is primarily dominated by Computed Tomography (CT) and Magnetic Resonance Imaging (MRI)\cite{hussain2022modern}. CT is faster and preferred in emergency cases, while MRI provides superior contrast for soft tissues such as the brain. Unlike CT, which is a standardized modality that produces quantitative measurements in Hounsfield units, MRI is generally not calibrated and can generate a wide range of imaging contrasts (e.g., T1w, T2w, FLAIR) to visualize different tissues and abnormalities. This diversity in contrast and the lack of standardization complicate the quantitative analysis of MRI scans. As a result, most existing MRI analysis methods are contrast-specific and often suffer from performance degradation when voxel size or MRI contrast differs between training and testing datasets\cite{wang2018deep}. This limits the generalizability of machine learning models and leads to redundant data collection and training efforts for new datasets. 
Recent contrast-agnostic models that leverage synthetic data~\cite{Iglesias2020JointSA, Iglesias2023SynthSRAP, Billot2021SynthSegSO, Hoffmann2020SynthMorphLC, Liu2024BrainID} have demonstrated impressive results, significantly extending their applicability to diverse clinical acquisition protocols. However, these models are primarily designed for analyzing \textit{healthy} brain anatomy and typically struggle to produce reliable results in the presence of extensive abnormalities~(\cref{fig: synth,fig: real}).

To the best of our knowledge, the recently proposed \texttt{PEPSI}~\cite{liu2024pepsi} is the only contrast-agnostic brain MRI analysis method that is compatible with extensive pathology. \texttt{PEPSI} leverages synthetic data to estimate T1w and FLAIR MRI from input scans containing pathology. 
%This approach exhibits promising performance in handling images with abnormalities. 
However, it has several limitations: 
\textit{(i)}~It relies on \textit{paired} pathology segmentation map associated with each brain anatomy during training, which limits its application to datasets that provide pathology annotations;
\textit{(ii)}~It requires access to \textit{pre-trained} pathology segmentation models to compute the implicit pathology segmentation loss; 
%\textit{(iii)}~\texttt{PEPSI} brightens abnormalities in its FLAIR image synthesis, which may \textit{misrepresent} certain abnormal tissue (e.g., necrotic tissue) and potentially \textit{mislead} clinicians~(\cref{fig: real}). 
and \textit{(iii)}~It requires additional \textit{fine-tuning} to detect anomalies.

%I would emphasize either here or in the abstract the benefits of healthy image reconstruction: for disease --> healthy, often image processing pipelines won't work on images with pathology/healthy reconstructions allow for anomaly detection, and healthy --> healthy allows for the generation of missing modalities (great as there's often limited/missing data in medical datasets)

Here, we introduce \texttt{UNA}, the first modality-agnostic learning method for \textbf{U}nraveling \textbf{N}ormal \textbf{A}natomy. \texttt{UNA} leverages the power of synthetic data, and can be applied to real images (CT \textit{and} MRI) of both healthy and diseased populations, \textit{without} the need for fine-tuning~(\cref{showcase}).
%[COMMENT: maybe you should say here what the ouputs of UNA are?]
\begin{itemize}
    \item[1)] We propose fluid-driven anomaly randomization~(\cref{sec: fluid}) to overcome the scarcity of pathology segmentation annotations. Using only limited existing pathology segmentations as initial conditions, our fluid-driven anomaly generator generates \textit{unlimited} new pathology profiles on-the-fly through advection-diffusion partial differential equations (PDEs). This formulation offers a continuous and controllable trajectory for pathology evolution and also naturally enforces \textit{realistic constraints} on brain abnormalities through boundary conditions~(\cref{showcase} (left)).
    \item[2)] We introduce a modality-agnostic learning framework to reconstruct healthy brain anatomy from images with potential pathology~(\cref{sec: framework}). Our framework leverages symmetry priors of brain anatomy and incorporates subject-specific anatomical features from contralateral healthy tissue in a self-contrastive learning fashion.
    \item[3)] We extensively evaluate the healthy anatomy reconstruction performance of \texttt{UNA} on simulated and real images with stroke lesions, in both CT and different MR contrasts (T1w, T2w, and FLAIR) (\cref{exp: simulation,exp: real}). We further demonstrate the direct application of \texttt{UNA} to anomaly detection, without fine-tuning (\cref{exp: anomaly_detection}). \texttt{UNA} achieves state-of-the-art performance in all tasks and modalities.
\end{itemize}

\noindent By bridging the gap between healthy and diseased anatomy \texttt{UNA} enables the use of general-purpose models for images containing pathology, unlocking the tremendous potential for analyzing clinical images with pathology.

%% file: sec/related_work.tex
%\vspace{-0.2cm}

\section{Related Work}
\label{sec: related_work}

%\vspace{-0.1 cm}
\paragraph{Foundation Models in Medical Imaging.}
\label{related: foundation_models} 
Large-scale datasets in medical imaging require significantly more effort to compile than those in natural imaging or language due to varying acquisition protocols and privacy requirements across institutions. Consequently, medical foundation models are not as well developed as their natural image counterparts. There have been, nevertheless, some notable efforts. SAM-Med3D-MoE~\cite{wang2024sam} provides a 3D foundation model for medical image segmentation, trained on 22,000 scans. The MONAI~\cite{monai} project includes a model zoo with pre-trained models, which are highly task-specific and sensitive to particular image contrasts. Zhou et al.~\cite{Zhou2023AFM} constructed a medical foundation model designed for detecting eye and systemic health conditions from retinal scans. Still, it only functions with color fundus photography and optical coherence tomography modalities. 
Recently, generalist biomedical AI systems, e.g., GMAI~\cite{Moor2023FoundationMF} and Med-PaLM M~\cite{Singhal2022LargeLM, Tu2023TowardsGB}, have demonstrated significant potential in biomedical tasks within a vision-language context, including visual question answering, image classification, and radiology report generation. However, they have not tackled more complex dense 3D prediction tasks such as reconstruction, segmentation, and registration.

\vspace{-0.4cm}
%\vspace{-0.5cm}
\paragraph{Contrast-Agnostic Learning for MRI.}
\label{related: contrast_agnostic} 
MRI scans acquired across sites vary substantially in appearance due to differences in contrast, resolution, and orientation.  This heterogeneity leads to duplicate training efforts for approaches that are sensitive to specific MR contrast. Classical approaches in brain segmentation used Bayesian inference for contrast robustness~\cite{Leemput2003AUF, Fischl2002WholeBS}, but require long processing times and struggle with resolutions that are not high and isotropic~\cite{Puonti2016FastAS,Iglesias2023SynthSRAP}. SynthSeg~\cite{Billot2021SynthSegSO,billot2023robust} achieves contrast- and resolution-agnostic segmentation with a synthetic generator that simulates widely diverse contrasts and resolutions. The same generator has been used to achieve contrast invariance in tasks like image registration~\cite{Hoffmann2020SynthMorphLC,dey2024learning}, super-resolution~\cite{Iglesias2020JointSA}, or skull stripping~\cite{Hoopes2022SynthStripSF}. Brain-ID~\cite{Liu2024BrainID} explored contrast-agnostic feature representations that generalize across various fundamental medical image analysis tasks, including image synthesis, segmentation, and super-resolution. However, all these general-purpose methods are either trained exclusively on healthy anatomical labels, or require paired anatomy-pathology annotations, which limits their application primarily to healthy subjects or every specific pathology (e.g., white matter lesions) -- as opposed to previously unseen pathology profiles~(\cref{fig: synth,fig: real}).

%%%%%%%%%%%%%%%%%%%%%%%%%%%%%%%%

\vspace{-0.4cm}
\paragraph{Fluid-Based Dynamics Modeling.}
\label{related: fluid_based} 

Fluid dynamics is a fundamental topic in physics and plays a crucial role in various real-world applications such as weather forecasting, airflow analysis~\cite{emmanuel2018}, optical flow~\cite{sun2018pwc,teed2020raft}, image registration~\cite{shen2021accurate,yang2017quicksilver,tian2020fluid}, and perfusion analysis~\cite{liu2021piano}. In fluid dynamics, advection-diffusion PDEs are commonly employed to describe the fluid transport processes. 
Liu et al.~\cite{liu2021yeti} introduced regularization-free representations to ensure the compressibility and positive semi-definiteness of estimated velocity and diffusion fields. Franz et al.~\cite{franz2023learning} simulated 3D density and velocity fields from single-view data without 3D supervision. Xing et al.~\cite{xing2024helmfluid} proposed to learn the velocity field from past physical observations using Helmholtz dynamics, eliminating the need for ground truth velocity. 
In these studies, the inverse problem of velocity estimation provides interpretable insights for predicting future fluid behavior. We build upon the concept of fluid flow simulation and frame anomaly pattern randomization as a \textit{forward} process of advection-diffusion PDEs. This formulation naturally enables us to ensure that simulated anomaly outcomes are well posed, through controllable velocity fields and established boundary conditions~(\cref{setup}). 

%% file: sec/method/fluid.tex
%\vspace{-0.2cm}
\section{Fluid-Driven Anomaly Randomization}
\label{sec: fluid}
\vspace{-0.1cm}

%\subsection{Motivation and Problem Setup}
%\label{sec: setup}

%\input{sec/method/fig/isles_case}

%%%%%%%%%%%%%%%%%%%%%%%%%%%%%%

%\paragraph{Motivation}~
Manually annotating pathology to create gold-standard segmentation is extremely costly, particularly for 3D medical images. This process not only requires specialized expertise from clinicians, but is also highly time-consuming and not reproducible. Consequently, large-scale datasets with gold-standard pathology annotations are almost inexistent (BraTS~\cite{menze2014multimodal} being a notable exception).
%is impractical s\footnote{Some 3D datasets exist for specific diseases and modalities, such as BraTS~\cite{menze2014multimodal}. However, the image quality varies greatly and most of the scans are of large slice spacing.}
%, demanding significantly more effort than is required for natural images. 
In addition, discrepancies often arise among the gold-standard pathology segmentation maps provided by different datasets. %For instance, \cref{fig: isles_example} shows a FLAIR image from the ISLES~\cite{Hernandez2022ISLES} stroke dataset. While this FLAIR image indicates white matter hyperintensities (WMH) (circled in red), the corresponding gold-standard pathology segmentation \textit{only} annotates the areas affected by stroke lesions. %Given a limited number of available pathology segmentation maps collected from datasets with potential discrepancies, 
To address these issues, we seek to design an anomaly randomization approach that is:

\begin{itemize}
\item[\textit{i.}] \textit{Expressive}: the generated anomaly profiles should exhibit diverse and expressive shapes and intensities that sufficiently reflect the variety of pathological appearances encountered in clinical practice.

\item[\textit{ii.}] \textit{Realistic}: the randomized abnormalities must conform to realistic constraints. For example, abnormalities in white matter should not appear in other tissue structures, brain tumors should be localized within the brain region.
\end{itemize}

\noindent To achieve these two aims, we propose randomizing unlimited, diverse anomaly profiles by formulating the generation as a forward mass transport process, with realistic constraints naturally guaranteed by boundary conditions. 
Our anomaly randomization consists of three steps (Alg.~\ref{alg: fluid}): \textit{(i)}~Initializations of random anomaly ($P_0$), velocity (${\mathbf{V}}$), and diffusion ($D$) for anomaly transport; \textit{(ii)}~Forward transport of abnormal intensities for random time steps; \textit{(iii)}~Appearance encoding of the generated anomaly on healthy images of any modality. \cref{sec: pde} below describes the generation of abnormal profiles (\textit{i-ii}), and \cref{sec: anomaly_encode} introduces the encoding of abnormalities on healthy images (\textit{iii}).

%%%%%%%%%%%%%%%%%%%%%%%%%%%%%%

\subsection{Anomaly Profile Randomization}
\label{sec: pde}
%\subsection{Background and Overview}
%\label{sec: pde}

%%%%%%%%%%%%%%%%%%%%%%%%%%%%%% 
%\input{sec/method/fig/demo_anomaly} 
\input{sec/method/fig/pseudocode}

%%%%%%%%%%%%%%%%%%%%%%%%%%%%%%

%You could further / more clearly motivate the advection-diffusion approach / fluid-driven approach as follows.
%1. You want to augment realistic pathological maps Dpatol.
%2. You want to go beyond random deformation as that doesn't really change the "relative" shape of the pathology with respect to the surrounding tissue.
%3. Inspired by lesion growth models (eg "Image guided personalization of reaction-diffusion type tumor growth models using modified anisotropic eikonal equations"~\cite{konukoglu2009image} but also many other papers), you propose this fluid-driven approach that mimics realistic lesion growth (and maybe also shrinkage?)

\paragraph{Background.}\label{background}~Advection-diffusion PDEs describe a large family of fluid dynamics processes, e.g., heat conduction, wind dynamics, and blood flow~\cite{emmanuel2018,liu2021piano,xing2024helmfluid}. In general, the advection term refers to the mass transport driven by fluid flow, while the diffusion term refers to the gradient of mass concentration. Inspired by the advection-diffusion process, which computes the natural progression of mass intensities, we propose to randomize an unlimited variety of anomaly profiles by formulating the generation as a \textit{forward} advection-diffusion, starting from either a single realistic pathology annotation map or a random shape.

\vspace{-0.35cm}
\paragraph{Problem Setup.}\label{setup}~Let $P({\mathbf{x}},\, t)$ denote the pathology probability at location ${\mathbf{x}}$ in a bounded domain of interest $\Omega\subset \mathbb{R}^3$ (e.g., brain), at time $t$. The local pathology probability changes of an anomaly randomization process are described by the advection-diffusion PDE:
\vspace{-0.15cm}
\begin{equation}
\frac{\partial P({\mathbf{x}},\,t)}{\partial t} = \underbrace{-  \nabla \left({\mathbf{V}}({\mathbf{x}}) \cdot P({\mathbf{x}},\,t) \right)}_{\text{Flow}} + \underbrace{\nabla \cdot \left(D({\mathbf{x}})\, \nabla P({\mathbf{x}},\,t)\right)}_{\text{Diffusion}},
\label{eq: full_adv_diff} 
\vspace{-0.75 cm}
\end{equation}
\begin{equation}
s.t. ~ \underbrace{P({\mathbf{x}}, \, 0) = P_0({\mathbf{x}})}_{\text{Initial Condition}},\, \underbrace{\frac{P({\mathbf{x}}, \, t)}{\partial \mathbf{n}}\bigg|_{\partial \Omega_p} = 0}_{\text{Zero-Neumann}},\, t \leq T_{\text{max}} \,,
\label{eq: constraint}
\vspace{-0.15cm}
\end{equation}
where $t$ ($T_{\text{max}}$) refers to the (maximum) time steps used for the generation of new anomaly profiles. The spatially varying velocity field ${\mathbf{V}}(\mathbf{x}) \in \mathbb{R}^3$ and diffusion scalar field $D(\mathbf{x}) \in \mathbb{R}$ govern the advection and diffusion process of an initial anomaly, $P_0({\mathbf{x}})$. The zero Neumann boundary condition ensures that the randomization process of $P_0$ satisfies pre-assumed bounds of the anomaly developing regions. To ensure that the dynamics of anomaly changes are well posed, we impose the incompressible flow and non-negative diffusion constraints on ${\mathbf{V}}$ and $D$~\cite{liu2021yeti}, and rewrite the advection-diffusion process in~\cref{eq: full_adv_diff} as:
\vspace{-0.13cm}
\begin{equation}
\begin{aligned}
\frac{\partial P({\mathbf{x}},\, t)}{\partial t} &= - {\mathbf{V}}({\mathbf{x}})\cdot\nabla P({\mathbf{x}},\, t) + \nabla \cdot \left({{D}}({\mathbf{x}})\, \nabla P({\mathbf{x}}, \, 
 t)\right) \vspace*{-0.2 cm} \\ 
&= \underbrace{- \nabla \times \boldsymbol{\Psi}({\mathbf{x}})\cdot\nabla P({\mathbf{x}}, t)}_{\text{Incompressible Flow}} + \underbrace{\nabla \cdot \left({\Phi}^2({\mathbf{x}})\, \nabla P({\mathbf{x}}, t)\right)}_{\text{Non-Negative Diffusion}},  
\label{eq: divfree_adv_diff} 
\end{aligned} 
\end{equation}
where $\boldsymbol{\Psi} \in L^3(\Omega)^{3}$ and ${\Phi} \in \mathbb{R}^+(\Omega)$ refer to the potential fields for representing ${\mathbf{V}}$ and $D$, respectively, such that the resulting flow and diffusion will be incompressible and non-negative \textit{by construction}. %Such representations are subjective; the complete proof is provided in Appendix~\hyperlink{app: repre}{A}.

%%%%%%%%%%%%%%%%%%%%%%%%%%%%%%

%%%%%%%%%%%%%%%%%%%%%%%%%%%%%%
\input{sec/method/fig/fw}

%\input{sec/method/fig/fw_draft} 
%%%%%%%%%%%%%%%%%%%%%%%%%%%%%%

\vspace{-0.35cm}
\paragraph{Initializations of $P_0$, ${\mathbf{V}}$, $D$.}\label{initialization}~To enrich the diversity of abnormal profiles, we initialize the anomaly ($P_0$ in \cref{eq: constraint}) from two sources: 
{\textit{(i)}}~Publicly available pathology annotations from the ATLAS~\cite{Liew2017ATLAS} and ISLES~\cite{Hernandez2022ISLES} stroke datasets, which include high-quality gold-standard segmentation of stroke lesions. %Beginning with real pathology segmentation provides us with better clinical reality. 
{\textit{(ii)}}~Random shapes using randomly thresholded Perlin noise, a widely used procedural generation algorithm known for creating rich textures. We further generate random Perlin noise for creating random potentials $\boldsymbol{\Psi}$ for ${\mathbf{V}}$, and $\Phi$ for $D$. %Details are provided in Appendix~\hyperlink{app: perlin}{B}.

\vspace{-0.35cm}
\paragraph{Forward Scheme.}\label{forward}~We employ a first-order upwind scheme~\cite{leveque2002} to approximate the differential operators associated with the advection term, and a nested central-forward-backward difference scheme for the diffusion term in \cref{eq: divfree_adv_diff}. Discretizing the spatial derivatives leads to a system of ordinary differential equations that can be solved with numerical integration. To enhance numerical stability and ensure compliance with the Courant-Friedrichs-Lewy (CFL) condition \cite{gottlieb2000ssp,leveque2002}, we apply the RK45 method for adaptive time-stepping ($\delta t$) in advancing to $P^{t+\delta t}$. %Detailed information on the numerical discretization and stability considerations can be found in Appendix~\hyperlink{app: numericals}{C}.

As shown in \cref{showcase} (left), we can generate infinite variations from a single pathology profile via the introduced fluid-driven anomaly transport, while naturally satisfying boundary conditions imposed by the brain contour. %Below, we will introduce how \texttt{UNA} incorporates the generated pathology profiles into healthy anatomies of \textit{randomized} modality. 

%%%%%%%%%%%%%%%%%%%%%%%%%%%%%%

\subsection{Anomaly Apprearance Randomization}
\label{sec: anomaly_encode}
%\vspace{-0.1cm}

%%%%%%%%%%%%%%%%%%%%%%%%%%%%%%
%\input{sec/method/fig/fw_gen} 
%%%%%%%%%%%%%%%%%%%%%%%%%%%%%%

As mentioned in \cref{related: foundation_models}, large-scale annotation of 3D medical imaging data requires tremendous effort. \texttt{UNA} is instead trained on a combination of synthetic and real images (many of them labeled automatically). Specifically, we encode the generated pathology profiles, $P$, into normal anatomy of healthy control scans, enabling the generation of diverse images with \textit{random} modalities, each exhibiting a distinct appearance introduced by $P$.

\vspace{-0.35cm}
\paragraph{Random Modality Generation.}\label{modality}~To generate healthy images with complex structural details, we first leverage domain randomization~\cite{Liu2024BrainID} to synthesize images of random modality and resolution with healthy anatomy (\cref{fw}~(left)). Specifically, we randomly sample intensities on 3D neuroanatomical segmentation (label maps $L$), where the intensities are conditioned on the label at each location: 
\vspace{-0.2cm}
\begin{equation} 
    \begin{cases} 
        I_0({\mathbf{x}}) \sim \mathcal{N}(\mu_l,\, \sigma_l ) \,, \quad l \in L\,,\\ \mu_l \sim \mathcal{U}(0,\, 1 \, \vert\, \theta_{\mu}, \, \theta_{l}) \,,~
        \sigma_l \sim \mathcal{U}(0,\, 1 \, \vert \, \theta_{\sigma}, \, \theta_{l}) \,,
    \end{cases}
    \label{eq: contrast} 
    \vspace{-0.2cm}
\end{equation}
where $\mu_l$ and $\sigma_l$ refer to the mean and variance of the uniform distribution of each label $l$. $\theta_{l},\, \theta_{\mu},\, \theta_{\sigma} \in \Theta$ control the shifts and scales. A random deformation field is then generated for augmentation purposes, comprising linear and non-linear transformations~\cite{Iglesias2020JointSA,Liu2024BrainID}.

%%%%%%%%%%%%%%%%%%%%

\vspace{-0.4cm}
\paragraph{Anomaly Profile Encoding.}\label{encode}~We encode the random anomaly profiles from \cref{sec: pde} into the generated healthy anatomy $I_0$, based on \textit{a priori} knowledge on the white and gray matter intensities of $I_0$~\cite{liu2024pepsi,laso2024quantifying}:
\vspace{-0.15 cm}
\begin{equation}  
I({\mathbf{x}}) = I_0({\mathbf{x}}) + \Delta I({\mathbf{x}}) * P({\mathbf{x}})\,,
\label{eq: pathol_encode} 
\end{equation}
\vspace{-0.85cm}
\begin{equation}  
s.t. ~ \Delta I({\mathbf{x}}) \sim    
\begin{cases} 
    \{ 0 \} \,, & x \notin \Omega_{P}\\ 
    \mathcal{N}(-\mu_{\text{w}} / 2 ,\, \mu_{\text{w}} / 2 ) \,,& x \in \Omega_{P}\,, \mu_{\text{w}} > \mu_{\text{g}}\\
    \mathcal{N}(\mu_{\text{w}} / 2 ,\, \mu_{\text{w}} / 2 ) \,, & x \in \Omega_{P}\,, \mu_{\text{w}} \leq \mu_{\text{g}} \\ 
\end{cases}
\label{eq: pathol_condition} 
\end{equation}
$\mu_{\text{w}}$ ($\mu_{\text{g}}$) is the mean of $I_0$'s white (gray) matter intensities. A higher $\mu_{\text{w}}$ resembles T1w, where pathology appears darker, while a lower $\mu_{\text{w}}$ resembles T2w/FLAIR, where pathology is typically brighter. Considering extreme scenarios, 
% such as dead tissue which appears dark in all modalities, 
we randomly assign the sign of $\Delta I({\mathbf{x}})$ 20\% of the time. 
$I$ further undergoes a standard augmentation pipeline~\cite{Iglesias2023SynthSRAP}, introducing partial voluming~\cite{Billot2021SynthSegSO} and various resolutions, noise, scanning artifacts commonly found in clinical practice.

%%%%%%%%%%%%%%%%%%%%%%%%%%%%%%

%% file: sec/method/fig/pseudocode.tex
\begin{algorithm}[t]
\caption{Fluid-Driven Anomaly Randomization}
\label{alg: fluid}
{\small

  %\Output{(a) Estimated velocity field ${\bf{V}}$ and diffusion field ${\bf{D}}$ \\ ~~~~~~~~~~~~ (b) Predicted CA concentration maps after $t_0$ with time interval $\delta t$}
%\Settings{$\Omega,\, \Omega_p,\, \mathbf{n},\, T_{\text{max}} $ in \cref{eq: constraint}}
\Dataset{Healthy images with anatomy labels ($\mathbb{D}_{\text{Synth}}$); Gold standard pathology annotations ($\mathbb{D}_{\text{Pathol}})$}
%\Initialization{$\Gamma_1({\bf{x}}),\, \Gamma_2({\bf{x}}),\, L({\bf{x}}) \sim 0.001\times\mathcal{N}(0,\,1),\quad \forall {\bf{x}} \in \Omega$}
\Settings{$\Omega,\, \Omega_p,\, \mathbf{n},\, T_{\text{max}} $ in \cref{eq: constraint}; $\theta_{l},\, \theta_{\mu},\, \theta_{\sigma}$ in \cref{eq: contrast}}

\Input{Anatomy label $L$, or, real image $I_0$}

\Output{Image ($I$) which is encoded with the randomized pathology profile ($P$)}

\tcc{Initialization}
Randomly select $P_0 \in \mathbb{D}_{\text{Pathol}}$\\
Randomly select label $L$ or image $I$ $\in \mathbb{D}_{\text{Synth}}$ \\

\tcc{Fluid-Driven Forward Randomization}
Randomly sample potential fields $\boldsymbol{\Psi}$ and $\Phi$ in \cref{eq: divfree_adv_diff} \\
\While{$t \leq T_{\text{max}}$}
   {
   Randomly pick anomaly transport time $T \leq T_{\text{max}}$ \\  %\tcp*{Add comment} %,\quad i = 0,\, 1,\, \ldots,\, T - T_\text{pd}
   Reconstruct $\mathbf{V}$ and $D$ via \cref{eq: divfree_adv_diff} \\
   Compute forward scheme via \cref{eq: full_adv_diff,eq: constraint}
   }
Obtain randomized $P = P(\mathbf{x},\, T_{\text{max}})$ \\

\tcc{Random Modality Generation}
\If{$L$ as input}{
    Synthesize random modality $I_0$ via \cref{eq: contrast} \\ % \tcp*{}
}
%\Else{
%    Jump to Anomaly Appearance Encoding 
%}

\tcc{Anomaly Profile Encoding}
Encode randomized $P$ into $I_0$ via \cref{eq: pathol_encode,eq: pathol_condition} 
}
\end{algorithm}

%% file: sec/method/fig/fw.tex
\begin{figure*}[t]
\centering 
\resizebox{\linewidth}{!}{
	\begin{tikzpicture}[lattice/.cd,spacing/.initial=4,superlattice
  period/.initial=12,amplitude/.initial=2]
		\tikzstyle{myarrows}=[line width=0.8mm,draw=blue!50,-triangle 45,postaction={draw, line width=0.05mm, shorten >=0.02mm, -}]
		\tikzstyle{mylines}=[line width=0.8mm]

	\pgfmathsetmacro{\cubex}{0.5*3}
	\pgfmathsetmacro{\cubey}{0.5*3} 
	\pgfmathsetmacro{\cubez}{0.12}

	% Main size box
	%\draw[thin, color = black] (-10, -5.5) -- (18, -5.5) -- (18, 3) -- (-10, 3) -- (-10, -5.5); 
  
	\draw[fill = hwy!5, draw = hwy!5, line width=0.6mm] (-10.75, -5.75) -- (4., -5.75) -- (4., 3) -- (-10.75, 3) -- (-10.75, -5.75);
	\draw[fill = hblue!5, draw = hblue!5, line width=0.6mm] (4.05, -5.75) -- (4.05, 3) -- (17.75, 3) -- (17.75, -5.75) -- (4.05, -5.75); 
 
        \node[right] at (0.65, 2.5) {\large\textbf{\textit{Fluid-Driven}}};
        \node[right] at (0.65, 1.9) {\large\textbf{\textit{Anomaly}}};
        \node[right] at (0.65, 1.3) {\large\textbf{\textit{Randomization}}};
  
        \node[right] at (4.35, 2.5) {\large\textbf{\textit{Modality-Agnostic Learning of Healthy Anatomy Beyond Gold Standards}}};
        
        \node[left] at (4.-0.1, -5.25) {\large\textbf{\cref{sec: fluid}}}; 
        \node[left] at (17.75-0.1, -5.25) {\large\textbf{\cref{sec: framework}}};

	%\draw[color = hblue!5, line width=1.6mm] (4., -5.79) -- (4., 3);

        %%%%%%% \Phi %%%%%%% 
 
	\pgfmathsetmacro{\x}{-8.5}
	\pgfmathsetmacro{\y}{2.5}
 
	\draw[black,fill=gray!35, line width = 0.02mm] (\x,\y,0.1*3) -- ++(-\cubex,0,0) -- ++(0,-\cubey,0) -- ++(\cubex,0,0) -- cycle; 
	\draw[black,fill=gray!35, line width = 0.02mm] (\x,\y,0.1*3) -- ++(0,0,-\cubez) -- ++(0,-\cubey,0) -- ++(0,0,\cubez) -- cycle;
	\draw[black,fill=gray!35, line width = 0.02mm] (\x,\y,0.1*3) -- ++(-\cubex,0,0) -- ++(0,0,-\cubez) -- ++(\cubex,0,0) -- cycle;
 
        \node at (\x-0.5*\cubex,\y-1.2*\cubey,0.1*3) {$\boldsymbol{\Psi}$};
        \node at (\x-0.5*\cubex,\y-0.5*\cubey,0.1*3) {\includegraphics[width=1.5cm]{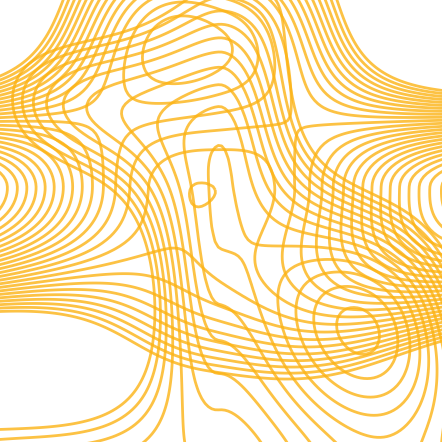}};

        \node at (\x+0.15 + 0.5*\cubex, \y-0.55*\cubey + 0.45) {\cref{eq: divfree_adv_diff}};
        \node at (\x+0.15 + 0.5*\cubex, \y-0.55*\cubey - 0.45) {\cref{initialization}};
        \draw [myarrows, color = hwy!50](\x+0.3, \y-0.55*\cubey) -- (\x + \cubex, \y-0.55*\cubey);

        %%%%%%% Bounding box %%%%%%% 

	\pgfmathsetmacro{\x}{-8.5+2.35*\cubex} 
 
	\draw[dashed, color = hwy!50, line width=0.6mm] (\x-1.2*\cubex, \y-1.45*\cubey) -- (\x+1.35*\cubex, \y-1.45*\cubey) -- (\x+1.35*\cubex, \y+0.15*\cubey) -- (\x-1.2*\cubex, \y+0.15*\cubey) -- (\x-1.2*\cubex, \y-1.45*\cubey);

        %%%%%%% V %%%%%%% 

	\draw[black,fill=gray!35, line width = 0.02mm] (\x,\y,0.1*3) -- ++(-\cubex,0,0) -- ++(0,-\cubey,0) -- ++(\cubex,0,0) -- cycle; 
	\draw[black,fill=gray!35, line width = 0.02mm] (\x,\y,0.1*3) -- ++(0,0,-\cubez) -- ++(0,-\cubey,0) -- ++(0,0,\cubez) -- cycle;
	\draw[black,fill=gray!35, line width = 0.02mm] (\x,\y,0.1*3) -- ++(-\cubex,0,0) -- ++(0,0,-\cubez) -- ++(\cubex,0,0) -- cycle;

        \node at (\x-0.5*\cubex,\y-1.2*\cubey,0.1*3) {$\mathbf{V}$};
        \node at (\x-0.5*\cubex,\y-0.5*\cubey,0.1*3) {\includegraphics[width=1.5cm]{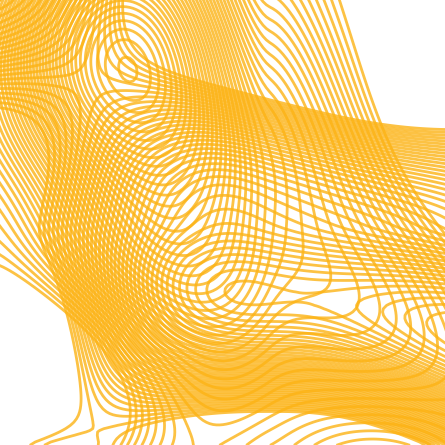}};
        
        %%%%%%%  D %%%%%%% 
        
	\pgfmathsetmacro{\x}{-8.5+2.4*\cubex+1.2*\cubex} 
 
	\draw[black,fill=gray!35, line width = 0.02mm] (\x,\y,0.1*3) -- ++(-\cubex,0,0) -- ++(0,-\cubey,0) -- ++(\cubex,0,0) -- cycle; 
	\draw[black,fill=gray!35, line width = 0.02mm] (\x,\y,0.1*3) -- ++(0,0,-\cubez) -- ++(0,-\cubey,0) -- ++(0,0,\cubez) -- cycle;
	\draw[black,fill=gray!35, line width = 0.02mm] (\x,\y,0.1*3) -- ++(-\cubex,0,0) -- ++(0,0,-\cubez) -- ++(\cubex,0,0) -- cycle;

        \node at (\x-0.5*\cubex,\y-1.2*\cubey,0.1*3) {$D$};
        \node at (\x-0.5*\cubex,\y-0.5*\cubey,0.1*3) {\includegraphics[width=1.5cm]{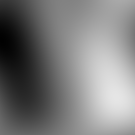}};

        %%%%%%% Phi %%%%%%% 

        \node at (\x+0.2 + 0.5*\cubex + 0.05, \y-0.55*\cubey + 0.45) {\cref{eq: divfree_adv_diff}};
        \node at (\x+0.2 + 0.5*\cubex + 0.05, \y-0.55*\cubey - 0.45) {\cref{initialization}};
        \draw [myarrows, color = hwy!50](\x +\cubex + 0.1, \y-0.55*\cubey) -- (\x + 0.4, \y-0.55*\cubey);

	\pgfmathsetmacro{\x}{-8.5+2.4*\cubex+1.2*\cubex + 2.35*\cubex} 

	\draw[black,fill=gray!35, line width = 0.02mm] (\x,\y,0.1*3) -- ++(-\cubex,0,0) -- ++(0,-\cubey,0) -- ++(\cubex,0,0) -- cycle; 
	\draw[black,fill=gray!35, line width = 0.02mm] (\x,\y,0.1*3) -- ++(0,0,-\cubez) -- ++(0,-\cubey,0) -- ++(0,0,\cubez) -- cycle;
	\draw[black,fill=gray!35, line width = 0.02mm] (\x,\y,0.1*3) -- ++(-\cubex,0,0) -- ++(0,0,-\cubez) -- ++(\cubex,0,0) -- cycle;
  
        \node at (\x-0.5*\cubex,\y-1.2*\cubey,0.1*3) {$\Phi$};
        \node at (\x-0.5*\cubex,\y-0.5*\cubey,0.1*3) {\includegraphics[width=1.5cm]{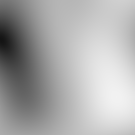}};

        \draw [->, color = hwy!50, line width = 0.1cm](\x - 3.5*\cubex, \y-1.43*\cubey) -- (\x - 3.5*\cubex, \y-1.85*\cubey);

        %%%%%%% PDE %%%%%%% 

        %%%%%%% P0 %%%%%%% 
        
	\pgfmathsetmacro{\x}{-8.5} 
	\pgfmathsetmacro{\y}{2.5-2*\cubey} 
 
	\draw[black,fill=gray!35, line width = 0.02mm] (\x,\y,0.1*3) -- ++(-\cubex,0,0) -- ++(0,-\cubey,0) -- ++(\cubex,0,0) -- cycle; 
	\draw[black,fill=gray!35, line width = 0.02mm] (\x,\y,0.1*3) -- ++(0,0,-\cubez) -- ++(0,-\cubey,0) -- ++(0,0,\cubez) -- cycle;
	\draw[black,fill=gray!35, line width = 0.02mm] (\x,\y,0.1*3) -- ++(-\cubex,0,0) -- ++(0,0,-\cubez) -- ++(\cubex,0,0) -- cycle;

        \node at (\x-0.5*\cubex,\y-1.2*\cubey,0.1*3) {$P_0$};
        \node at (\x-0.5*\cubex,\y-0.5*\cubey,0.1*3) {\includegraphics[width=1.5cm]{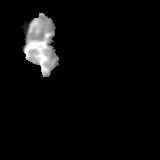}};

        \node at (\x+0.15 + 0.5*\cubex, \y-0.55*\cubey + 0.45) {\cref{eq: constraint}};
        \node at (\x+0.15 + 0.5*\cubex, \y-0.55*\cubey - 0.45) {\cref{setup}};
        \draw [myarrows, color = hwy](\x+0.3, \y-0.55*\cubey) -- (\x + \cubex, \y-0.55*\cubey);

        %%%%%%% P1 %%%%%%% 

	\pgfmathsetmacro{\x}{-8.5+2.4*\cubex+1.2*\cubex + 2.35*\cubex} 

	\draw[black,fill=gray!35, line width = 0.02mm] (\x,\y,0.1*3) -- ++(-\cubex,0,0) -- ++(0,-\cubey,0) -- ++(\cubex,0,0) -- cycle; 
	\draw[black,fill=gray!35, line width = 0.02mm] (\x,\y,0.1*3) -- ++(0,0,-\cubez) -- ++(0,-\cubey,0) -- ++(0,0,\cubez) -- cycle;
	\draw[black,fill=gray!35, line width = 0.02mm] (\x,\y,0.1*3) -- ++(-\cubex,0,0) -- ++(0,0,-\cubez) -- ++(\cubex,0,0) -- cycle;
 
        \node at (\x-0.5*\cubex,\y-1.2*\cubey,0.1*3) {$P_{T_{\text{max}}}$};
        \node at (\x-0.5*\cubex,\y-0.5*\cubey,0.1*3) {\includegraphics[width=1.5cm]{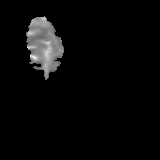}};

	\pgfmathsetmacro{\x}{-8.5+2.4*\cubex+1.2*\cubex} 
 
        \node at (\x+0.2 + 0.5*\cubex + 0.05, \y-0.55*\cubey + 0.45) {\cref{eq: divfree_adv_diff}};
        \node at (\x+0.2 + 0.5*\cubex + 0.05, \y-0.55*\cubey - 0.45) {\cref{forward}};
        \draw [myarrows, color = hwy] (\x + 0.4, \y-0.55*\cubey) -- (\x +\cubex + 0.1, \y-0.55*\cubey);

        %%%%%%% PDE %%%%%%% 
        
		%\node at (7.5+\dx + \x, -2.5+\dy + \y){$\mathcal{T}^{(0)}:$};  
		
		\pgfmathsetmacro{\x}{-15.9-0.5}
		\pgfmathsetmacro{\y}{0.2}

		\pgfmathsetmacro{\dx}{0.7}
		\pgfmathsetmacro{\dy}{0.7}

		%% ODE box
		
		\pgfmathsetmacro{\recth}{2.25}
		\pgfmathsetmacro{\rectw}{3.8} 
		\draw[gray,fill=hwy!30, dashed]  (9.65 + \x, -1.6 - \recth/2 + \y) rectangle (9.65 + \rectw + \x, -1.6 + \recth/2 + \y);  
		
		\pgfmathsetmacro{\dy}{0.4} 
		 %\node at (11.55 + \x, -2.35 + \y + 0.63){\cref{forward}};
		 \node at (11.55 + \x, -2.35 + \y + 0.1){Forward Scheme};
   
  		 \draw[-stealth] (9.65 + 0.1 + \x, -2.35+\dy + \y)--(9.65 + \rectw - 0.1 + \x,-2.35+\dy + \y); 
		 
   		 \draw plot [smooth] coordinates {(10+\x,-1.6+\dy + \y) (10.25 + \x,-1.8+\dy + \y) (10.5 + \x, -1.9+\dy + \y)  (10.9 + \x, -1.832+\dy + \y) (11.2 + \x, -1.6+\dy + \y) (11.5 + \x, -1.35+\dy + \y) (11.8 + \x, -1.35+\dy + \y) (12. + \x, -1.5+\dy + \y) (12.2 + \x, -1.58+\dy + \y) (12.5 + \x, -1.59+\dy + \y) (13 + \x, -1.45+\dy + \y) (13.1 + \x, -1.42+\dy + \y)};  
   		 \node (a0) at (10 + \x,-1.6+\dy + \y) {};
   		 \draw[fill=hr!40] (a0) circle [radius=5pt];       
   		 %\node at (10 + \x,-1.2+\dy + \y) {\small $P_0$};   
   		 \node (a1) at (10.5 + \x,-1.9+\dy + \y) {};
   		 \draw[fill=hr!40] (a1) circle [radius=5pt];    
   		 \node (a2) at (11.2 + \x,-1.6+\dy + \y) {};
   		 \draw[fill=hr!40] (a2) circle [radius=5pt];   
   		 \node (a3) at (11.8 + \x,-1.35+\dy + \y) {};
   		 \draw[fill=hr!40] (a3) circle [radius=5pt];    
  		  \node (a4) at (12.5 + \x,-1.59+\dy + \y) {};
  		  \draw[fill=hr!40] (a4) circle [radius=5pt];    
  		  \node (a5) at (13.1 + \x,-1.42+\dy + \y) {};
  		  \draw[fill=hr!40] (a5) circle [radius=5pt];  
   		 %\node at (13.1 + \x,-1.8+\dy + \y) {\small $P_{T_{\text{max}}}$};   

        %%%%%%% L & I %%%%%%% 

	\pgfmathsetmacro{\x}{-8.5} 
	\pgfmathsetmacro{\y}{2.5-4*\cubey} 
 
	\draw[dashed, color = hwy, line width=0.6mm] (\x-1.2*\cubex, \y-1.15*\cubey) -- (\x+1.45*\cubex, \y-1.15*\cubey) -- (\x+1.45*\cubex, \y+0.15*\cubey) -- (\x-1.2*\cubex, \y+0.15*\cubey) -- (\x-1.2*\cubex, \y-1.15*\cubey); 
 
        \node at (\x+0.125*\cubex, \y-0.1) {{\color{hwy}\textbf{\textit{Healthy}}} Subjects Pool};
        \node[right] at (\x-1.15*\cubex, \y-0.1-0.45) {\texttt{HCP}~\cite{Essen2012TheHC},  \texttt{ADHD200}~\cite{Brown2012ADHD200GC},};
        \node[right] at (\x-1.15*\cubex, \y-0.1-2*0.45) {\texttt{ADNI}~\cite{Jack2008TheAD}, \texttt{ADNI3}~\cite{Weiner2017TheAD},};
        \node[right] at (\x-1.15*\cubex, \y-0.1-3*0.45) {\texttt{OASIS3}~\cite{LaMontagne2018OASIS3LN}, \texttt{AIBL}~\cite{Fowler2021FifteenYO}};

        %%%%%%% L %%%%%%% 

	\pgfmathsetmacro{\x}{-8.5+2.4*\cubex-1.2*\cubex + 2.35*\cubex+0.3} 
	\pgfmathsetmacro{\y}{2.5-4*\cubey} 

	\draw[black,fill=gray!35, line width = 0.02mm] (\x,\y,0.1*3) -- ++(-\cubex,0,0) -- ++(0,-\cubey,0) -- ++(\cubex,0,0) -- cycle; 
	\draw[black,fill=gray!35, line width = 0.02mm] (\x,\y,0.1*3) -- ++(0,0,-\cubez) -- ++(0,-\cubey,0) -- ++(0,0,\cubez) -- cycle;
	\draw[black,fill=gray!35, line width = 0.02mm] (\x,\y,0.1*3) -- ++(-\cubex,0,0) -- ++(0,0,-\cubez) -- ++(\cubex,0,0) -- cycle;
 
        \node at (\x-0.5*\cubex,\y-1.2*\cubey,0.1*3) {$L$};
        \node at (\x-0.5*\cubex,\y-0.5*\cubey,0.1*3) {\includegraphics[width=1.5cm]{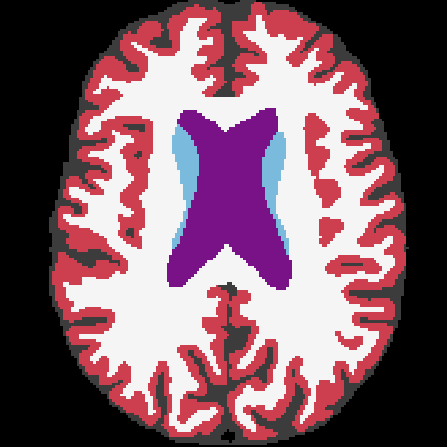}};

	\pgfmathsetmacro{\x}{-8.5+2.4*\cubex-1.2*\cubex+0.3} 
        \draw [myarrows, color = hwy] (\x + 0.4, \y-0.55*\cubey) -- (\x +\cubex + 0.1, \y-0.55*\cubey);

        %%%%%%% I0 %%%%%%% 

	\pgfmathsetmacro{\x}{-8.5+2.4*\cubex+1.2*\cubex + 2.35*\cubex} 
	\pgfmathsetmacro{\y}{2.5-4*\cubey} 

	\draw[black,fill=gray!35, line width = 0.02mm] (\x,\y,0.1*3) -- ++(-\cubex,0,0) -- ++(0,-\cubey,0) -- ++(\cubex,0,0) -- cycle; 
	\draw[black,fill=gray!35, line width = 0.02mm] (\x,\y,0.1*3) -- ++(0,0,-\cubez) -- ++(0,-\cubey,0) -- ++(0,0,\cubez) -- cycle;
	\draw[black,fill=gray!35, line width = 0.02mm] (\x,\y,0.1*3) -- ++(-\cubex,0,0) -- ++(0,0,-\cubez) -- ++(\cubex,0,0) -- cycle;
 
        \node at (\x-0.5*\cubex,\y-1.2*\cubey,0.1*3) {$I_0$};
        \node at (\x-0.5*\cubex,\y-0.5*\cubey,0.1*3) {\includegraphics[width=1.5cm]{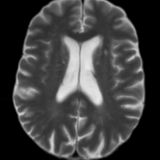}};

	\pgfmathsetmacro{\x}{-8.5+2.4*\cubex+1.2*\cubex} 
 
        \node at (\x+0.2 + 0.5*\cubex + 0.05, \y-0.55*\cubey + 0.45) {\cref{eq: contrast}};
        \node at (\x+0.2 + 0.5*\cubex + 0.05, \y-0.55*\cubey - 0.45) {\cref{modality}};
        \draw [myarrows, color = hwy] (\x + 0.4, \y-0.55*\cubey) -- (\x +\cubex + 0.1, \y-0.55*\cubey); 

        %%%%%%%%%%%%%%%%%%%%%
        %%%%%%%   Is  %%%%%%% 
        %%%%%%%%%%%%%%%%%%%%%
 
	\pgfmathsetmacro{\x}{-8.3+2.4*\cubex+1.2*\cubex-0.4} 
    \draw [myarrows, color = hwy] (\x+2.*\cubex+0.4 + 0.725, \y+0.45*\cubey) -- (\x+4.*\cubex + 0.1 + 0.725 + 0.65, \y+0.45*\cubey); 
    
    \node at (\x+3.*\cubex+0.25 + 0.725 + 0.2, \y+0.45*\cubey + 0.9) {\color{hwy}\textbf{\textit{Anomaly Profile}}};
    \node at (\x+3.*\cubex + 0.725 + 0.2, \y+0.45*\cubey + 0.45) {{\color{hwy}\textbf{\textit{Encoding}}} via};
    \node at (\x+3.*\cubex+0.25 + 0.725 + 0.2, \y+0.45*\cubey - 0.45) {\color{black}\cref{eq: pathol_encode,eq: pathol_condition}};
    
    \draw [decorate,decoration={brace,amplitude=5pt,mirror,raise=6ex},line width=0.65mm,color = hwy] (\x+2.*\cubex, \y-\cubey) -- (\x+2.*\cubex, \y + 1.9*\cubey);

        %%%%%%% I %%%%%%% 
        
	\pgfmathsetmacro{\x}{-8+2.4*\cubex+2.2*\cubex + 2.45*\cubex + 2.35*\cubex} 
 
	\pgfmathsetmacro{\y}{2.5-4*\cubey+0.22*\cubey+0.75*\cubey} 
	\draw[black,fill=gray!35, line width = 0.02mm] (\x,\y,0.1*3) -- ++(-\cubex,0,0) -- ++(0,-\cubey,0) -- ++(\cubex,0,0) -- cycle; 
	\draw[black,fill=gray!35, line width = 0.02mm] (\x,\y,0.1*3) -- ++(0,0,-\cubez) -- ++(0,-\cubey,0) -- ++(0,0,\cubez) -- cycle;
	\draw[black,fill=gray!35, line width = 0.02mm] (\x,\y,0.1*3) -- ++(-\cubex,0,0) -- ++(0,0,-\cubez) -- ++(\cubex,0,0) -- cycle;
 
        \node at (\x-0.5*\cubex,\y-1.2*\cubey,0.1*3) {$I$};
        \node at (\x-0.5*\cubex,\y-0.5*\cubey,0.1*3) {\includegraphics[width=1.5cm]{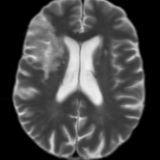}};

   % \draw [decorate,decoration={brace,amplitude=5pt,mirror,raise=6ex},line width=0.65mm,color = hc] (\x-0.5*\cubex, \y-1.4*\cubey) -- (\x-0.5*\cubex, \y + 1.65*\cubey);
    
        %%%%%%% I_flip %%%%%%% 
        
	\pgfmathsetmacro{\y}{2.5-2*\cubey-0.25*\cubey+0.75*\cubey} 
	\draw[black,fill=gray!35, line width = 0.02mm] (\x,\y,0.1*3) -- ++(-\cubex,0,0) -- ++(0,-\cubey,0) -- ++(\cubex,0,0) -- cycle; 
	\draw[black,fill=gray!35, line width = 0.02mm] (\x,\y,0.1*3) -- ++(0,0,-\cubez) -- ++(0,-\cubey,0) -- ++(0,0,\cubez) -- cycle;
	\draw[black,fill=gray!35, line width = 0.02mm] (\x,\y,0.1*3) -- ++(-\cubex,0,0) -- ++(0,0,-\cubez) -- ++(\cubex,0,0) -- cycle;
 
        \node at (\x-0.5*\cubex,\y-1.2*\cubey,0.1*3) {$\phi^{-1}_{I\to\overline{I}} \circ \overline{I}$};
        \node at (\x-0.5*\cubex,\y-0.5*\cubey,0.1*3) {\includegraphics[width=1.5cm]{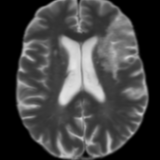}};

    %%% contrastive loss %%%
    \draw [dashed, <->, line width = 0.6mm, color = hgrey] (-8+8.85*\cubex, \y+0.1*\cubey) -- 
 (-8+8.85*\cubex, \y+0.45*\cubey) -- (-8+12.75*\cubex + 5.1, \y+0.45*\cubey) -- (-8+12.75*\cubex + 5.1, \y-0.75*\cubey); 
 
    \node at ((-8+10.8*\cubex + 2.55, \y+0.45*\cubey+0.45) { \textbf{\textit{\color{hblue}$\uparrow$~Intra-Subject Similarity}} in Healthy Contralateral Anatomy via \cref{eq: loss_contrast}};

    \draw [dashed, <->, line width = 0.6mm, color = hgrey] (-8+8.85*\cubex, \y-3.*\cubey) -- 
 (-8+8.85*\cubex, \y-3.*\cubey-0.35*\cubey) -- (-8+12.75*\cubex + 5.1, \y-3.*\cubey-0.35*\cubey) -- (-8+12.75*\cubex + 5.1, \y-2.25*\cubey); 

    \node at ((-8+10.8*\cubex + 3, \y-3.*\cubey-0.35*\cubey+0.45) {\textbf{\color{hblue}\textit{$\uparrow$~Intra-Subject Distinctiveness}}: Healthy v.s. Diseased via \cref{eq: loss_contrast}};

        %%%%% Box %%%%%
	\pgfmathsetmacro{\x}{-8+2.4*\cubex+2.2*\cubex + 2.45*\cubex + 2.35*\cubex} 
	\pgfmathsetmacro{\y}{2.5-4*\cubey+0.2*\cubey+0.75*\cubey} 
        %\node at (\x-0.5925*\cubex, \y+0.15*\cubey + 1.5*\cubey + 0.4) {\color{hc}$\large( I,\, \phi^{-1}_{I\to\overline{I}} \circ \overline{I} \large)$ in \cref{input}};
        %\node at (\x-0.5925*\cubex, \y+0.15*\cubey + 1.5*\cubey + 0.4+0.5) {\color{hc}\textbf{Contralateral-Paired Input}};
        
	\draw[dashed, color = hc, line width=0.6mm] (\x-1.2*\cubex, \y-1.45*\cubey) -- (\x+0.05*\cubex, \y-1.45*\cubey) -- (\x+0.05*\cubex, \y+0.15*\cubey + 1.5*\cubey) -- (\x-1.2*\cubex, \y+0.15*\cubey + 1.5*\cubey) -- (\x-1.2*\cubex, \y-1.45*\cubey); 

        %%%%% Encoder-Decoder %%%%%

	\pgfmathsetmacro{\y}{2.5-4*\cubey+0.75*\cubey} 
    \draw [myarrows, color = hc] (\x+0.*\cubex+0.4, \y+0.45*\cubey) -- (\x+1.*\cubex + 0.1 + 1.5*\cubex, \y+0.45*\cubey); 
        
    \node at (\x+1.25*\cubex+0.2, \y+0.45*\cubey + 0.9) {\color{hc}\textbf{Contralateral-Paired}};
    \node at (\x+1.25*\cubex+0.25, \y+0.45*\cubey + 0.45) {\color{hc}\textbf{Input: } $\left( I,\, \phi^{-1}_{I\to\overline{I}} \circ \overline{I} \right)$};
    \node at (\x+1.25*\cubex+0.25, \y+0.45*\cubey - 0.45) {\cref{input}};

    % Backbone 
    % Sigma-Net 
    \pgfmathsetmacro{\sx}{8.3+1.5*\cubex}
    \pgfmathsetmacro{\sy}{5.2+0.5*\cubey}
    \pgfmathsetmacro{\dx}{\sx+0}
    \pgfmathsetmacro{\dy}{\sy+0.2} % y of  vdm_net
    \pgfmathsetmacro{\ddy}{-3.5} % ddy for Sigma-Net arrows
    \pgfmathsetmacro{\dxt}{\dx+0.5} % dx for encoders
    
    %\node at (1.5+\sx, -6.+\dy+\ddy){\large $\mathcal{F}$};
    % Encoder %
    \pgfmathsetmacro{\dy}{\dy+\ddy} % update y for Sigma-Net blocks
    % Block gap: 0.2
    \networkLayer{2.4}{0.05}{5.-\dy+\dxt}{13-2.6*\dy}{color=hc!75}{}
    \networkLayer{2}{0.1}{5.1-\dy+\dxt}{13-2.6*\dy}{color=hc!60}{}
    \networkLayer{1.6}{0.2}{5.2-\dy+\dxt}{13-2.6*\dy}{color=hc!45}{}
    \networkLayer{1.2}{0.4}{5.4-\dy+\dxt}{13-2.6*\dy}{color=hc!30}{}
    \networkLayer{0.8}{0.8}{5.8-\dy+\dxt}{13-2.6*\dy}{color=hc!15}{}
    
    % Decoder %
    % Local location control
    \pgfmathsetmacro{\dxs}{\dx+0}
    \pgfmathsetmacro{\dys}{\dy+0} % y of v_net
    % Block gap: 0.4 
    \networkLayer{0.8}{0.8}{7.2-\dys+\dxs}{13-2.6*\dys}{color=hblue!15}{}
    \networkLayer{1.2}{0.4}{7.6-\dys+\dxs}{13-2.6*\dys}{color=hblue!30}{}
    \networkLayer{1.6}{0.2}{8.-\dys+\dxs}{13-2.6*\dys}{color=hblue!45}{}
    \networkLayer{2}{0.1}{8.3-\dys+\dxs}{13-2.6*\dys}{color=hblue!60}{}
    \networkLayer{2.4}{0.05}{8.7-\dys+\dxs}{13-2.6*\dys}{color=hblue!60}{}

    \pgfmathsetmacro{\x}{-8+2.4*\cubex+2.2*\cubex + 2.45*\cubex + 2.35*\cubex + 5.2 + 1.5*\cubex} 
    \draw [myarrows, color = hblue] (\x+0.*\cubex+0.4, \y+0.45*\cubey) -- (\x+1.*\cubex + 0.1, \y+0.45*\cubey); 
        
        %%%%%%% I_Out %%%%%%% 
        
    \pgfmathsetmacro{\x}{-8+2.4*\cubex+2.2*\cubex + 2.45*\cubex + 2.35*\cubex + 5.2 + 2.35*\cubex + 1.5*\cubex} 
	\pgfmathsetmacro{\y}{2.5-3*\cubey+0.75*\cubey} 
	\draw[black,fill=gray!35, line width = 0.02mm] (\x,\y,0.1*3) -- ++(-\cubex,0,0) -- ++(0,-\cubey,0) -- ++(\cubex,0,0) -- cycle; 
	\draw[black,fill=gray!35, line width = 0.02mm] (\x,\y,0.1*3) -- ++(0,0,-\cubez) -- ++(0,-\cubey,0) -- ++(0,0,\cubez) -- cycle;
	\draw[black,fill=gray!35, line width = 0.02mm] (\x,\y,0.1*3) -- ++(-\cubex,0,0) -- ++(0,0,-\cubez) -- ++(\cubex,0,0) -- cycle;
 
        \node at (\x-0.5*\cubex,\y-1.2*\cubey,0.1*3) {$\widetilde{I}$};
        \node at (\x-0.5*\cubex,\y-0.5*\cubey,0.1*3) {\includegraphics[width=1.5cm]{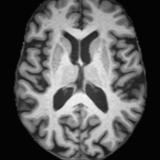}};

	%%%%%%%%%%%%%%%%%%%%%%%%%%%%%%%%%%%%%%%%%%%%%%%%%
       
 \vspace*{-0.6cm}
	\end{tikzpicture}
	}   
 \vspace*{-0.6cm}
\caption{\texttt{UNA}'s framework overview for modality-agnostic learning of healthy anatomy, supported by fluid-driven anomaly randomization.} 
	 \label{fw}
\end{figure*}

%% file: sec/method/trainer.tex
%\vspace{-0.1cm}
%%%%%%%%%%%%%%%%%%%%%%%%%%%%%%

\section{Learning Anatomy Beyond Gold Standards}
\label{sec: framework}
\vspace{-0.1cm}

In this section, we present \texttt{UNA}’s end-to-end training framework, which learns to unravel normal anatomy from images of random modality containing potential pathology.

%\subsection{Learning with Symmetry Priors}
%\label{sec: loss}

%\vspace{-0.2cm}

%%%%%%%%%%%%%%%%%%%%%%%%%%%%%%

%This paragraph needs a bit more detail on 2 things:
%1. What do you do when pathology is present at the same location on both hemispheres? -- it is still inputted but will be ignored during training loss computation
%2. How do you compute registration in the presence of pathology?

\vspace{-0.35cm}
\paragraph{Contralateral-Paired Input.}\label{input}~Healthy human brain anatomy typically exhibits a high degree of symmetry in structure. Based on this fact, we combine the original input image ($I$) with its contralateral-mirrored image ($\overline{I}$) to create paired inputs for \texttt{UNA}'s healthy anatomy reconstruction learning. This approach allows our model to ``borrow'' healthy information from the contralateral counterpart, thereby enhancing subject-specific healthy anatomy reconstruction. 
To ensure structural correspondence and minimize computational complexity during training, we pre-compute the deformation ($\phi_{I\to\overline{I}}$) between each training subject's scan and its axial-flipped image using NiftyReg~\cite{rueckert1999nonrigid,modat2010fast}. As a result, the contralateral-paired input for each subject sample is represented as $\left( I,\, \phi^{-1}_{I\to\overline{I}} \circ \overline{I} \right)$. %The detailed processing steps, including left-right registration in the presence of pathology, can be found in Appendix~\hyperlink{app: dataset}{D}.

%By leveraging contralateral symmetry, we aim to improve the robustness of our model in reconstructing healthy anatomy, particularly in cases where data may be limited or suboptimal. Through this methodology, we hope to enhance the accuracy and reliability of our anatomical reconstructions, ultimately contributing to better outcomes in medical imaging and analysis.

%%%%%%%%%%%%%%%%%%%%%%%%%%%%%%
\vspace{-0.35cm}
\paragraph{Modality-Agnostic Healthy Anatomy Reconstruction.}\label{recon}~To enhance model generalizability, \texttt{UNA} is trained on both real datasets containing pathology ($\mathbb{D}_{\text{Real}}$) and synthetic images ($\mathbb{D}_{\text{Synth}}$) generated from fluid-driven anomaly randomization (\cref{sec: fluid}), featuring varying simulated modalities and abnormality conditions. During training, we define the following healthy anatomy reconstruction loss, which takes into account both the subject-level and the voxel-level abnormality  of the input image ($I$):
\vspace{-0.25cm}
\begin{equation}
\mathcal{L}_{\text{Recon}} = \int_{\Omega}k(\mathbf{x})~\{ \vert \widetilde{I}(\mathbf{x}) - I(\mathbf{x}) \vert + \lambda_{\nabla} \vert \nabla \widetilde{I}(\mathbf{x}) - \nabla I(\mathbf{x}) \vert\} \, d\mathbf{x} \nonumber
\label{eq: loss_recon} 
\vspace{-0.8 cm}
\end{equation}
\begin{equation}
\hspace*{-0.2cm}s.t.\quad k(\mathbf{x}) = 
\begin{cases} 
    1 - d \cdot p(\mathbf{x}) \,, & \mathbf{x} \in \Omega_{P} \,, \\ 
    (1 + \lambda_{p}) \cdot (1-d) \cdot p(\mathbf{x}) \,, & \mathbf{x} \notin \Omega_{P} \,, \\
\end{cases}
\vspace{-0.2cm}
\end{equation}
where $d = \{1: I \in \mathbb{D}_{\text{Real}} \,; 0: I \in \mathbb{D}_{\text{Synth}}\} $ indicates whether the current image is sourced from real datasets ($\mathbb{D}_{\text{Real}}$) or generated synthetically ($\mathbb{D}_{\text{Synth}}$). The parameters $\lambda_{\nabla}$ and $\lambda_{p}$ control the training weights for gradient \texttt{L1} loss and attention to pathology, respectively. Specifically: \textit{(i)} if the current training input image ($I$) is generated by \texttt{UNA}, i.e., the ground truth healthy anatomy of the entire brain region is accessible, we compute the anatomy reconstruction loss across the whole brain ($\Omega$). \textit{(ii)} Conversely, if $I$ is sourced from real datasets, the ground truth healthy anatomy of the entire brain is not available. In this case, we compute the voxel-wise reconstruction loss \textit{exclusively} for the healthy regions, while masking out any abnormalities.

%%%%%%%%%%%%%%%%%%%%%%%%%%%%%%
\vspace{-0.35cm}
\paragraph{Intra-Subject Self-Contrastive Learning.}\label{contrast}~In \cref{eq: loss_recon}, the anatomy reconstruction in abnormal regions is not supervised when dealing with real images containing pathology. To enhance the performance of learning healthy anatomy, we propose an intra-subject learning strategy 
% inspired by the decision-making processes of clinicians. Radiologists rely heavily on their domain knowledge, with their understanding of anatomy and the contextual nuances of medical images. They can identify abnormalities by comparing the diseased hemisphere with its contralateral counterpart within specific anatomical regions. Drawing on this observation, our contrastive learning loss aims to encourage the following two properties: 
that exploits the (approximate) symmetry of the brain with a contrastive loss that encourages two properties:

\begin{itemize}
\item[\textit{i.}] \textit{Similarity} in appearance between the reconstructed healthy anatomy and its contralateral healthy counterpart.

\item[\textit{ii.}] \textit{Distinctiveness} between the reconstructed anatomy and the original regions that exhibit abnormalities.
\end{itemize}

\noindent Specifically, we define this intra-subject contrastive loss as:
\vspace{-0.35 cm}
\begin{equation}
\mathcal{L}_{\text{Contrast}} = - ~log~ \frac{\displaystyle \int_{\Omega_{p\setminus\overline{p}}} 
 e^{ \widetilde{I} \cdot \left(\phi^{-1}_{I\to\overline{I}} \circ \overline{I}\right) / \alpha} \, d\mathbf{x} }{\displaystyle \int_{\Omega_{p\setminus\overline{p}}}
 e^{ \widetilde{I} \cdot \left(\phi^{-1}_{I\to\overline{I}} \circ \overline{I}\right) / \beta} + e^{ \widetilde{I} \cdot I / \gamma} \, d\mathbf{x} } \,,
    \label{eq: loss_contrast}
\vspace{-0.2 cm}
\end{equation}
where $\Omega_{p\setminus\overline{p}} = \Omega_p \setminus \left(\Omega_p \cap \Omega_{\phi^{-1}_{I\to\overline{I}} \circ \overline{P}}\right),$ ensuring that we exclude pathologies that appear at the same contralateral location on both hemispheres. $\alpha,\, \beta,\, \gamma$ represent the corresponding temperature scaling factors of each term.

%%%%%%%%%%%%%%%%%%%%%%%%%%%%%%
%%%%%%%%%%%%%%%%%%%%%%%%%%%%%%

Thus, \texttt{UNA}'s end-to-end healthy anatomy reconstruction training loss is obtained by the sum of \cref{eq: loss_recon,eq: loss_contrast}:
\vspace{-0.23cm}
\begin{equation}
    \mathcal{L} = \mathcal{L}_{\text{Recon}} + \lambda_{\text{Contrast}}\, \mathcal{L}_{\text{Contrast}} \,,
    \label{eq: loss}
\vspace{-0.23cm}
\end{equation}
where $\lambda_{\text{Contrast}}$ is the weight of self-contrastive learning loss. 

As shown in \cref{showcase}, as a general model for healthy anatomy reconstruction, \texttt{UNA} also addresses the following tasks: \textit{(i)} Given an input image without any abnormalities, \texttt{UNA} performs anatomy reconstruction; \textit{(ii)} Given a T1w MRI of any resolution, \texttt{UNA} performs super-resolution.

%% file: sec/exp/main.tex
%\vspace{-0.15cm}

\section{Experiments}
\label{sec: exp}
\vspace{-0.1cm}

We evaluate \texttt{UNA}'s performance and demonstrate its impact from three perspectives. \textit{(i)}~The reconstruction of anatomy from healthy images. This enables analysis with standard tools made for high-resolution T1w MRI, such as segmentation and parcellation using FreeSurfer~\cite{fischl2002freesurfer}, registration with NiftyReg~\cite{rueckert1999nonrigid,modat2010fast}, ANTs~\cite{avants2009ants}, etc.  
\textit{(ii)}~The synthesis of healthy anatomy from images with pathology. This allows for the application of well-established general-purpose models to images with extensive pathology.
For a more comprehensive assessment, we test on both synthetic data  -- where ground truth healthy images are available~(\cref{exp: simulation}) --  and real images from two public stroke datasets -- where the ground truth healthy anatomy is unknown~(\cref{exp: real}). \textit{(iii)}~We further demonstrate \texttt{UNA}'s \textit{direct} application to anomaly detection~(\cref{exp: anomaly_detection}). Our test data includes CT and various MRI modalities (T1w, T2w, FLAIR).

%%%%%%%%%%%%%%%%%%%% 
\input{sec/exp/setup}

%%%%%%%%%%%%%%%%%%%% 

%%%%%%%%%%%%%%%%%%%%

\input{sec/exp/fig/tab_synth}

\input{sec/exp/fig/fig_synth}

%%%%%%%%%%%%%%%%%%%%

%\vspace{-0.05cm}
\subsection{Simulations with Ground Truth Anatomy}
\vspace{-0.1cm}
\label{exp: simulation}

To better evaluate \texttt{UNA}'s performance in healthy anatomy reconstruction, we first conduct experiments using 1,000 healthy images encoded with simulated pathologies, for which ground truth segmentations are available for quantitative assessment. %Details on the generation and pre-processing of these 1,000 test samples are provided in \cref{data: synth} and Appendix~\hyperlink{app: dataset}{D}. 
To explicitly assess the model performance in pathology regions, we report reconstruction scores not only for the entire brain but also separately for areas that are originally healthy and diseased in the input image.

\cref{tab: synth} reports the quantitative comparison results between \texttt{UNA} and the state-of-the-art modality-agnostic synthesis models. \texttt{UNA} yields the best performance across \textit{all} metrics, modalities, and regions of interest -- including the full brain, healthy anatomy, and pathological regions. %\footnote{Since \texttt{PEPSI}~\cite{liu2024pepsi} is designed to emphasize the abnormalities, its scores in the diseased region are expected to be worse on healthy anatomy reconstruction. We still report their scores in \cref{tab: synth} for completeness.} 
Remarkably, \texttt{UNA} outperforms competing models by a large margin in anatomy reconstruction within diseased tissue. 
Visualization results for each test modality are provided in \cref{fig: synth}. \texttt{UNA} demonstrates consistent performance across modality and resolution. Notably, other models either fail to capture any anatomy (\texttt{SynthSR}~\cite{Iglesias2023SynthSRAP}) or generate unrealistic patterns around the pathology (\texttt{Brain-ID}~\cite{Liu2024BrainID} and \texttt{PEPSI}~\cite{liu2024pepsi}) when
%%%% CAREFUL WITH WHAT YOU WROTE HERE, the 4th ROW OF  FIGURE 4 IS NOT A CT SCAN
%%% I THINK THE PROBLEM IS THAT YOU SHOULD BE REFERRING TO FIGURE 3 AND NOT 4
given a noisy CT scan (4$^{\text{th}}$ row in \cref{fig: synth}), 
%%%
whereas \texttt{UNA} successfully reconstructs plausible healthy anatomy.

%%%%%%%%%%%%%%%%%%%%
%\vspace{-0.05cm}
\subsection{Real-World Datasets with Potential Pathology}
\vspace{-0.1cm}
\label{exp: real}

%%%%%%%%%%%%%%%%%%%%

\input{sec/exp/fig/fig_real}
\input{sec/exp/fig/tab_real}

%%%%%%%%%%%%%%%%%%%%

We further evaluate \texttt{UNA}'s performance on all the real datasets as introduced in \cref{datasets}, among which \texttt{ATLAS}~\cite{Liew2017ATLAS} and \texttt{ISLES}~\cite{Hernandez2022ISLES} contain stroke patients. \cref{tab: real} reports the reconstruction scores over all datasets and their available modalities: 
\textit{(i)}~For anatomy reconstruction of originally healthy subjects, \texttt{UNA} achieves the highest scores across most datasets, with the remaining scores on par with \texttt{Brain-ID}~\cite{Liu2024BrainID}, which is specifically designed for healthy anatomy; \textit{(ii)}~On the \texttt{ATLAS} stroke dataset, \texttt{UNA} outperforms competing models by a larger margin~($\approx 10\%$).

As shown in \cref{fig: real}, other models tend to generate unrealistic patterns within and around abnormalities, whereas \texttt{UNA}'s reconstructions are notably more visually coherent. Additionally, we present a failure case (4$^{\text{th}}$ row in \cref{fig: real}), where we observe that \texttt{UNA} tends to ``over-distinguish'' the reconstructed healthy anatomy from the diseased regions, particularly in challenging scenarios where the pathology pattern
% heavily disagree (see comment)
completely \textit{occludes} the underlying anatomy.
% Move to discussion
%This issue will be more closely investigated in future work.

%%%%%%%%%%%%%%%%%%%%
%%%%%%%%%%%%%%%%%%%%
%\vspace{-0.05cm} 
\subsection{Direct Application: Anomaly Detection}
\vspace{-0.1cm}
\label{exp: anomaly_detection}

%%%%%%%%%%%%%%%%%%%%
\input{sec/exp/fig/fig_anomaly}
\input{sec/exp/fig/tab_anomaly}

%%%%%%%%%%%%%%%%%%%%

\texttt{UNA}'s ability to synthesize diseased-to-healthy anatomy naturally equips it with the potential for application to anomaly detection. To demonstrate its effectiveness, we \textit{directly} use the reconstructed healthy anatomy from \texttt{UNA} to detect abnormalities. 
Specifically, we follow the standard evaluation pipeline for unsupervised anomaly detection in medical images~\cite{baur2021autoencoders, graham2023unsupervised} and compute \texttt{UNA}'s anomaly estimation maps by calculating the voxel-wise absolute differences between the diseased input and the reconstructed output. The anomaly detection \texttt{Dice} scores are then obtained by comparing the ground truth pathology segmentations with the computed anomaly estimation maps, scaled to the range $[0,\, 1]$ such that they represent the normalized abnormality. The same procedure is applied to other competing models.

As shown in \cref{fig: anomaly}, \texttt{UNA}'s difference maps clearly identify anomalies with varying shapes and sizes. 
Quantitative comparisons are provided in \cref{tab: anomaly}, where \texttt{UNA}:
\textit{(i)} outperforms other modality-agnostic synthesis models, and the state-of-the-art anomaly detection models; and \textit{(ii)} demonstrates consistent performance across various datasets.

%%%%%%%%%%%%%%%%%%%%
%%%%%%%%%%%%%%%%%%%%
%\vspace{-0.05cm} 
\subsection{Ablation Study}
\label{exp: ablation}
\vspace{-0.1cm}

%%%%%%%%%%%%%%%%%%%%
\input{sec/exp/fig/fig_ablat}
\input{sec/exp/fig/tab_ablat}

%%%%%%%%%%%%%%%%%%%%

To assess the contributions of \texttt{UNA}'s individual components, we perform an ablation study with several variants: \hypertarget{ablat: a}{\textit{(a)}}~Training without fluid-driven anomaly randomization, i.e., training exclusively with real images with pathology; \hypertarget{ablat: b}{\textit{(b)}}~Training with fluid-driven anomaly randomization, but initializing the anomaly profiles with random noise; \hypertarget{ablat: c}{\textit{(c)}}~Training without contralateral-paired input, i.e., using only a single image without its contralateral counterpart; \hypertarget{ablat: d}{\textit{(d)}}~Training without the intra-subject self-contrastive loss. 

As shown in \cref{fig: ablat,tab: ablat}, training without fluid-driven anomaly randomization (\texttt{UNA}-(\hyperlink{ablat: a}{\textit{a}})) results in the largest performance drop, showing only slight improvement over \texttt{Brain-ID}~\cite{Liu2024BrainID} (reported in  \cref{fig: synth}), which does not train on diseased inputs at all. Introducing fluid-driven anomaly randomization improves overall performance, but performance gaps remain evident when compared to the proposed \texttt{UNA} when no real pathology profiles are used for initialization (\texttt{UNA}-(\hyperlink{ablat: b}{\textit{b}})). Leveraging subject-specific contralateral information (\texttt{UNA}-(\hyperlink{ablat: c}{\textit{c}}), \texttt{UNA}-(\hyperlink{ablat: d}{\textit{d}})) further enhances reconstruction results, particularly within diseased regions.

%% file: sec/exp/setup.tex
%\vspace{-0.1cm}
%\subsection{Datasets and Models for Comparison}
%\label{sec: setup}

%%%%%%%%%%%%%%%%%%%%%%%%%%%%%%%%%%%%%%%% 

\vspace{-0.5cm}
\paragraph{Datasets.}\label{datasets}~We conducted experiments using eight public datasets: \texttt{ADNI}~\cite{Jack2008TheAD}, \texttt{ADNI3}~\cite{Weiner2017TheAD}, \texttt{HCP}~\cite{Essen2012TheHC}, \texttt{ADHD200}~\cite{Brown2012ADHD200GC}, \texttt{AIBL}~\cite{Fowler2021FifteenYO}, \texttt{OASIS3}~\cite{LaMontagne2018OASIS3LN}, \texttt{ATLAS}~\cite{Liew2017ATLAS}, \texttt{ISLES}~\cite{Hernandez2022ISLES}. \texttt{ATLAS} and \texttt{ISLES} include stroke patients, associated with gold-standard manual segmentations of stroke lesions (referred to as $\mathbb{D}_{\text{Stroke}}$ hereafter). The other datasets contain subjects with healthy anatomy ($\mathbb{D}_{\text{Healthy}}$). These datasets cover both MR (T1w, T2w, FLAIR) and CT images. The train/test subject splits for each dataset are listed in \cref{tab: real}.

%%%%%%%%%%%%%%%%%%%%%%%%%%%%%%%%%%%%%%%% 

\vspace{-0.5cm}
\paragraph{Synthetic Data Generation.}\label{data: synth}~We use the anatomical labels of training subjects from $\mathbb{D}_{\text{Healthy}}$ for random modality generation (\cref{modality}). The synthetic abnormal profiles are generated using \texttt{UNA}'s fluid-driven anomaly randomization (\cref{sec: fluid}), with initial profiles either sampled from the gold standard lesion segmentation maps of training subjects in $\mathbb{D}_{\text{Stroke}}$, or Perlin noise (\cref{initialization}). 
For evaluation on simulated data in \cref{exp: simulation}, we employ our synthetic generator to create 1,000 testing samples from $\mathbb{D}_{\text{Healthy}}$, encoded with random anomaly profiles from $\mathbb{D}_{\text{Stroke}}$. This generation is solely for providing ground truth healthy anatomy; therefore, we encode random anomaly profiles without applying any additional deformation and corruption.

%%%%%%%%%%%%%%%%%%%%%%%%%%%%%%%%%%%%%%%% 

\vspace{-0.5cm}
\paragraph{Metrics.}\label{metrics}~%We use various metrics to evaluate different aspects. 
For anatomy reconstruction and synthesis, we use \texttt{L1} distance, \texttt{PSNR}, and \texttt{SSIM}. For anomaly detection, we assess performance using \texttt{Dice} scores.

%%%%%%%%%%%%%%%%%%%%%%%%%%%%%%%%%%%%%%%% 

%\vspace{-0.5cm}
%\paragraph{Random-Modality Image Synthesis}
%Detailed description regarding the random-modality image synthesis and pathology encoding is provided in Appendix~\hyperlink{app: synthesis}{D}. 

%%%%%%%%%%%%%%%%%%%%%%%%%%%%%%%%%%%%%%%% 

\vspace{-0.5cm}
\paragraph{Implementation Details.}
\label{implement}
%As a general feature representation model, \texttt{UNA} can use any backbone for the reconstruction. 
For fair comparisons, we adopt the same 3D UNet~\cite{Ronneberger2015UNetCN} as utilized in the models~\cite{Iglesias2023SynthSRAP,Liu2024BrainID,liu2024pepsi} we compare with. The training sample images are sized at $160^3$, with a batch size of 4. We use the AdamW optimizer, beginning with a learning rate of $10^{-4}$ for the first 300,000 iterations, which is then reduced to $10^{-5}$ for the subsequent 100,000 iterations. The additional attention parameter ($\lambda_p$ in \cref{eq: loss_recon}) is set to 1 for healthy anatomy reconstruction in pathological regions. The intra-subject contrastive learning weight ($\lambda_{\text{contrast}}$ in \cref{eq: loss}) is set to 2. The training process took approximately 14 days on an NVIDIA A100 GPU.
% [COMMENT: one line justifying the choice of lambdas would be super useful (e.g., do they come from a pilot experiment? From a previous paper?)] -- added them in the Appendix

%%%%%%%%%%%%%%%%%%%%%%%%%%%%%%%%%%%%%%%% 

\vspace{-0.5cm}
\paragraph{Competing Models.}\label{models}~\texttt{UNA} is the first model achieving modality-agnostic healthy anatomy synthesis and reconstruction. We compare \texttt{UNA} with the closest state-of-the-art modality-agnostic models for image reconstruction and anomaly detection: \textit{(i)}~\texttt{SynthSR}~\cite{Iglesias2023SynthSRAP}, a modality-agnostic super-resolution model; % We provide dataset-specific fine-tuned results for fair comparison; 
\textit{(ii)}~\texttt{Brain-ID}~\cite{Liu2024BrainID}, a  modality-agnostic feature representation and T1w synthesis model; 
\textit{(iii)}~\texttt{PEPSI}~\cite{liu2024pepsi}, a modality-agnostic pathology representation model for T1w and FLAIR MRI synthesis. Note that \texttt{PEPSI} does \emph{not} synthesize healthy tissue in regions of pathology; 
\textit{(iv)}~\texttt{VAE}~\cite{baur2021autoencoders}, an unsupervised anomaly detection variational autoencoder model for brain MRI; 
\textit{(v)}~\texttt{LDM}~\cite{graham2023unsupervised}, an out-of-distribution detection model for 3D medical images using latent diffusion.

Appendices~\hyperlink{app: dataset}{D} and \hyperlink{app: implement}{E} provide further details on metrics, datasets, pre-processing, and implementations.

%% file: sec/exp/fig/tab_synth.tex
\begin{table}[t]
\caption{Quantitative comparisons of healthy anatomy reconstruction performance between \texttt{UNA} and state-of-the-art contrast-agnostic T1w synthesis models, using images with simulated pathology. \texttt{PEPSI}~\cite{liu2024pepsi} is designed to emphasize the abnormalities, therefore we do not report its scores within diseased regions. (F: full brain; H: healthy region; D: diseased region.)
    } 
    \label{tab: synth}
\resizebox{\linewidth}{!}{
\centering 
    \begin{tabular}{clccccccccc}
    %\begin{tabular}{cccccccccccc}
       \toprule \\[-3ex] 
      \multicolumn{1}{c}{\multirow{2}{*}{\footnotesize\textbf{Modality}}} & \multicolumn{1}{c}{\multirow{2}{*}{\footnotesize\textbf{Method}}} & \multicolumn{3}{c}{\multirow{1}{*}{\footnotesize\textbf{\texttt{L1} ($\downarrow$)}}} & \multicolumn{3}{c}{\multirow{1}{*}{\footnotesize\textbf{\texttt{PSNR} ($\uparrow$)}}} & \multicolumn{3}{c}{\multirow{1}{*}{\footnotesize\textbf{\texttt{SSIM}  ($\uparrow$)}}} \\ [-0.ex]
        %\cmidrule(lr){2-3}% \\ [-4.1ex]
        \cmidrule(lr){3-5}
        \cmidrule(lr){6-8}
        \cmidrule(lr){9-11} \\ [-3.ex]
          & & {\footnotesize{F}} & {\footnotesize{H}} & {\footnotesize{D}} & {\footnotesize{F}} & {\footnotesize{H}} & {\footnotesize{D}} & {\footnotesize{F}} & {\footnotesize{H}} & {\footnotesize{D}}  \\ [-0.2ex]
     \midrule\\[-2.5ex]

        \multicolumn{1}{c}{\multirow{4}{*}{\thead{\footnotesize{T1w}\\MRI}}} & \texttt{SynthSR}~\cite{Iglesias2023SynthSRAP} & 0.0285 & 0.0253 & 0.0010 & 20.71 & 22.90 & 36.59 & 0.823 & 0.879 & 0.895 \\ %[0.7ex]  
        & \texttt{Brain-ID}~\cite{Liu2024BrainID} & 0.0231 & 0.0219 & 0.0007 & 22.86 & 23.71 & 40.22 & 0.859 & 0.890 & 0.904 \\ %[-0.5ex] 
        & \texttt{PEPSI}~\cite{liu2024pepsi} & 0.0257 & 0.0194 & N/A & 21.78 & 23.21 & N/A & 0.831 & 0.872 & N/A \\ %[-0.5ex] 
        & \texttt{UNA} & \textbf{0.0147} & \textbf{0.0143} & \textbf{0.0003} & \textbf{31.98} & \textbf{33.25} & \textbf{45.61} & \textbf{0.981} & \textbf{0.992} & \textbf{0.998} \\ %[-0.5ex] 
         \hline \\[-2.5ex]

        \multicolumn{1}{c}{\multirow{4}{*}{\thead{\footnotesize{T2w}\\MRI}}} & \texttt{SynthSR}~\cite{Iglesias2023SynthSRAP} & 0.0362 & 0.0337 & 0.0016 & 18.25 & 20.66 & 35.47 & 0.816 & 0.864 & 0.880  \\ %[0.7ex]  
        & \texttt{Brain-ID}~\cite{Liu2024BrainID} & 0.0277 & 0.0269 & 0.0008 & 20.98 & 22.31 & 39.62 & 0.844 & 0.881 & 0.892 \\ %[-0.5ex] 
        & \texttt{PEPSI}~\cite{liu2024pepsi} & 0.0295 & 0.0279 & N/A & 19.33 & 23.18 & N/A & 0.820 & 0.845 & N/A  \\ %[-0.5ex] 
        & \texttt{UNA} & \textbf{0.0184} & \textbf{0.0182} & \textbf{0.0003} & \textbf{25.14} & \textbf{26.22} & \textbf{45.69} & \textbf{0.938} & \textbf{0.981} & \textbf{0.998} \\ %[-0.5ex] 
         \hline \\[-2.5ex]
     
        \multicolumn{1}{c}{\multirow{4}{*}{\thead{\footnotesize{FLAIR}\\MRI}}} & \texttt{SynthSR}~\cite{Iglesias2023SynthSRAP} & 0.0327 & 0.0300 & 0.0016 & 19.30 & 21.04 & 34.88 & 0.823 & 0.869 & 0.895 \\ %[0.7ex]  
        & \texttt{Brain-ID}~\cite{Liu2024BrainID} & 0.0285 & 0.0242 & 0.0010 & 19.98 & 20.32 & 38.76 & 0.840 & 0.879 & 0.907 \\ %[-0.5ex] 
        & \texttt{PEPSI}~\cite{liu2024pepsi} & 0.0301 & 0.0287 & N/A & 19.82 & 21.59 & N/A & 0.842 & 0.850 & N/A  \\ %[-0.5ex] 
        & \texttt{UNA} & \textbf{0.0202} & \textbf{0.0194} & \textbf{0.0007} & \textbf{28.34} & \textbf{28.93} & \textbf{42.91} & \textbf{0.921} & \textbf{0.982} & \textbf{0.996} \\ %[-0.5ex] 
         \hline \\[-2.5ex]
     
     \multicolumn{1}{c}{\multirow{4}{*}{\footnotesize{CT}}} & \texttt{SynthSR}~\cite{Iglesias2023SynthSRAP} & 0.0541 & 0.0536 & 0.0029 & 13.97 & 13.13 & 28.50 & 0.712 & 0.763 & 0.725  \\ %[0.7ex]  
        & \texttt{Brain-ID}~\cite{Liu2024BrainID} & 0.0339 & 0.0357 & 0.0018 & 20.15 & 21.20 & 32.87 & 0.811 & 0.824 & 0.843  \\ %[-0.5ex] 
        & \texttt{PEPSI}~\cite{liu2024pepsi} & 0.0473 & 0.0420 & N/A & 16.72 & 16.90 & N/A & 0.723 & 0.782 & N/A \\ %[-0.5ex] 
        & \texttt{UNA} & \textbf{0.0259} & \textbf{0.0266} & \textbf{0.0010} & \textbf{25.63} & \textbf{25.70} & \textbf{42.53} & \textbf{0.883} & \textbf{0.897} & \textbf{0.895} \\ %[-0.5ex]  
         %\hline %\\[-2.3ex]

\bottomrule  \\ [-4ex]  
    \end{tabular} 
}
\end{table}

%% file: sec/exp/fig/fig_synth.tex
\begin{figure}[t]

\centering 

\resizebox{1\linewidth}{!}{
	\begin{tikzpicture}
        %\hspace*{-0.5cm}
        
		\tikzstyle{myarrows}=[line width=0.8mm,draw=blue!50,-triangle 45,postaction={draw, line width=0.05mm, shorten >=0.02mm, -}]
		\tikzstyle{mylines}=[line width=0.8mm]
  
		% Main size box
		%\draw[thin, color = black] (20, -8) -- (-2, -8) -- (-2, 3.42) -- (20, 3.42) -- (20, -8); 

	%%%%%%%%%%%%%%%%%%%%%%%%%%%%%%%%%%%%%%%%%%%%%%%%%

	\pgfmathsetmacro{\shift}{-0.9}

        %\node at (-0.2+ \shift+0.1*3, 3.3+0.15*3, 0.75*3) {Input};
	%\node at (-0.2+ \shift+0.1*3, 2+0.15*3, 0.75*3) {\includegraphics[width=0.22\textwidth]{fig/gen/def2_3.png}};
 
	\node at (\shift+1.45, 5.2+0.15*3, 0.75*3) {T1w};
	\node at (\shift+1.45, 4.8+0.15*3, 0.75*3) {(MRI)};
	\node at (\shift+1.45, 3.2+0.15*3, 0.75*3) {T2w};
	\node at (\shift+1.45, 2.8+0.15*3, 0.75*3) {(MRI)};
	\node at (\shift+1.45, 1.2+0.15*3, 0.75*3) {FLAIR};
	\node at (\shift+1.45, 0.8+0.15*3, 0.75*3) {(MRI)};
	\node at (\shift+1.45, -1+0.15*3, 0.75*3) {CT};
	%\node at (\shift+1.45, -1+0.15*3, 0.75*3) {BF};
        
	\pgfmathsetmacro{\shift}{-3.2}
	\node at (5.2+\shift+0.1*3, 6.2+0.15*3, 0.75*3) {Input};
	%\node at (7.2+\shift+0.1*3, 6.6+0.15*3, 0.75*3) {\texttt{SynthSR}};
	\node at (7.2+\shift+0.1*3, 6.2+0.15*3, 0.75*3) {\texttt{SynthSR}};
	\node at (9.2+\shift+0.1*3, 6.2+0.15*3, 0.75*3) {\texttt{Brain-ID}};
	\node at (11.2+\shift+0.1*3, 6.2+0.15*3, 0.75*3) {\texttt{PEPSI}};
	\node at (13.2+\shift+0.1*3, 6.2+0.15*3, 0.75*3) {\texttt{UNA}};
	\node at (15.2+\shift+0.1*3, 6.2+0.15*3, 0.75*3) {Ground Truth};

	\node at (5.2+\shift+0.1*3, 5+0.15*3, 0.75*3) {\includegraphics[width=0.22\textwidth]{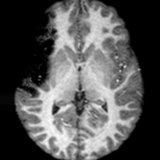}};
	\node at (7.2+\shift+0.1*3, 5+0.15*3, 0.75*3) {\includegraphics[width=0.22\textwidth]{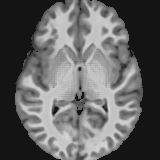}}; 
	\node at (9.2+\shift+0.1*3, 5+0.15*3, 0.75*3) {\includegraphics[width=0.22\textwidth]{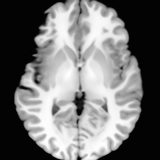}}; 
	\node at (11.2+\shift+0.1*3, 5+0.15*3, 0.75*3) {\includegraphics[width=0.22\textwidth]{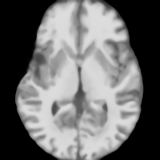}};
	\node at (13.2+\shift+0.1*3, 5+0.15*3, 0.75*3) {\includegraphics[width=0.22\textwidth]{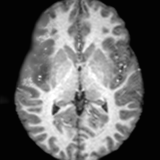}}; 
	\node at (15.2+\shift+0.1*3, 5+0.15*3, 0.75*3) {\includegraphics[width=0.22\textwidth]{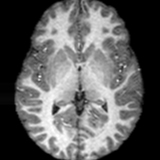}};

        \foreach \i in {5.2, 7.2, 9.2, 11.2, 13.2, 15.2}
        {
        \draw[fill=none, draw=hc, line width=0.6mm] (\i+\shift+0.1*3-0.45, 5+0.15*3+0.25, 0.75*3) ellipse (3mm and 6mm);
        }

	%%%%%%%%%%%%%%%%%%%%%%%%%%%%%%%%%%%%%%%%%%%%%%%%%

	\node at (5.2+\shift+0.1*3, 3+0.15*3, 0.75*3) {\includegraphics[width=0.22\textwidth]{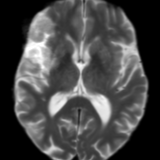}};
	\node at (7.2+\shift+0.1*3, 3+0.15*3, 0.75*3) {\includegraphics[width=0.22\textwidth]{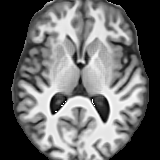}}; 
	\node at (9.2+\shift+0.1*3, 3+0.15*3, 0.75*3) {\includegraphics[width=0.22\textwidth]{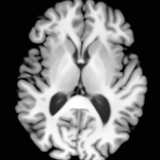}}; 
	\node at (11.2+\shift+0.1*3, 3+0.15*3, 0.75*3) {\includegraphics[width=0.22\textwidth]{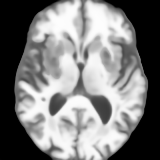}};
	\node at (13.2+\shift+0.1*3, 3+0.15*3, 0.75*3) {\includegraphics[width=0.22\textwidth]{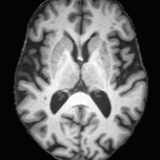}}; 
	\node at (15.2+\shift+0.1*3, 3+0.15*3, 0.75*3) {\includegraphics[width=0.22\textwidth]{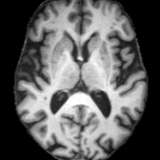}};

        \foreach \i in {5.2, 7.2, 9.2, 11.2, 13.2, 15.2}
        {
        \draw[fill=none, draw=hc, line width=0.6mm] (\i+\shift+0.1*3-0.45, 3+0.15*3+0.3, 0.75*3) ellipse (3mm and 5.2mm);
        }

	%%%%%%%%%%%%%%%%%%%%%%%%%%%%%%%%%%%%%%%%%%%%%%%%%

	\node at (5.2+\shift+0.1*3, 1+0.15*3, 0.75*3) {\includegraphics[width=0.22\textwidth]{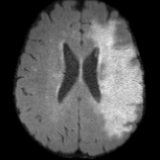}};
	\node at (7.2+\shift+0.1*3, 1+0.15*3, 0.75*3) {\includegraphics[width=0.22\textwidth]{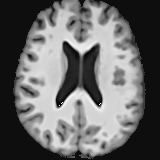}}; 
	\node at (9.2+\shift+0.1*3, 1+0.15*3, 0.75*3) {\includegraphics[width=0.22\textwidth]{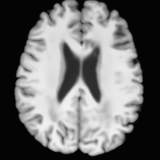}}; 
	\node at (11.2+\shift+0.1*3, 1+0.15*3, 0.75*3) {\includegraphics[width=0.22\textwidth]{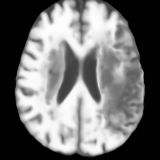}};
	\node at (13.2+\shift+0.1*3, 1+0.15*3, 0.75*3) {\includegraphics[width=0.22\textwidth]{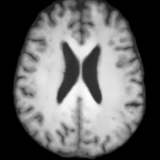}}; 
	\node at (15.2+\shift+0.1*3, 1+0.15*3, 0.75*3) {\includegraphics[width=0.22\textwidth]{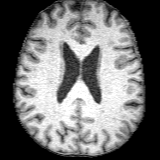}};

        \foreach \i in {5.2, 7.2, 9.2, 11.2, 13.2, 15.2}
        {
        \draw[fill=none, draw=hc, line width=0.6mm] (\i+\shift+0.1*3+0.4, 1+0.15*3+0., 0.75*3) ellipse (3.5mm and 7.3mm);
        }

	%%%%%%%%%%%%%%%%%%%%%%%%%%%%%%%%%%%%%%%%%%%%%%%%%

	\node at (5.2+\shift+0.1*3, -1+0.15*3, 0.75*3) {\includegraphics[width=0.22\textwidth]{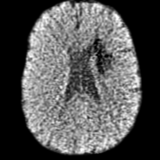}};
	\node at (7.2+\shift+0.1*3, -1+0.15*3, 0.75*3) {\includegraphics[width=0.22\textwidth]{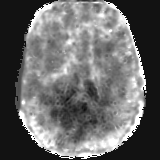}}; 
	\node at (9.2+\shift+0.1*3, -1+0.15*3, 0.75*3) {\includegraphics[width=0.22\textwidth]{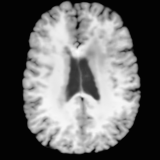}}; 
	\node at (11.2+\shift+0.1*3, -1+0.15*3, 0.75*3) {\includegraphics[width=0.22\textwidth]{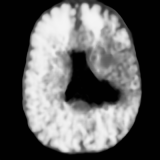}};
	\node at (13.2+\shift+0.1*3, -1+0.15*3, 0.75*3) {\includegraphics[width=0.22\textwidth]{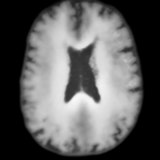}}; 
	\node at (15.2+\shift+0.1*3, -1+0.15*3, 0.75*3) {\includegraphics[width=0.22\textwidth]{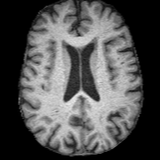}};

        \foreach \i in {5.2, 7.2, 9.2, 11.2, 13.2, 15.2}
        {
        \draw[fill=none, draw=hc, line width=0.6mm] (\i+\shift+0.1*3+0.25, -1+0.15*3+0.27, 0.75*3) ellipse (2.5mm and 3.5mm);
        }

	%%%%%%%%%%%%%%%%%%%%%%%%%%%%%%%%%%%%%%%%%%%%%%%%%

	\end{tikzpicture}
	}   
    \vspace{-0.6cm}
\caption{Qualitative comparisons on healthy anatomy reconstruction, between \texttt{UNA}, and the state-of-the-art modality-agnostic T1w synthesis method. %: \texttt{SynthSR}~\cite{Iglesias2023SynthSRAP}, \texttt{Brain-ID}~\cite{Liu2024BrainID}, \texttt{PEPSI}~\cite{liu2024pepsi}. 
Testing images are generated from real healthy subjects encoded with randomly simulated pathology profiles. Pathology regions are circled in {\color{hc}red}.
%The \textcolor{mint!200}{mint} circles highlight some details.
} 

	 \label{fig: synth} 
\end{figure}

%% file: sec/exp/fig/fig_real.tex
\begin{figure}[t]

\centering 

\resizebox{1\linewidth}{!}{
	\begin{tikzpicture}
        %\hspace*{-0.5cm}
        
		\tikzstyle{myarrows}=[line width=0.8mm,draw=blue!50,-triangle 45,postaction={draw, line width=0.05mm, shorten >=0.02mm, -}]
		\tikzstyle{mylines}=[line width=0.8mm]
  
		% Main size box
		%\draw[thin, color = black] (20, -8) -- (-2, -8) -- (-2, 3.42) -- (20, 3.42) -- (20, -8); 

	%%%%%%%%%%%%%%%%%%%%%%%%%%%%%%%%%%%%%%%%%%%%%%%%%

	\pgfmathsetmacro{\shift}{-0.9}

        %\node at (-0.2+ \shift+0.1*3, 3.3+0.15*3-0.1, 0.75*3) {Input};
	%\node at (-0.2+ \shift+0.1*3, 2+0.15*3, 0.75*3) {\includegraphics[width=0.22\textwidth]{fig/gen/def2_3.png}};
 
	\node at (\shift+1.45, 3.2+0.15*3-0.1, 0.75*3) {FLAIR};
	\node at (\shift+1.45, 2.8+0.15*3-0.1, 0.75*3) {(\texttt{ISLES})};
	\node at (\shift+1.45, 1.2+0.15*3-0.2, 0.75*3) {FLAIR};
	\node at (\shift+1.45, 0.8+0.15*3-0.2, 0.75*3) {(\texttt{ISLES})};
	\node at (\shift+1.45, -0.8+0.15*3-0.25, 0.75*3) {T1w};
	\node at (\shift+1.45, -1.2+0.15*3-0.25, 0.75*3) {(\texttt{ATLAS})};
	\node at (\shift+1.45, -2.6+0.15*3-0.325, 0.75*3) {T1w};
	\node at (\shift+1.45, -3+0.15*3-0.325, 0.75*3) {(\texttt{ATLAS})};
	\node at (\shift+1.45, -3.4+0.15*3-0.325, 0.75*3) {{\color{red}\xmark\textit{Failed}}};
        
	\pgfmathsetmacro{\shift}{-3.2}
	\node at (5.2+\shift+0.1*3, 4.2+0.15*3, 0.75*3) {Input};
	%\node at (7.2+\shift+0.1*3, 6.6+0.15*3, 0.75*3) {\texttt{SynthSR}};
	\node at (7.2+\shift+0.1*3, 4.2+0.15*3, 0.75*3) {\texttt{SynthSR}};
	\node at (9.2+\shift+0.1*3, 4.2+0.15*3, 0.75*3) {\texttt{Brain-ID}};
	\node at (11.2+\shift+0.1*3, 4.2+0.15*3, 0.75*3) {\texttt{PEPSI}};
	\node at (13.2+\shift+0.1*3, 4.2+0.15*3, 0.75*3) {\texttt{UNA}};
	\node at (15.2+\shift+0.1*3, 4.2+0.15*3, 0.75*3) {Lesion};

	%%%%%%%%%%%%%%%%%%%%%%%%%%%%%%%%%%%%%%%%%%%%%%%%%

	\node at (5.2+\shift+0.1*3, 3+0.15*3-0.1, 0.75*3) {\includegraphics[width=0.22\textwidth]{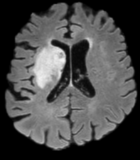}};
	\node at (7.2+\shift+0.1*3, 3+0.15*3-0.1, 0.75*3) {\includegraphics[width=0.22\textwidth]{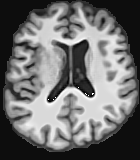}}; 
	\node at (9.2+\shift+0.1*3, 3+0.15*3-0.1, 0.75*3) {\includegraphics[width=0.22\textwidth]{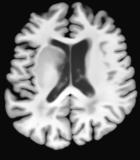}}; 
	\node at (11.2+\shift+0.1*3, 3+0.15*3-0.1, 0.75*3) {\includegraphics[width=0.22\textwidth]{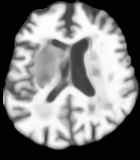}};
	\node at (13.2+\shift+0.1*3, 3+0.15*3-0.1, 0.75*3) {\includegraphics[width=0.22\textwidth]{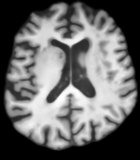}}; 
	\node at (15.2+\shift+0.1*3, 3+0.15*3-0.1, 0.75*3) {\includegraphics[width=0.22\textwidth]{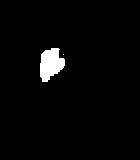}};

        \foreach \i in {5.2, 7.2, 9.2, 11.2, 13.2}
        {
        \draw[fill=none, draw=hc, line width=0.6mm] (\i+\shift+0.1*3-0.25, 3+0.15*3-0.1+0.2, 0.75*3) ellipse (4mm and 4mm);
        }

	%%%%%%%%%%%%%%%%%%%%%%%%%%%%%%%%%%%%%%%%%%%%%%%%%

	\node at (5.2+\shift+0.1*3, 1+0.15*3-0.2, 0.75*3) {\includegraphics[width=0.22\textwidth]{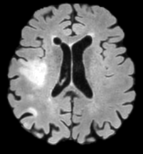}};
	\node at (7.2+\shift+0.1*3, 1+0.15*3-0.2, 0.75*3) {\includegraphics[width=0.22\textwidth]{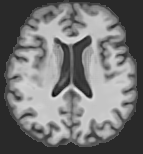}}; 
	\node at (9.2+\shift+0.1*3, 1+0.15*3-0.2, 0.75*3) {\includegraphics[width=0.22\textwidth]{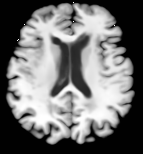}};
	\node at (11.2+\shift+0.1*3, 1+0.15*3-0.2, 0.75*3) {\includegraphics[width=0.22\textwidth]{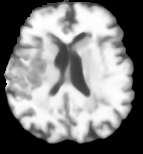}}; 
	\node at (13.2+\shift+0.1*3, 1+0.15*3-0.2, 0.75*3) {\includegraphics[width=0.22\textwidth]{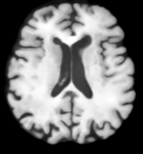}};
	\node at (15.2+\shift+0.1*3, 1+0.15*3-0.2, 0.75*3) {\includegraphics[width=0.22\textwidth]{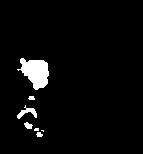}}; 

        \foreach \i in {5.2, 7.2, 9.2, 11.2, 13.2}
        {
        \draw[fill=none, draw=hc, line width=0.6mm] (\i+\shift+0.1*3-0.45, 1+0.15*3-0.15, 0.75*3) ellipse (3.mm and 3.mm);
        }

	%%%%%%%%%%%%%%%%%%%%%%%%%%%%%%%%%%%%%%%%%%%%%%%%%

	\node at (5.2+\shift+0.1*3, -1+0.15*3-0.25, 0.75*3) {\includegraphics[width=0.22\textwidth]{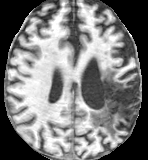}};
	\node at (7.2+\shift+0.1*3, -1+0.15*3-0.25, 0.75*3) {\includegraphics[width=0.22\textwidth]{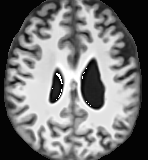}}; 
	\node at (9.2+\shift+0.1*3, -1+0.15*3-0.25, 0.75*3) {\includegraphics[width=0.22\textwidth]{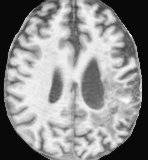}}; 
	\node at (11.2+\shift+0.1*3, -1+0.15*3-0.25, 0.75*3) {\includegraphics[width=0.22\textwidth]{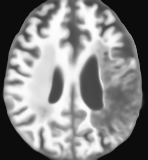}};
	\node at (13.2+\shift+0.1*3, -1+0.15*3-0.25, 0.75*3) {\includegraphics[width=0.22\textwidth]{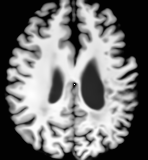}}; 
	\node at (15.2+\shift+0.1*3, -1+0.15*3-0.25, 0.75*3) {\includegraphics[width=0.22\textwidth]{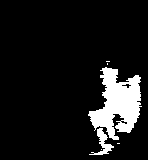}};

        \foreach \i in {5.2, 7.2, 9.2, 11.2, 13.2}
        {
        \draw[fill=none, draw=hc, line width=0.6mm] (\i+\shift+0.1*3+0.5, -1+0.15*3-0.3255, 0.75*3) ellipse (4mm and 7mm);
        }

	%%%%%%%%%%%%%%%%%%%%%%%%%%%%%%%%%%%%%%%%%%%%%%%%%

	\node at (5.2+\shift+0.1*3, -3+0.15*3-0.325, 0.75*3) {\includegraphics[width=0.22\textwidth]{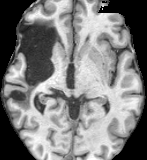}};
	\node at (7.2+\shift+0.1*3, -3+0.15*3-0.325, 0.75*3) {\includegraphics[width=0.22\textwidth]{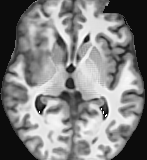}}; 
	\node at (9.2+\shift+0.1*3, -3+0.15*3-0.325, 0.75*3) {\includegraphics[width=0.22\textwidth]{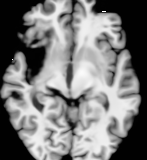}}; 
	\node at (11.2+\shift+0.1*3, -3+0.15*3-0.325, 0.75*3) {\includegraphics[width=0.22\textwidth]{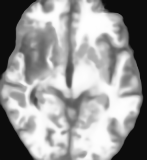}};
	\node at (13.2+\shift+0.1*3, -3+0.15*3-0.325, 0.75*3) {\includegraphics[width=0.22\textwidth]{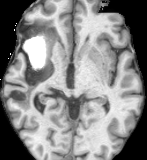}}; 
	\node at (15.2+\shift+0.1*3, -3+0.15*3-0.325, 0.75*3) {\includegraphics[width=0.22\textwidth]{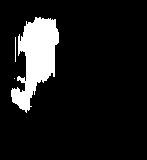}};

        \foreach \i in {5.2, 7.2, 9.2, 11.2, 13.2}
        {
        \draw[fill=none, draw=hc, line width=0.6mm] (\i+\shift+0.1*3-0.45, -3+0.15*3-0.1+0.1, 0.75*3) ellipse (4mm and 6mm);
        }

	%%%%%%%%%%%%%%%%%%%%%%%%%%%%%%%%%%%%%%%%%%%%%%%%%

	\end{tikzpicture}
	}    
    \vspace{-0.6cm}

\caption{Qualitative comparisons on healthy anatomy reconstruction between \texttt{UNA} and state-of-the-art modality-agnostic synthesis models. %: \texttt{SynthSR}~\cite{Iglesias2023SynthSRAP}, \texttt{Brain-ID}~\cite{Liu2024BrainID}, \texttt{PEPSI}~\cite{liu2024pepsi}. 
Testing images are from real stroke datasets (\texttt{ISLES}~\cite{Hernandez2022ISLES} and \texttt{ATLAS}~\cite{Liew2017ATLAS}), where the stroke lesion annotations are provided, yet the ground truth healthy anatomy is unavailable. The last row shows a failure case of \texttt{UNA}, where it ``over-corrects'' the diseased anatomy. Pathology regions are circled in {\color{hc}red}.
} 

	 \label{fig: real} 
\end{figure}

%% file: sec/exp/fig/tab_real.tex
\begin{table}[t]
\caption{Quantitative comparisons of healthy anatomy reconstruction performance between \texttt{UNA} and state-of-the-art, contrast-agnostic T1w synthesis models, evaluated on real images. Since we do not have ground truth anatomy for the stroke datasets, we only report the reconstruction performance within healthy regions. 
(\texttt{ISLES}~\cite{Hernandez2022ISLES} stroke dataset does not provide T1w MRI scans, therefore we only show qualitative results on \texttt{ISLES} in \cref{fig: real}.)
    } 
    \label{tab: real}

\resizebox{\linewidth}{!}{
\centering 
    \begin{tabular}{cclccc}
    %\begin{tabular}{cccccccccccc}
       \toprule \\[-3ex] 
      \multicolumn{1}{c}{\multirow{2}{*}{\textbf{Modality}}} & \multicolumn{1}{c}{\multirow{2}{*}{\thead{\normalsize \textbf{Dataset}\\\normalsize (Train/Test)}}} & \multicolumn{1}{c}{\multirow{2}{*}{\textbf{Method}}} & \multicolumn{3}{c}{\textbf{Reconstruction} (\textit{on Healthy})}  \\ %[-0.2ex]
         & & & \texttt{\textbf{L1}} ($\downarrow$) & \texttt{\textbf{PSNR}} ($\uparrow$) & \texttt{\textbf{SSIM}} ($\uparrow$) \\ [-0.2ex]
     \midrule[\heavyrulewidth]\\[-3ex]

%\thead{$\mu^r$\\($\downarrow$)
        \multicolumn{1}{c}{\multirow{24}{*}{\thead{\normalsize{T1w}\\ \vspace*{-0.2cm}~\\ \normalsize{MRI}}}} & {\multirow{4}{*}{\thead{\normalsize \texttt{ADNI}~\cite{Jack2008TheAD} \\\vspace*{-0.2cm}~\\ \normalsize (1841/204)}}} 
        & \texttt{SynthSR}~\cite{Iglesias2023SynthSRAP} & 0.014 & 26.78 & 0.984     \\ %[0.7ex] 
        & & \texttt{Brain-ID}~\cite{Liu2024BrainID} & \textbf{0.012} & \textbf{33.82} & 0.993 \\ %[0.7ex]  
        & & \texttt{PEPSI}~\cite{liu2024pepsi} & 0.014 & 31.25 & 0.989 \\ % 
        & & {\texttt{UNA}} & \textbf{0.012} & 32.96 &  \textbf{0.995}  \\ %
        %\hline\\[-2.3ex]
        \cmidrule(lr){2-6}
        
         %& \parbox[t]{2mm}{\multirow{5}{*}{\rotatebox[origin=c]{90}{\texttt{HCP}}}}
        & \multirow{4}{*}{\thead{\normalsize \texttt{HCP}~\cite{Essen2012TheHC} \\\vspace*{-0.2cm}~\\  \normalsize (808/87)}}
        & \texttt{SynthSR}~\cite{Iglesias2023SynthSRAP} & 0.033 & 22.13 & 0.854     \\ %[0.7ex] 
        & & \texttt{Brain-ID}~\cite{Liu2024BrainID} & 0.020 & 27.47 & 0.957 \\ %[0.7ex]  
        & & \texttt{PEPSI}~\cite{liu2024pepsi} & 0.023 & 28.20 & 0.971 \\ % 
        & & {\texttt{UNA}} & \textbf{0.017} & \textbf{31.61} &  \textbf{0.986}  \\ %
        %\hline\\[-2.3ex]
        \cmidrule(lr){2-6}
        
         %& \parbox[t]{2mm}{\multirow{5}{*}{\rotatebox[origin=c]{90}{\texttt{(ADNI3)}}}}
        & \multirow{4}{*}{\thead{\normalsize \texttt{ADNI3}~\cite{Weiner2017TheAD} \\ \vspace*{-0.2cm}~\\ \normalsize (298/33)}}
        & \texttt{SynthSR}~\cite{Iglesias2023SynthSRAP} & 0.023 & 23.60 & 0.928     \\ %[0.7ex] 
        & & \texttt{Brain-ID}~\cite{Liu2024BrainID} & 0.021 & 29.89 & 0.966 \\ %[0.7ex]  
        & & \texttt{PEPSI}~\cite{liu2024pepsi} & 0.020 & 26.67 & 0.935 \\ % 
        & & {\texttt{UNA}} & \textbf{0.019} & \textbf{30.01} & \textbf{0.975} \\ %
        %\hline\\[-2.3ex]
        \cmidrule(lr){2-6}
        
         %& \parbox[t]{2mm}{\multirow{5}{*}{\rotatebox[origin=c]{90}{\texttt{ADHD200}}}}
        & \multirow{4}{*}{\thead{\normalsize \texttt{ADHD200}~\cite{Brown2012ADHD200GC} \\ \vspace*{-0.2cm}~\\ \normalsize (865/96)}}
        & \texttt{SynthSR}~\cite{Iglesias2023SynthSRAP} & 0.035 & 21.67 & 0.882     \\ %[0.7ex] 
        & & \texttt{Brain-ID}~\cite{Liu2024BrainID} & \textbf{0.011} & 32.48 & 0.996 \\ %[0.7ex]  
        & & \texttt{PEPSI}~\cite{liu2024pepsi} & 0.015 & 29.87 & 0.976 \\ %  
        & & {\texttt{UNA}} & 0.012 & \textbf{30.12} & \textbf{0.980}  \\ %
        %\hline\\[-2.3ex]
        \cmidrule(lr){2-6}
        
         %& \parbox[t]{2mm}{\multirow{5}{*}{\rotatebox[origin=c]{90}{\texttt{ADHD200}}}}
        & \multirow{4}{*}{\thead{\normalsize \texttt{AIBL}~\cite{Fowler2021FifteenYO} \\ \vspace*{-0.2cm}~\\ \normalsize (601/67)}}
        & \texttt{SynthSR}~\cite{Iglesias2023SynthSRAP} & 0.026 & 22.95 & 0.916   \\ %[0.7ex] 
        & & \texttt{Brain-ID}~\cite{Liu2024BrainID} & \textbf{0.009} & \textbf{33.73} & \textbf{0.972} \\ %[0.7ex]  
        & & \texttt{PEPSI}~\cite{liu2024pepsi} & 0.012 & 29.86 & 0.950 \\ %  
        & & {\texttt{UNA}} & 0.010 & 32.89 & 0.964 \\ %
        %\hline\\[-2.3ex]
        \cmidrule(lr){2-6}
        
         %& \parbox[t]{2mm}{\multirow{5}{*}{\rotatebox[origin=c]{90}{\texttt{AIBL}}}}
        & \multirow{4}{*}{\thead{\normalsize \textbf{\color{hc}*~Stroke~*}\\\vspace*{-0.2cm}~\\{\normalsize\texttt{ATLAS}~\cite{Liew2017ATLAS}}\\ \vspace*{-0.2cm}~\\ \normalsize ({590/65})}}
        & \texttt{SynthSR}~\cite{Iglesias2023SynthSRAP} & 0.030 & 23.50 & 0.881   \\ %[0.7ex] 
        & & \texttt{Brain-ID}~\cite{Liu2024BrainID} & 0.027 & 26.09 & 0.892 \\ %[0.7ex]  
        & & \texttt{PEPSI}~\cite{liu2024pepsi} & 0.025 & 26.73 & 0.905 \\ % 
        & & {\texttt{UNA}} & \textbf{0.020} & \textbf{29.10} & \textbf{0.974} \\ %
        %\hline\\[-2.3ex]
        %\cmidrule(lr){2-12}
        \hline\\[-2.3ex] 

        %%%%%%%%%%%%%%%%%%%%%%%%%%%%%%%%%%%%%%%%%%%%%%%%%%%

        \multicolumn{1}{c}{\multirow{8}{*}{\thead{\normalsize{T2w}\\ \vspace*{-0.2cm}~\\ \normalsize{MRI}}}} & 
        %\parbox[t]{2mm}{\multirow{5}{*}{\rotatebox[origin=c]{90}{\texttt{HCP}}}}
        \multirow{4}{*}{\thead{\normalsize \texttt{HCP}~\cite{Essen2012TheHC} \\ \vspace*{-0.2cm}~\\ \normalsize (808/87)}}
        & \texttt{SynthSR}~\cite{Iglesias2023SynthSRAP} & 0.034 & 21.46 & 0.833     \\ %[0.7ex] 
        & & \texttt{Brain-ID}~\cite{Liu2024BrainID} & \textbf{0.016} & 28.10 & 0.934 \\ %[0.7ex]  
        & & \texttt{PEPSI}~\cite{liu2024pepsi} & 0.018 & 26.45 & 0.915 \\ %  
        & & {\texttt{UNA}} & \textbf{0.016} & \textbf{28.62} & \textbf{0.949} \\ %
        %\hline\\[-2.3ex]
        \cmidrule(lr){2-6}
        
        % & \parbox[t]{2mm}{\multirow{5}{*}{\rotatebox[origin=c]{90}{\texttt{AIBL}}}}
        & \multirow{4}{*}{\thead{\normalsize \texttt{AIBL}~\cite{Fowler2021FifteenYO} \\ \vspace*{-0.2cm}~\\ \normalsize (272/30)}}
        & \texttt{SynthSR}~\cite{Iglesias2023SynthSRAP} & 0.033 & 20.08 & 0.805     \\ %[0.7ex] 
        & & \texttt{Brain-ID}~\cite{Liu2024BrainID} & 0.022 & 23.99 & 0.861  \\ %[0.7ex]  
        & & \texttt{PEPSI}~\cite{liu2024pepsi} & 0.024 & 22.93 & 0.859 \\ % 
        & & {\texttt{UNA}} & \textbf{0.021} & \textbf{24.76} & \textbf{0.892}  \\ %
        %\hline\\[-2.3ex]
        %\cmidrule(lr){2-12}
        \hline\\[-2.3ex]

        %%%%%%%%%%%%%%%%%%%%%%%%%%%%%%%%%%%%%%%%%%%%%%%%%%%

        \multicolumn{1}{c}{\multirow{8}{*}{\thead{\normalsize{FLAIR}\\ \vspace*{-0.2cm}~\\ \normalsize{MRI}}}} & 
        %\parbox[t]{2mm}{\multirow{5}{*}{\rotatebox[origin=c]{90}{\texttt{(ADNI3)}}}}
        \multirow{4}{*}{\thead{\normalsize \texttt{ADNI3}~\cite{Weiner2017TheAD} \\ \vspace*{-0.2cm}~\\ \normalsize (298/33)}}
        & \texttt{SynthSR}~\cite{Iglesias2023SynthSRAP} & 0.026 & 22.77 & 0.919    \\ %[0.7ex] 
        & & \texttt{Brain-ID}~\cite{Liu2024BrainID} & 0.017 & 26.44 & 0.927  \\ %[0.7ex]  
        & & \texttt{PEPSI}~\cite{liu2024pepsi} & 0.023 & 25.62 & 0.929 \\ % 
        & & {\texttt{UNA}} & \textbf{0.015} & \textbf{27.43} & \textbf{0.965}  \\ %
        %\hline\\[-2.3ex]
        \cmidrule(lr){2-6}
        
         %& \parbox[t]{2mm}{\multirow{5}{*}{\rotatebox[origin=c]{90}{\texttt{AIBL}}}}
        & \multirow{4}{*}{\thead{\normalsize \texttt{AIBL}~\cite{Fowler2021FifteenYO} \\ \vspace*{-0.2cm}~\\ \normalsize (302/34)}}
        & \texttt{SynthSR}~\cite{Iglesias2023SynthSRAP} & 0.029 & 21.77 & 0.902    \\ %[0.7ex] 
        & & \texttt{Brain-ID}~\cite{Liu2024BrainID} & 0.019 & 27.25 & 0.936  \\ %[0.7ex]  
        & & \texttt{PEPSI}~\cite{liu2024pepsi} & 0.021 & 25.43 & 0.914 \\ %
        & & {\texttt{UNA}} & \textbf{0.017} & \textbf{27.76} & \textbf{0.967} \\ %
        %\hline\\[-2.3ex]
        %\cmidrule(lr){2-6}
        \hline\\[-2.3ex]

        %%%%%%%%%%%%%%%%%%%%%%%%%%%%%%%%%%%%%%%%%%%%%%%%%%%
        
        \multicolumn{1}{c}{\multirow{4}{*}{CT}} 
        %& \parbox[t]{2mm}{\multirow{5}{*}{\rotatebox[origin=c]{90}{\texttt{(OASIS3)}}}}
        & \multirow{4}{*}{\thead{\normalsize \texttt{OASIS3}~\cite{LaMontagne2018OASIS3LN} \\ \vspace*{-0.2cm}~\\ \normalsize  (795/88)}}
        & \texttt{SynthSR}~\cite{Iglesias2023SynthSRAP} & 0.041 & 20.93 & 0.758    \\ %[0.7ex] 
        & & \texttt{Brain-ID}~\cite{Liu2024BrainID} & 0.023 & 25.49 & 0.891 \\ %[0.7ex]  
        & & \texttt{PEPSI}~\cite{liu2024pepsi} & 0.027 & 22.98 & 0.842 \\ %
        & & {\texttt{UNA}} & \textbf{0.022} & \textbf{25.68} & \textbf{0.897} \\ %
        %\hline\\[-2.3ex]
        %\cmidrule(lr){2-12}
    
%\midrule[\heavyrulewidth] \\ [-3.2ex]
%\multicolumn{8}{l}{\small (1) } \\ [-0.5ex]
%\multicolumn{8}{l}{\small (2) } \\ [-0.5ex] 

\bottomrule  \\ [-3.6ex]  
    \end{tabular}   
}

\end{table}

%% file: sec/exp/fig/fig_anomaly.tex
\begin{figure}[t]

\centering 

\resizebox{1\linewidth}{!}{
	\begin{tikzpicture}
        %\hspace*{-0.5cm}
        
		\tikzstyle{myarrows}=[line width=0.8mm,draw=blue!50,-triangle 45,postaction={draw, line width=0.05mm, shorten >=0.02mm, -}]
		\tikzstyle{mylines}=[line width=0.8mm]
  
		% Main size box
		%\draw[thin, color = black] (20, -8) -- (-2, -8) -- (-2, 3.42) -- (20, 3.42) -- (20, -8); 

	%%%%%%%%%%%%%%%%%%%%%%%%%%%%%%%%%%%%%%%%%%%%%%%%%

	\pgfmathsetmacro{\shift}{-0.9}

	\node at (\shift+1.45, 5+0.15*3, 0.75*3) {Input}; 
	\node at (\shift+1.45, 3.2+0.15*3, 0.75*3) {\texttt{UNA}}; 
	\node at (\shift+1.45, 2.8+0.15*3, 0.75*3) {Output}; 
	\node at (\shift+1.45, 1.2+0.15*3, 0.75*3) {Estimated};
	\node at (\shift+1.45, 0.8+0.15*3, 0.75*3) {Anomaly}; 
	%\node at (\shift+1.45, -0.6+0.15*3, 0.75*3) {Ground};
	%\node at (\shift+1.45, -1+0.15*3, 0.75*3) {Truth};
	%\node at (\shift+1.45, -1.4+0.15*3, 0.75*3) {Anomaly};

	\pgfmathsetmacro{\shift}{-3.2}

	%\node at (5.20+\shift+0.1*3, 6.2+0.15*3, 0.75*3) {\textit{(i)}}; 
	%\node at (7.20+\shift+0.1*3, 6.2+0.15*3, 0.75*3) {\textit{(ii)}};
	%\node at (9.20+\shift+0.1*3, 6.2+0.15*3, 0.75*3) {\textit{(iii)}};
	%\node at (11.2+\shift+0.1*3, 6.2+0.15*3, 0.75*3) {\textit{(iv)}};
	%\node at (13.2+\shift+0.1*3, 6.2+0.15*3, 0.75*3) {\textit{(v)}};
	%\node at (15.2+\shift+0.1*3, 6.2+0.15*3, 0.75*3) {\textit{(vi)}};

	\node at (9.2+\shift+0.1*3, 5+0.15*3, 0.75*3) {\includegraphics[width=0.22\textwidth]{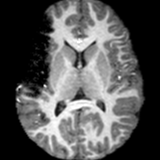}};
	\node at (13.20+\shift+0.1*3, 5+0.15*3, 0.75*3) {\includegraphics[width=0.22\textwidth]{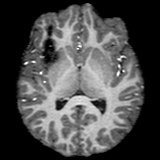}};
	\node at (11.20+\shift+0.1*3, 5+0.15*3, 0.75*3) {\includegraphics[width=0.22\textwidth]{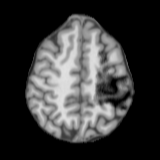}};
	\node at (15.2+\shift+0.1*3, 5+0.15*3, 0.75*3) {\includegraphics[width=0.22\textwidth]{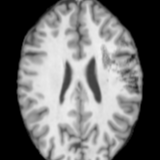}};
	\node at (5.2+\shift+0.1*3, 5+0.15*3, 0.75*3) {\includegraphics[width=0.22\textwidth]{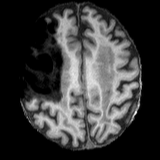}};
	\node at (7.2+\shift+0.1*3, 5+0.15*3, 0.75*3) {\includegraphics[width=0.22\textwidth]{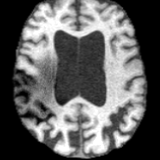}};

	%%%%%%%%%%%%%%%%%%%%%%%%%%%%%%%%%%%%%%%%%%%%%%%%%

	\node at (9.2+\shift+0.1*3, 3+0.15*3, 0.75*3) {\includegraphics[width=0.22\textwidth]{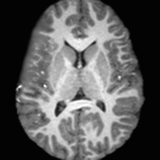}};
	\node at (13.20+\shift+0.1*3, 3+0.15*3, 0.75*3) {\includegraphics[width=0.22\textwidth]{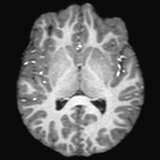}};
	\node at (11.20+\shift+0.1*3, 3+0.15*3, 0.75*3) {\includegraphics[width=0.22\textwidth]{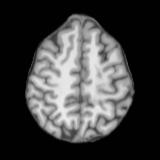}};
	\node at (15.2+\shift+0.1*3, 3+0.15*3, 0.75*3) {\includegraphics[width=0.22\textwidth]{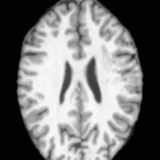}};
	\node at (5.2+\shift+0.1*3, 3+0.15*3, 0.75*3) {\includegraphics[width=0.22\textwidth]{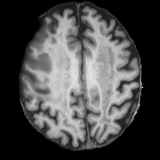}};
	\node at (7.2+\shift+0.1*3, 3+0.15*3, 0.75*3) {\includegraphics[width=0.22\textwidth]{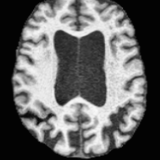}};

	%%%%%%%%%%%%%%%%%%%%%%%%%%%%%%%%%%%%%%%%%%%%%%%%%

	\node at (9.2+\shift+0.1*3, 1+0.15*3, 0.75*3) {\includegraphics[width=0.22\textwidth]{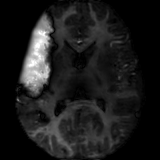}};
	\node at (13.20+\shift+0.1*3, 1+0.15*3, 0.75*3) {\includegraphics[width=0.22\textwidth]{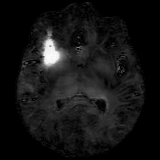}};
	\node at (11.20+\shift+0.1*3, 1+0.15*3, 0.75*3) {\includegraphics[width=0.22\textwidth]{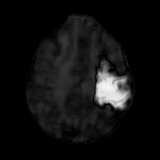}};
	\node at (15.20+\shift+0.1*3, 1+0.15*3, 0.75*3) {\includegraphics[width=0.22\textwidth]{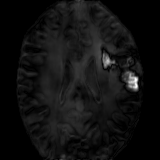}};
	\node at (5.2+\shift+0.1*3, 1+0.15*3, 0.75*3) {\includegraphics[width=0.22\textwidth]{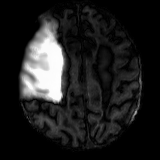}};
	\node at (7.2+\shift+0.1*3, 1+0.15*3, 0.75*3) {\includegraphics[width=0.22\textwidth]{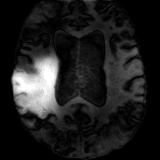}};

	%%%%%%%%%%%%%%%%%%%%%%%%%%%%%%%%%%%%%%%%%%%%%%%%%

	\end{tikzpicture}
	}   
    \vspace{-0.6cm}
\caption{Visualizations of directly applying \texttt{UNA}'s healthy anatomy reconstruction for anomaly detection. The estimated anomaly is computed as the \textit{absolute difference} between diseased T1w MRI scans and \texttt{UNA}'s reconstructed healthy anatomy. 
%The testing images are generated from real healthy T1w MRI encoded with simulated pathology.
} 

	 \label{fig: anomaly} 
\end{figure}

%% file: sec/exp/fig/tab_anomaly.tex
\begin{table}[t]
\caption{\texttt{Dice} scores ($\uparrow$) of downstream anomaly detection performance based on the voxel-wise absolute differences between the diseased input and the reconstruction. The testing images include healthy T1w MRI scans with simulated pathology, and real T1w MRI images from stroke patients in \texttt{ATLAS}~\cite{Liew2017ATLAS} dataset.
    } 
    \label{tab: anomaly}

\resizebox{\linewidth}{!}{
\centering 
    \begin{tabular}{clccccccccc}
    %\begin{tabular}{cccccccccccc}
       \toprule \\[-3ex] 
      \multicolumn{1}{c}{\multirow{1}{*}{\footnotesize\textbf{Image Source}}} & \multicolumn{1}{c}{\multirow{1}{*}{\footnotesize\textbf{Dataset}}} & \multicolumn{1}{c}{\multirow{1}{*}{\footnotesize{\texttt{SynthSR}~\cite{Iglesias2023SynthSRAP}}}} & \multicolumn{1}{c}{\multirow{1}{*}{\footnotesize{\texttt{Brain-ID}~\cite{Liu2024BrainID}}}} & 
      {\footnotesize{\texttt{VAE}~\cite{baur2021autoencoders}}} & {\footnotesize{\texttt{LDM}~\cite{graham2023unsupervised}}} & 
      {\footnotesize{\texttt{UNA}}} \\ [-0.2ex]
     \midrule\\[-2.4ex]

        \multicolumn{1}{c}{\multirow{5}{*}{\thead{\footnotesize{Healthy T1w}\\\footnotesize{with}\\\footnotesize{Simulated}\\\footnotesize{Pathology}}}} & \texttt{ADNI}~\cite{Jack2008TheAD} & 0.27 & 0.26 & 0.18 & 0.23 & \textbf{0.36}  \\ %[0.7ex]  
        & \texttt{HCP}~\cite{Essen2012TheHC} & 0.28 & 0.28 & 0.13 & 0.21 & \textbf{0.33} \\ %[-0.5ex] 
        & \texttt{ADHD200}~\cite{Brown2012ADHD200GC} & 0.23 & 0.25 & 0.15 & 0.23 & \textbf{0.34} \\ %[-0.5ex] 
        & \texttt{ADNI3}~\cite{Weiner2017TheAD} & 0.27 & 0.28 & 0.17 & 0.24 & \textbf{0.37} \\ %[-0.5ex] 
        & \texttt{AIBL}~\cite{Fowler2021FifteenYO} & 0.25 & 0.24 & 0.12 & 0.20 & \textbf{0.32} \\ [0.2ex] 
         \hline \\[-2.ex]

     \multicolumn{1}{c}{\multirow{1}{*}{\footnotesize{Stroke T1w}}} & \texttt{ATLAS}~\cite{Liew2017ATLAS} & 0.24 & 0.24 & 0.11 & 0.22 & \textbf{0.31}  \\ %[0.7ex]   
         %\hline %\\[-2.3ex]

\bottomrule  \\ [-3.6ex]  
    \end{tabular} 
}
\end{table}

%% file: sec/exp/fig/fig_ablat.tex
\begin{figure}[t]

\centering 

\resizebox{1\linewidth}{!}{
	\begin{tikzpicture}
        %\hspace*{-0.5cm}
        
		\tikzstyle{myarrows}=[line width=0.8mm,draw=blue!50,-triangle 45,postaction={draw, line width=0.05mm, shorten >=0.02mm, -}]
		\tikzstyle{mylines}=[line width=0.8mm]
  
		% Main size box
		%\draw[thin, color = black] (20, -8) -- (-2, -8) -- (-2, 3.42) -- (20, 3.42) -- (20, -8); 

	%%%%%%%%%%%%%%%%%%%%%%%%%%%%%%%%%%%%%%%%%%%%%%%%%

	\pgfmathsetmacro{\shift}{-3.2}
	\node at (5.2+\shift+0.1*3, 6.2+0.15*3, 0.75*3) {Input};
	%\node at (7.2+\shift+0.1*3, 6.6+0.15*3, 0.75*3) {\texttt{SynthSR}};
	\node at (7.2+\shift+0.1*3, 6.2+0.15*3, 0.75*3) {\texttt{UNA}-(\hyperlink{ablat: a}{\textit{a}})};
	\node at (9.2+\shift+0.1*3, 6.2+0.15*3, 0.75*3) {\texttt{UNA}-(\hyperlink{ablat: b}{\textit{b}})};
	\node at (11.2+\shift+0.1*3, 6.2+0.15*3, 0.75*3) {\texttt{UNA}-(\hyperlink{ablat: c}{\textit{c}})};
	\node at (13.2+\shift+0.1*3, 6.2+0.15*3, 0.75*3) {\texttt{UNA}-(\hyperlink{ablat: d}{\textit{d}})};
	\node at (15.2+\shift+0.1*3, 6.2+0.15*3, 0.75*3) {\texttt{UNA}};
	\node at (17.2+\shift+0.1*3, 6.2+0.15*3, 0.75*3) {Ground Truth};

	\node at (5.2+\shift+0.1*3, 5+0.15*3, 0.75*3) {\includegraphics[width=0.22\textwidth]{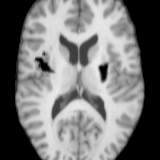}};
	\node at (7.2+\shift+0.1*3, 5+0.15*3, 0.75*3) {\includegraphics[width=0.22\textwidth]{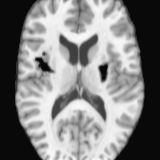}}; 
	\node at (9.2+\shift+0.1*3, 5+0.15*3, 0.75*3) {\includegraphics[width=0.22\textwidth]{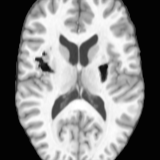}}; 
	\node at (11.2+\shift+0.1*3, 5+0.15*3, 0.75*3) {\includegraphics[width=0.22\textwidth]{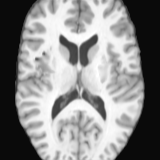}};
	\node at (13.2+\shift+0.1*3, 5+0.15*3, 0.75*3) {\includegraphics[width=0.22\textwidth]{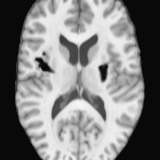}}; 
	\node at (15.2+\shift+0.1*3, 5+0.15*3, 0.75*3) {\includegraphics[width=0.22\textwidth]{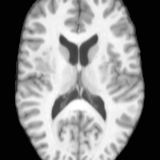}};
	\node at (17.2+\shift+0.1*3, 5+0.15*3, 0.75*3) {\includegraphics[width=0.22\textwidth]{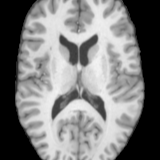}};

        \foreach \i in {5.2, 7.2, 9.2, 11.2, 13.2, 15.2, 17.2}
        {
        \draw[fill=none, draw=hc, line width=0.6mm] (\i+\shift+0.1*3-0.4, 5+0.15*3+0.2, 0.75*3) ellipse (2.5mm and 2.5mm);
        \draw[fill=none, draw=hc, line width=0.6mm] (\i+\shift+0.1*3+0.3, 5+0.15*3+0.1, 0.75*3) ellipse (2.5mm and 2.5mm);
        }

	%%%%%%%%%%%%%%%%%%%%%%%%%%%%%%%%%%%%%%%%%%%%%%%%%

	\end{tikzpicture}
	}   
    \vspace{-0.6cm}
\caption{Ablations on \texttt{UNA}'s healthy anatomy reconstruction.} %Models (\hyperlink{ablat: a}{\textit{a}})-(\hyperlink{ablat: d}{\textit{d}}) are described in \cref{exp: ablation}. The testing images are generated from real healthy T1w MRI encoded with randomly simulated pathology profiles. %The \textcolor{mint!200}{mint} circles highlight some details.} 

	 \label{fig: ablat} 
\end{figure}

%% file: sec/exp/fig/tab_ablat.tex
\begin{table}[t]
\caption{Ablation study on \texttt{UNA}. Testing images are real T1w MRI encoded with simulated pathology (same as first-row group in \cref{tab: synth}). %Details on models (\hyperlink{ablat: a}{\textit{a}})-(\hyperlink{ablat: d}{\textit{d}}) are described in \cref{exp: ablation}. 
(F: full brain; H: healthy region; D: diseased region.)} 
    \label{tab: ablat}
\resizebox{\linewidth}{!}{
\centering 
    \begin{tabular}{lccccccccc}
    %\begin{tabular}{cccccccccccc}
       \toprule \\[-3ex] 
      %\multicolumn{1}{c}{\multirow{2}{*}{\footnotesize\textbf{Modality}}} & 
      \multicolumn{1}{c}{\multirow{2}{*}{\footnotesize\textbf{Method}}} & \multicolumn{3}{c}{\multirow{1}{*}{\footnotesize\textbf{\texttt{L1} ($\downarrow$)}}} & \multicolumn{3}{c}{\multirow{1}{*}{\footnotesize\textbf{\texttt{PSNR} ($\uparrow$)}}} & \multicolumn{3}{c}{\multirow{1}{*}{\footnotesize\textbf{\texttt{SSIM}  ($\uparrow$)}}} \\ [-0.1ex]
        %\cmidrule(lr){2-3}% \\ [-4.1ex]
        \cmidrule(lr){2-4}
        \cmidrule(lr){5-7}
        \cmidrule(lr){8-10} \\ [-3.2ex]
          %& 
          & {\footnotesize{F}} & {\footnotesize{H}} & {\footnotesize{D}} & {\footnotesize{F}} & {\footnotesize{H}} & {\footnotesize{D}} & {\footnotesize{F}} & {\footnotesize{H}} & {\footnotesize{D}}  \\ [-0.4ex]
     \midrule\\[-2.5ex]

        %\multicolumn{1}{c}{\multirow{5}{*}{\thead{\footnotesize{T1w}\\MRI}}} & 
        \texttt{UNA}-(\hyperlink{ablat: a}{\textit{a}}) & 0.0229 & 0.0193 & 0.0008 & 23.71 & 25.09 & 38.92 & 0.859 & 0.890 & 0.904  \\ %[0.7ex]  
        %& 
        \texttt{UNA}-(\hyperlink{ablat: b}{\textit{b}}) & 0.0195 & 0.0182 & 0.0005 & 25.79 & 27.30 & 42.35 & 0.903 & 0.925 & 0.950  \\ %[-0.5ex] 
        %& 
        \texttt{UNA}-(\hyperlink{ablat: c}{\textit{c}}) & 0.0155 & 0.0163 & 0.0004 & 30.00 & 31.92 & 43.61 & 0.959 & 0.977 & 0.982  \\ %[-0.5ex] 
        %& 
        \texttt{UNA}-(\hyperlink{ablat: d}{\textit{d}}) & 0.0195 & 0.0182 & 0.0005 & 27.13 & 28.04 & 42.97 & 0.931 & 0.950 & 0.969  \\ %[-0.5ex] 
        %& 
        \texttt{UNA} & \textbf{0.0147} & \textbf{0.0143} & \textbf{0.0003} & \textbf{31.98} & \textbf{33.25} & \textbf{45.61} & \textbf{0.981} & \textbf{0.992} & \textbf{0.998} \\ %[-0.5ex] 
         %\hline %\\[-2.3ex] 

\bottomrule  \\ [-3.6ex]  
    \end{tabular} 
}
\end{table}

%% file: sec/con.tex
\vspace{-0.1cm}
\section{Limitations and Future Work}
\label{sec: future}
\vspace{-0.1cm}

\paragraph{Handling Extreme Cases.}~As discussed in \cref{exp: real}, \texttt{UNA} appears to ``over-correct'' its reconstructed healthy anatomy, especially in extreme cases where the pathology in the input image heavily \textit{occludes} the underlying anatomy. This issue will be further investigated in our future work.

\vspace{-0.45cm}
\paragraph{Broader Applications.}~By bridging the gap between healthy and diseased anatomy, \texttt{UNA} opens up a wide range of applications beyond anomaly detection. For example, it could enable modality-agnostic image registration in the presence of pathology, as well as stroke treatment outcome prediction based on \texttt{UNA}'s reconstructed healthy anatomy. We plan to further explore these applications of \texttt{UNA}.

\section{Conclusion}
\label{sec: con}
\vspace{-0.1cm}

We introduce \texttt{UNA}, a modality-agnostic model for reconstructing healthy anatomy that works both with healthy subjects and images with varying degrees of pathology. Our fluid-driven anomaly randomization approach enables the generalization of an unlimited number of anomaly profiles from just a few real pathology segmentations. \texttt{UNA} can be directly applied to real images containing pathologies without fine-tuning. We demonstrate \texttt{UNA}'s superior performance across eight public datasets, including MR and CT images from healthy subjects and stroke patients. Additionally, we showcase \texttt{UNA}'s direct applicability to anomaly detection tasks. By bridging the gap between different modalities and the underlying anatomy, as well as between healthy and diseased images, we believe \texttt{UNA} opens up exciting opportunities for general image analysis in clinical practice, particularly for images with diverse pathologies.

%% file: sec/appendix/main.tex
\clearpage
\onecolumn

\nolinenumbers

{\centering
\Large
\textbf{Unraveling Normal Anatomy via Fluid-Driven Anomaly Randomization} \\\vspace{0.05cm}{(Appendix)} \\
\vspace{1.5em}
}

\setcounter{table}{0}
\setcounter{figure}{0}
\setcounter{equation}{0}

\renewcommand\thetable{\thesection.\arabic{table}}
\renewcommand\thefigure{\thesection.\arabic{figure}}
\renewcommand\theequation{\thesection.\arabic{equation}}

\appendix

%\thispagestyle{empty} % no page number

%\twocolumn
\noindent This Appendix provides additional context regarding: 

\noindent\textbf{\ref{app: perlin}:} Computing Derivatives of Perlin Noise;

\noindent\textbf{\ref{app: dataset}:} Datasets and Metrics;

\noindent\textbf{\ref{app: implement}:} Implementation Details.

%%%%%%%%%%%%%%%%%%%%%%%%%%%

\input{sec/appendix/perlin}
\input{sec/appendix/dataset}
\input{sec/appendix/implement}

%% file: sec/appendix/perlin.tex
%{\hypertarget{app: perlin}{\color{white}xx}}
%\vspace{-0.4cm}
\section{Computing Derivatives of Perlin Noise}
\label{app: perlin}

Perlin noise is a gradient noise function invented by Ken Perlin~\cite{perlin2002improving}. Unlike traditional random noise, which produces entirely chaotic, discontinuous patterns, Perlin noise is smooth and continuous, it generates a field of smoothly varying values that appear random but maintain a continuous flow, without abrupt jumps or visible seams. These properties make Perlin noise ideal for generating natural-looking patterns that have rich textures. In \texttt{UNA}, we resort to Perlin noise for generating random shape profiles for anomaly probability initialization as well as the incompressible flow and non-negative diffusion fields in Sec.~\hyperlink{sec: fluid}{3}.

\subsection{Perlin Noise and Random Anomaly Initialization}

Here, we present the implementation details of our Perlin noise generation and thresholding for random anomaly shape synthesis. As shown in the code below, the generation of the random anomaly probability map can be summarized into six steps:
\begin{enumerate}
    \item Generate a grid of random gradients at lattice points (Line 25-30).
    \item Compute the relative position of the point inside the grid cell (Line 31-44).
    \item Calculate the dot product of the gradients and the relative position vectors (45-52).
    \item Apply the fade function to smooth the interpolation (Line 54-61).
    \item Interpolate between dot products to get a smooth value (Line 62-69). 
    \item Threshold to get a random shape of anomaly profile (Line 71-78). 
\end{enumerate}

%%%%%%%%%%%%%%%%%%%%%%%%%%%%%%%%%%%%%%%

\input{sec/appendix/fig/code_perlin}

%%%%%%%%%%%%%%%%%%%%%%%%%%%%%%%%%%%%%%%
%%%%%%%%%%%%%%%%%%%%%%%%%%%%%%%%%%%%%%%

\subsection{Flow and Diffusion Initialization}

As discussed in Sec.~\hyperlink{initialization}{3.1}, we further utilize Perlin noise for creating the random potentials $\boldsymbol{\Psi}$ for ${\mathbf{V}}$, and ${\Phi}$ for $D$. The random map of ${L}$ initialization, as a scalar field, could be directly obtained from the above function ``generate$\_$perlin$\_$noise$\_$3d''. Here, we show details on the implementation of $\boldsymbol{\Psi}$ initialization, which is a 3-dimensional vector field. As shown in the code below, the generation of the random incompressible flow fields can be summarized into three steps:
\begin{enumerate}
    \item Generate three individual Perlin noise maps for the potential ($\boldsymbol{\Psi}$) construction (Line 6-9).
    \item Reshape the noise map to match the current subject sample's patch size (Line 11-23).
    \item Surjectively map the random potential to its corresponding incompressible flow space via \cref{app: repre} (Line 25-28).
\end{enumerate}

\input{sec/appendix/fig/code_shape}

%% file: sec/appendix/fig/code_perlin.tex
\begin{lstlisting}[language=Python]
import os, time 
import numpy as np  

def interpolant(t):
    return t*t*t*(t*(t*6 - 15) + 10)

def generate_perlin_noise_3d(shape, res, tileable=(False, False, False), interpolant=interpolant, percentile=None,):
    """Generate a 3D numpy array of perlin noise.

    Args:
        shape: The shape of the generated array (tuple of three ints). This must be a multiple of res.
        res: The number of periods of noise to generate along each axis (tuple of three ints). Note shape must be a multiple of res.
        tileable: If the noise should be tileable along each axis (tuple of three bools). Defaults to (False, False, False).
        interpolant: The interpolation function, defaults to t*t*t*(t*(t*6 - 15) + 10).
        percentile: The percentile for random shape thresholding.

    Returns:
        A numpy array of shape with the generated noise.
        (Optional) A numpy array of thresholded noise given an input percentile.
    """
    seed = int(time.time())
    os.environ['PYTHONHASHSEED'] = str(seed)
    np.random.seed(seed) 

    # Initialize the grid
    delta = (res[0] / shape[0], res[1] / shape[1], res[2] / shape[2])
    d = (shape[0] // res[0], shape[1] // res[1], shape[2] // res[2])
    grid = np.mgrid[0:res[0]:delta[0],0:res[1]:delta[1],0:res[2]:delta[2]]
    grid = np.mgrid[0:res[0]:delta[0],0:res[1]:delta[1],0:res[2]:delta[2]]
    grid = grid.transpose(1, 2, 3, 0) % 1
    # Gradients
    theta = 2*np.pi*np.random.rand(res[0] + 1, res[1] + 1, res[2] + 1)
    phi = 2*np.pi*np.random.rand(res[0] + 1, res[1] + 1, res[2] + 1)
    gradients = np.stack(
        (np.sin(phi)*np.cos(theta), np.sin(phi)*np.sin(theta), np.cos(phi)),
        axis=3
    )
    if tileable[0]:
        gradients[-1,:,:] = gradients[0,:,:]
    if tileable[1]:
        gradients[:,-1,:] = gradients[:,0,:]
    if tileable[2]:
        gradients[:,:,-1] = gradients[:,:,0]
    gradients = gradients.repeat(d[0], 0).repeat(d[1], 1).repeat(d[2], 2)
    g000 = gradients[    :-d[0],    :-d[1],    :-d[2]]
    g100 = gradients[d[0]:     ,    :-d[1],    :-d[2]]
    g010 = gradients[    :-d[0],d[1]:     ,    :-d[2]]
    g110 = gradients[d[0]:     ,d[1]:     ,    :-d[2]]
    g001 = gradients[    :-d[0],    :-d[1],d[2]:     ]
    g101 = gradients[d[0]:     ,    :-d[1],d[2]:     ]
    g011 = gradients[    :-d[0],d[1]:     ,d[2]:     ]
    g111 = gradients[d[0]:     ,d[1]:     ,d[2]:     ]
    # Ramps 
    n000 = np.sum(np.stack((grid[:,:,:,0]  , grid[:,:,:,1]  , grid[:,:,:,2]  ), axis=3) * g000, 3)
    n100 = np.sum(np.stack((grid[:,:,:,0]-1, grid[:,:,:,1]  , grid[:,:,:,2]  ), axis=3) * g100, 3)
    n010 = np.sum(np.stack((grid[:,:,:,0]  , grid[:,:,:,1]-1, grid[:,:,:,2]  ), axis=3) * g010, 3)
    n110 = np.sum(np.stack((grid[:,:,:,0]-1, grid[:,:,:,1]-1, grid[:,:,:,2]  ), axis=3) * g110, 3)
    n001 = np.sum(np.stack((grid[:,:,:,0]  , grid[:,:,:,1]  , grid[:,:,:,2]-1), axis=3) * g001, 3)
    n101 = np.sum(np.stack((grid[:,:,:,0]-1, grid[:,:,:,1]  , grid[:,:,:,2]-1), axis=3) * g101, 3)
    n011 = np.sum(np.stack((grid[:,:,:,0]  , grid[:,:,:,1]-1, grid[:,:,:,2]-1), axis=3) * g011, 3)
    n111 = np.sum(np.stack((grid[:,:,:,0]-1, grid[:,:,:,1]-1, grid[:,:,:,2]-1), axis=3) * g111, 3)
    # Interpolation
    t = interpolant(grid)
    n00 = n000*(1-t[:,:,:,0]) + t[:,:,:,0]*n100
    n10 = n010*(1-t[:,:,:,0]) + t[:,:,:,0]*n110
    n01 = n001*(1-t[:,:,:,0]) + t[:,:,:,0]*n101
    n11 = n011*(1-t[:,:,:,0]) + t[:,:,:,0]*n111
    n0 = (1-t[:,:,:,1])*n00 + t[:,:,:,1]*n10
    n1 = (1-t[:,:,:,1])*n01 + t[:,:,:,1]*n11

    noise = ((1-t[:,:,:,2])*n0 + t[:,:,:,2]*n1)
    if percentile is None:
        return noise
    shres = np.percentile(noise, percentile) 
    mask = np.zeros_like(noise) 
    mask[noise >= shres] = 1.
    noise *= mask
    return noise, mask
\end{lstlisting}

%% file: sec/appendix/fig/code_shape.tex
\begin{lstlisting}[language=Python]
import torch 

def generate_velocity_3d(shape, perlin_res, V_multiplier, device, save_orig_for_visualize = False):  
    pad_shape = [ 200, 200, 200 ] 

    # Generate random potentials (back to original shape)
    curl_a = generate_perlin_noise_3d(pad_shape, perlin_res, tileable=(True, False, False)) 
    curl_b = generate_perlin_noise_3d(pad_shape, perlin_res, tileable=(True, False, False)) 
    curl_c = generate_perlin_noise_3d(pad_shape, perlin_res, tileable=(True, False, False)) 

    # Back to original shape
    curl_a = curl_a[(pad_shape[0] - shape[0]) // 2 : (pad_shape[0] - shape[0]) // 2 + shape[0], \
                    (pad_shape[1] - shape[1]) // 2 : (pad_shape[1] - shape[1]) // 2 + shape[1], \
                    (pad_shape[2] - shape[2]) // 2 : (pad_shape[2] - shape[2]) // 2 + shape[2]
                    ]
    curl_b = curl_b[(pad_shape[0] - shape[0]) // 2 : (pad_shape[0] - shape[0]) // 2 + shape[0], \
                    (pad_shape[1] - shape[1]) // 2 : (pad_shape[1] - shape[1]) // 2 + shape[1], \
                    (pad_shape[2] - shape[2]) // 2 : (pad_shape[2] - shape[2]) // 2 + shape[2]
                    ]
    curl_c = curl_c[(pad_shape[0] - shape[0]) // 2 : (pad_shape[0] - shape[0]) // 2 + shape[0], \
                    (pad_shape[1] - shape[1]) // 2 : (pad_shape[1] - shape[1]) // 2 + shape[1], \
                    (pad_shape[2] - shape[2]) // 2 : (pad_shape[2] - shape[2]) // 2 + shape[2]
                    ]

    # Surjective mapping to incompressible flow space
    Vx, Vy, Vz = stream_3D(torch.from_numpy(curl_a).to(device), 
                            torch.from_numpy(curl_b).to(device), 
                            torch.from_numpy(curl_c).to(device))
    
    return {'Vx': (Vx * V_multiplier), 'Vy': (Vy * V_multiplier).to(device), 'Vz': (Vz * V_multiplier)}
\end{lstlisting}

%% file: sec/appendix/dataset.tex
%{\hypertarget{app: dataset}{\color{white}xx}}
%\vspace{-0.4cm}
\section{Datasets and Metrics}
\label{app: dataset}

\subsection{Datasets and Preprocessing}
We test and compare \texttt{UNA} over various datasets including modalities of MR and CT, the MR images further contain T1-weighted, T2-weighted, and FLAIR (fluid-attenuated inversion recovery) images. 

\begin{itemize}
\item \texttt{ADNI}~\cite{Jack2008TheAD}: we use T1-weighted (2045 cases) MRI scans from the Alzheimer’s Disease Neuroimaging Initiative (ADNI). All scans are acquired at 1 $mm$ isotropic resolution from a wide array of scanners and protocols. The dataset contains aging subjects, some diagnosed with mild cognitive impairment (MCI) or Alzheimer's Disease (AD). Many subjects present strong atrophy patterns and white matter lesions.

\item \texttt{HCP}~\cite{Essen2012TheHC}: we use T1-weighted (897 cases) and T2-weighted (897 cases) MRI scans of young subjects from the Human Connectome Project, acquired at 0.7 mm resolution.

\item \texttt{ADNI3}~\cite{Weiner2017TheAD}: we use T1-weighted (331 cases) and FLAIR (331 cases) MRI scans from ADNI3, which continues the previously funded ADNI1, ADNI-GO, and ADNI2 studies to determine the relationships between the clinical, cognitive, imaging, genetic and biochemical biomarker characteristics of the entire spectrum of sporadic late-onset AD.

\item \texttt{ADHD200}~\cite{Brown2012ADHD200GC}: we use T1-weighted (961 cases) MRI scans from ADH200 Sample, which is a grassroots initiative dedicated to the understanding of the neural basis of Attention Deficit Hyperactivity Disorder (ADHD).

\item \texttt{AIBL}~\cite{Fowler2021FifteenYO}: we use T1-weighted (668 cases), T2-weighted (302 cases) and FLAIR (336 cases) MRI scans from The Australian Imaging, Biomarkers and Lifestyle (AIBL) Study, which is a study of cognitive impairment (MCI) and Alzheimer’s disease dementia.

\item \texttt{OASIS3}~\cite{LaMontagne2018OASIS3LN}: we use CT (885 cases) scans from OASIS3, which is a longitudinal neuroimaging, clinical, and cognitive dataset for normal aging and AD. For our experiments, we use CT and T1-weighted MRI pair with the earliest date, from each subject.

\item \texttt{ATLAS}~\cite{Liew2017ATLAS}: we use T1-weighted (655 cases) MRI scans and the provided gold-standard stroke lesion segmentations, from Anatomical Tracings of Lesions After Stroke (ATLAS), which is a study of subacute/chronic stroke.

\item \texttt{ISLES}~\cite{Hernandez2022ISLES} we use FLAIR (152 cases) MRI scans and the provided gold-standard stroke lesion segmentation, from ISLES 2022, which is a MICCAI challenge in 2022 for acute/subacute stroke lesion detection and segmentation.

\end{itemize}

Among the above eight datasets, \texttt{ADNI}~\cite{Jack2008TheAD}, \texttt{ADNI3}~\cite{Weiner2017TheAD}, \texttt{HCP}~\cite{Essen2012TheHC}, \texttt{ADHD200}~\cite{Brown2012ADHD200GC}, \texttt{AIBL}~\cite{Fowler2021FifteenYO}, \texttt{OASIS3}~\cite{LaMontagne2018OASIS3LN} contain subjects with healthy anatomy. \texttt{ATLAS}~\cite{Liew2017ATLAS}, \texttt{ISLES}~\cite{Hernandez2022ISLES}. \texttt{ATLAS} and \texttt{ISLES} include stroke patients, with gold-standard manual segmentations of stroke lesions provided in both datasets.

For all datasets, we skull-strip all the images using SynthStrip~\cite{Hoopes2022SynthStripSF}, and resample them to 1 $mm$ isotropic resolution. For all the modalities other than T1-weighted MRI, we use NiftyReg~\cite{Modat2010FastFD} rigid registration to register all images to their same-subject T1-weighted MRI counterparts. The brain segmentation label maps are obtained by performing SynthSeg~\cite{Billot2021SynthSegSO} on the T1-weighted MR images of all the subjects.

To ease the computation burden during training, the deformations between gold-standard pathology segmentation maps and all healthy training subjects are pre-computed via NiftyReg~\cite{Modat2010FastFD} during data pre-processing. During training, the randomly selected anomaly profiles for \texttt{UNA}'s anomaly randomization initialization are registered to the current training subject on the fly, using the pre-computed deformation fields.

%%%%%%%%%%%%%%%%%%%%%%%%%%%%%%%%%%%%%%%%  

\subsection{Contrast Synthesis and Contralateral-Paired Input}
\label{app: synthesis}

\input{sec/appendix/fig/tab_setup}

\texttt{UNA}'s synthetic generator uses brain segmentation labels from FreeSurfer~\cite{fischl2002freesurfer}, for random-modality generation. In this work, we use the segmentation maps of training subjects from all the healthy datasets (\texttt{ADNI}~\cite{Jack2008TheAD}, \texttt{ADNI3}~\cite{Weiner2017TheAD}, \texttt{HCP}~\cite{Essen2012TheHC}, \texttt{ADHD200}~\cite{Brown2012ADHD200GC}, \texttt{AIBL}~\cite{Fowler2021FifteenYO}, \texttt{OASIS3}~\cite{LaMontagne2018OASIS3LN}). We follow Brain-ID~\cite{Liu2024BrainID}'s mild-to-severe data corruption strategy for enriching the training sample variations in resolution, orientation, and external artifacts. In \cref{app-tab: setup}, we list the generator parameters for mild, medium, and severe data corruption levels, respectively. Note that for each level, the setup parameters control the corruption value ranges, since the simulation is randomized, there could still be mildly corrupted samples generated under the ``severe'' settings. In addition, the random deformation fields are independent of data corruption levels.

To ease the burden of computing the deformation between original and hemisphere-flipped images for our contralateral-pair input, we preprocess the correspondence from the flipped to the original image for each subject during pre-processing. Specifically, to reduce the effects of the pathological regions for registration, we first use SynthSR~\cite{Iglesias2023SynthSRAP} to estimate the T1-weighted counterpart of both the hemisphere-flipped image and the original image. Then, we use NiftyReg~\cite{Modat2010FastFD} to compute the deformation fields from the hemisphere-flipped image to the original image. During training, the flipped sample is first registered to the domain of the original image using pre-computed deformations, then the contra-lateral paired inputs undergo the same deformation augmentations simultaneously.

%%%%%%%%%%%%%%%%%%%%%%%%%%%%%%%%%%%%%%%% 

\subsection{Metrics}
We resort to various metrics for evaluating individual tasks across multiple aspects: 

\begin{itemize}
\item \texttt{L1}: the average $L1$ distance, it is used for the voxel-wise prediction correctness of anatomy reconstruction.

\item \texttt{PSNR}: the peak signal-to-noise ratio (PSNR) that indicates the fidelity of predictions. It is used in anatomy reconstruction.

\item \texttt{SSIM}: the structural similarity scores between the generated and real images. It is used in anatomy reconstruction evaluation,

\item \texttt{Dice}: the similarity score between predicted and ground truth segmentations, and it is used in anomaly detection evaluation.

\end{itemize}

%% file: sec/appendix/fig/tab_setup.tex
\begin{table}[t]
%\hspace*{-0.48cm} 
\resizebox{0.6\linewidth}{!}{
\centering 
    \begin{tabular}{lcccc}
       \toprule \\[-3ex] 
        \multirow{2}{*}{\textbf{Category}} & \multirow{2}{*}{\textbf{Param}} & \multicolumn{3}{c}{\textbf{Corruption Level}} \\ [-0.2ex]
        \cmidrule(lr){3-5}
        & & Mild & Medium & Severe \\ [-0.2ex]
        \midrule\\[-3ex]
     
        \multirow{6}{*}{\thead{Deformation}} & affine-rotation$_{max}$ & 15 & $=$ & $=$ \\ 
        & affine-shearing$_{max}$ & 0.2 & $=$ & $=$ \\ 
        & affine-scaling$_{max}$ & 0.2 & $=$ & $=$ \\
        & nonlinear-scale $\mu_{min}$ & 0.03 & $=$ & $=$ \\
        & nonlinear-scale $\mu_{max}$ & 0.06 & $=$ & $=$ \\
        & nonlinear-scale $\sigma_{max}$ & 4 & $=$ & $=$ \\
         \hline\\[-2.3ex]

        \multirow{2}{*}{\thead{Resolution}} & $p_{\text{low-field}}$ & 0.1 & 0.3 & 0.5 \\ 
        & $p_{\text{anisotropic}}$ & 0 & 0.1 & 0.25 \\
         \hline\\[-2.3ex]

        \multirow{4}{*}{\thead{Bias Field}} & $\mu_{min}$ & 0.01 & 0.02 & 0.02 \\ 
        & $\mu_{max}$ & 0.02 & 0.03 & 0.04 \\
        & $\sigma_{min}$ & 0.01 & 0.05 & 0.1 \\ 
        & $\sigma_{max}$ & 0.05 & 0.3 & 0.6 \\
         \hline\\[-2.3ex]

        \multirow{2}{*}{\thead{Noises}} & $\sigma_{min}$ & 0.01 & 0.5 & 5 \\ 
        & $\sigma_{max}$ & 1 & 5 & 15 \\
        
\bottomrule %\\[-3.5ex]
    \end{tabular}  
    \caption{\texttt{UNA} synthetic generator setups: mild, medium, and severe levels. $p$ denotes probability, $\mu$ and $\sigma$ refer to the mean and variance of the Gaussian distributions, respectively.} 
    \label{app-tab: setup}
}
\end{table}

%% file: sec/appendix/implement.tex
%{\hypertarget{app: implement}{\color{white}xx}}
%\vspace{-0.4cm}
\section{Implementation Details and Additional Experiments}
\label{app: implement}

\input{sec/appendix/fig/tab_param}

%%%%%%%%%%%%%%%%%%%%%%%%%%%%%%%%%%%%%%%%%%%

%\subsection{Model Architecture}
As a general learning framework, \texttt{UNA} can use any backbone to extract brain features. For fair comparisons, we adopt the same 3D UNet~\cite{Ronneberger2015UNetCN} as utilized in themodels~\cite{Iglesias2023SynthSRAP,Liu2024BrainID,liu2024pepsi} we compare with, with 64 feature channels in the last layer. A linear regression layer is added following the feature outputs for anatomy reconstruction.

\texttt{UNA} is trained on the combination of synthetic and real data, with a probability of 50\% and 50\%, respectively. The training sample images are sized at $160^3$, with a batch size of 4. We use the AdamW optimizer, beginning with a learning rate of $10^{-4}$ for the first 300,000 iterations, which is then reduced to $10^{-5}$ for the subsequent 100,000 iterations. The entire training process took approximately 14 days on an NVIDIA A100 GPU.

The additional attention parameter ($\lambda_p$ in Eq.~\hyperlink{eq: loss_recon}{7}) is set to 1 for healthy anatomy reconstruction in pathological regions. The intra-subject contrastive learning weight ($\lambda_{\text{contrast}}$ in Eq.~\hyperlink{eq: loss}{9}) is set to 2. 
\cref{app-tab: param} provide additional experiments on hyperparameter search of the anomaly attention weight ($\lambda_p$) and the intra-subject contrastive learning weight ($\lambda_{\text{contrast}}$). Specifically, we observe that greater attention to anomalies helps improve the reconstruction of pathology regions, yet it harms the overall performance of originally healthy tissues.

%%%%%%%%%%%%%%%%%%%%%%%%%%%%%%%%%%%%%%%%%%%

%% file: sec/appendix/fig/tab_param.tex
\begin{table}[t]
\caption{Hyperparameter search of \texttt{UNA}. Testing images are real T1w MRI encoded with simulated pathology (same as the first-row group in \cref{tab: synth}). (F: full brain; H: healthy region; D: diseased region.)
}
\label{app-tab: param}
\resizebox{\linewidth}{!}{
\centering 
    \begin{tabular}{lccccccccc}
    %\begin{tabular}{cccccccccccc}
       \toprule \\[-3ex] 
      %\multicolumn{1}{c}{\multirow{2}{*}{\footnotesize\textbf{Modality}}} & 
      \multicolumn{1}{c}{\multirow{2}{*}{\footnotesize\textbf{Method}}} & \multicolumn{3}{c}{\multirow{1}{*}{\footnotesize\textbf{\texttt{L1} ($\downarrow$)}}} & \multicolumn{3}{c}{\multirow{1}{*}{\footnotesize\textbf{\texttt{PSNR} ($\uparrow$)}}} & \multicolumn{3}{c}{\multirow{1}{*}{\footnotesize\textbf{\texttt{SSIM}  ($\uparrow$)}}} \\ [-0.1ex]
        %\cmidrule(lr){2-3}% \\ [-4.1ex]
        \cmidrule(lr){2-4}
        \cmidrule(lr){5-7}
        \cmidrule(lr){8-10} \\ [-3.2ex]
          %& 
          & {\footnotesize{F}} & {\footnotesize{H}} & {\footnotesize{D}} & {\footnotesize{F}} & {\footnotesize{H}} & {\footnotesize{D}} & {\footnotesize{F}} & {\footnotesize{H}} & {\footnotesize{D}}  \\ [-0.4ex]
     \midrule\\[-2.5ex]

        $\lambda_p = 0$ & 0.0165 & 0.0153 & 0.0004 & 28.62 & 29.94 & 42.03 & 0.959 & 0.970 & 0.982  \\ %[0.7ex]   
        $\lambda_p = 0.5$ & 0.0150 & 0.0147 & 0.0004 & 31.01 & 32.85 & 44.20 & 0.973 & 0.989 & 0.995  \\ %[-0.5ex]  
        \texttt{UNA} ($\lambda_p = 1$) & \textbf{0.0147} & \textbf{0.0143} & \textbf{0.0003} & \textbf{31.98} & \textbf{33.25} & 45.61 & \textbf{0.981} & \textbf{0.992} & \textbf{0.998} \\ %[-0.5ex] 
        $\lambda_p = 1.5$ & 0.0149 & 0.0150 & \textbf{0.0003} & 30.61 & 31.27 & 45.73 & 0.979 & 0.986 & \textbf{0.998} \\ %[-0.5ex]  
        $\lambda_p = 2$ & 0.0152 & 0.0152 & \textbf{0.0003} & 30.29 & 32.43 & \textbf{45.78} & 0.973 & 0.989 & 0.995  \\
        \hline \\[-2.3ex] 
        \hline \\[-2.3ex] 

        $\lambda_{\text{contrast}} = 0$ & 0.0195 & 0.0182 & 0.0005 & 27.13 & 28.04 & 42.97 & 0.931 & 0.950 & 0.969  \\ %[0.7ex]   
        $\lambda_{\text{contrast}} = 1$ & 0.0158 & 0.0163 & 0.0004 & 30.78 & 31.82 & 44.05 & 0.953 & 0.961 & 0.981  \\ %[-0.5ex]  
        \texttt{UNA} ($\lambda_{\text{contrast}} = 2$) & \textbf{0.0147} & \textbf{0.0143} & 0.0003 & \textbf{31.98} & \textbf{33.25} & 45.61 & \textbf{0.981} & \textbf{0.992} & \textbf{0.998} \\ %[-0.5ex] 
        $\lambda_{\text{contrast}} = 3$ & 0.0150 & 0.0155 & \textbf{0.0002} & 31.82 & 32.59 & \textbf{45.63} & 0.974 & 0.984 & 0.996  \\ %[-0.5ex]  
        $\lambda_{\text{contrast}} = 4$ & 0.0154 & 0.0156 & 0.0003 & 31.76 & 32.40 & \textbf{45.63} & 0.970 & 0.981 & 0.996  \\
        
\bottomrule  %\\ [-3.6ex]  
    \end{tabular} 
}
\end{table}